\documentclass[10pt,aps,prd,showpacs,twocolumn,superscriptaddress,longbibliography,nofootinbib]{revtex4-1}

\usepackage{epsfig}
\usepackage{graphicx}
\usepackage[normalem]{ulem}
\usepackage{amssymb}
\usepackage{amsmath}
\usepackage{amsfonts}
\usepackage{latexsym}
\usepackage{verbatim}
\usepackage{mathtools}
\usepackage{setspace}
\usepackage{slashed}
\usepackage[all]{xypic}
\usepackage{psfrag}
\usepackage{comment}
\usepackage{array}
\usepackage{amssymb}
\usepackage{amsmath}
\usepackage{amsthm}
\usepackage{graphicx}
\usepackage{float}
\usepackage{amsfonts,wasysym,epsfig,verbatim,subfigure,bm,mathrsfs,lipsum}

\begin{document}
	\title{Stationary models of magnetized viscous tori around a Schwarzschild black hole}

	\author{Sayantani Lahiri}
	\affiliation{Institut fur Theoretische Physik, Goethe Universit\"at Frankfurt, Max-von-Laue-Str.1, 60438 Frankfurt am Main, Germany}
	\affiliation{University of Bremen, Center of Applied Space Technology and Microgravity (ZARM), 28359 Bremen.}
		
	\author{Sergio Gimeno-Soler}
	\affiliation{Departamento de Astronomia y Astrof\'{\i}sica, Universitat de Val\`encia, Dr.~Moliner 50, 46100 - Burjassot, Spain}
	
	\author{Jos\'{e} A.\ Font}
	\affiliation{Departamento de Astronomia y Astrof\'{\i}sica, Universitat de Val\`encia, Dr.~Moliner 50, 46100 - Burjassot, Spain}
	\affiliation{Observatori Astron\`omic, Universitat de Val\`encia, Catedr\'atico Jos\'e Beltr\'an 2, 46980, Paterna, Spain}
	
	\author{Alejandro Mus Mej\'{i}as}
	\affiliation{Departamento de Astronomia y Astrof\'{\i}sica, Universitat de Val\`encia, Dr.~Moliner 50, 46100 - Burjassot, Spain}
	\affiliation{Observatori Astron\`omic, Universitat de Val\`encia, Catedr\'atico Jos\'e Beltr\'an 2, 46980, Paterna, Spain}
	
	\begin{abstract}
We present stationary solutions of magnetized, viscous thick accretion disks around a Schwarzschild black hole. We assume that the tori are not self-gravitating, are endowed with a toroidal magnetic field and obey a constant angular momentum law. Our study focuses on the role of the black hole curvature in the shear viscosity tensor and in their potential combined effect on the stationary solutions. Those are built in the framework of a  causality-preserving,  second-order gradient expansion scheme of relativistic hydrodynamics in the Eckart frame description  which gives rise to hyperbolic equations of motion. The stationary models are constructed by numerically solving the general relativistic momentum conservation equation using the method of characteristics.
We place constraints in the range of validity of the second-order transport coefficients of the theory.
Our results reveal that the effects of the shear viscosity and curvature are particularly noticeable only close to the cusp of the disks. The surfaces of constant pressure are affected by viscosity and curvature and the self-intersecting iscocontour -- the cusp --
moves to smaller radii (i.e.~towards  the black hole horizon) as the effects become more significant.  For highly magnetized disks the shift in the cusp location is smaller. Our findings might have implications on the dynamical stability of constant angular momentum tori which, in the inviscid case, are affected by the runaway instability.
\end{abstract}

\maketitle
	
\section{Introduction}

One of the outstanding predictions of general relativity is the existence of black holes.  By their very nature, black holes can only be observed by the gravitational effects they produce in their environment.  An accretion disk embedded in the geometry of a black hole provides a natural framework for its indirect detection through the study of the gravitational influence it exerts on the disk. As a result of the black hole's gravity, the mass of an orbiting disk is pulled inwards resulting into an inward flow of its matter and the outward transport of angular momentum, a process accompanied by the conversion of gravitational energy into radiation and heat. This is one of the most efficient processes of energy release in the cosmos and it operates in systems as diverse as  proto-planetary disks, X-ray binaries, gamma-ray bursts, active galactic nuclei, and quasars~\cite{2002apa..book.....F}. 

Models of accretion disks around black holes are abundant in the scientific literature (see~\cite{Abramowicz_2013} and references therein). Among the various proposals, geometrically thick disks or tori (also referred to as ``Polish doughnuts'') are the simplest, relativistic, stationary configurations describing an ideal fluid orbiting around a rotating black hole under the assumption that the specific angular momentum of the disk is constant~\cite{Fishbone:1976,Abramowicz:1978,Kozlowski:1978}. Extensions of the original model to incorporate additional effects such as non-constant distributions of angular momentum, magnetic fields, or self-gravity, have also been put forward~\cite{Font:2002,Daigne:2004,Ansorg:2005,Komissarov:2006,Montero:2007,Shibata:2007,Qian:2009,Stergioulas:2011,Gimeno_Soler_2017,Pinmentel:2018,Mach:2019}.
 
In all stationary models the accretion torus is assumed to be composed of an ideal fluid and the effects of dissipation are neglected. However, the contribution of dissipative fluxes might not exactly vanish in an accretion disk, especially if it undergoes differential rotation, thus giving rise to shear viscous effects. It is well known that viscosity and magnetic fields play a key role in accretion disks to account for angular momentum transport, in particular through the magneto-rotational instability~\cite{Balbus:1991}. In this paper we discuss stationary models of magnetized viscous tori, assuming a toroidal distribution of the field and the presence of shear stresses.
 
The conservation laws of relativistic hydrodynamics of a non-ideal fluid involving dissipative effects like viscosity, developed  by Landau-Lifschitz and Eckart, do not give rise to hyperbolic equations of motion~\cite{ROMATSCHKE_2010}. Moreover, the corresponding equilibrium states are unstable under linear perturbations~\cite{PhysRevD.31.725}. 
 The pathological nature of the conservation laws is attributed to the existence of first-order gradients of hydrodynamical variables in the dissipative flux quantities. This limitation can be circumvented by including second-order gradients, a formalism first developed by M\"uller~\cite{Muller:1967zza} in the non-relativistic setup and later extended by Israel and Stewart~\cite{Israel:1976tn} for relativistic non-ideal fluids. The resulting conservation laws are hyperbolic and stable~\cite{2013rehy.book.....R}. 
 
Assuming that the shear viscosity is small and instils perturbative effects in the disk fluid, stationary solutions of constant angular momentum {\it unmagnetized} tori in the Schwarszchild geometry were first presented in~\cite{lahiri2019toy}.
This work showed that stationary models of viscous thick disks can only be constructed in the context of general relativistic causal approach by using the gradient expansion scheme~\cite{Lahiri_2020}. The imprints of the shear viscosity and of the curvature of the Schwarzschild geometry are clearly present on the isopressure surfaces of the tori. In particular, the  location  of  the cusps  of  such  surfaces  is  different  from  those  predicted  with  an  ideal  fluid model~\cite{Font:2002}. In the present paper the purely hydrodynamical solutions presented in~\cite{lahiri2019toy} are extended  by incorporating toroidal magnetic fields in the stationary solutions of the tori. Our new solutions are  built using the second-order gradient expansion scheme in the Eckart frame description~\cite{Lahiri_2020}, which keeps the same spirit of the  Israel-Stewart formalism and gives rise to hyperbolic equations of motion, hence preserving causality. Furthermore, we also adopt the test-fluid approximation, neglecting the self-gravity of the disk. As we show below, in our formalism the general form of the shear viscosity tensor contains additional curvature terms (as one of many second-order gradients) and, as a result, the curvature of the Schwarzschild geometry directly influences the isopressure surfaces, as in the hydrodynamical case considered in~\cite{lahiri2019toy}. The presence and strength of a toroidal magnetic field brings forth some  quantitatative differences with respect to the unmagnetized case, as we discuss below.

The paper is organised as follows: Section~\ref{framework} presents the mathematical framework of our approach introducing, in particular, the perturbation equations that characterize the stationary solutions. Those solutions are built following the procedure described in Section~\ref{methodology}. Our results are discussed in Section~\ref{results}. Finally Section~\ref{summary} summarizes our findings. Throughout the paper we use natural units where $c=G=1$. Greek indices in mathematical quantities run from 0 to 4 and Latin indices are purely spatial.
	
\section{Framework}
\label{framework}

\subsection{Basic equations}

Our framework assumes that the spacetime geometry is that of a Schwarzschild black hole of mass $M$ and that the disk is not self-gravitating, it has a constant distribution of specific angular momentum, and that the magnetic field only has a toroidal component. We neglect possible effects of heat flow and bulk viscosity and we further assume that the shear viscosity is small enough so as to act as a  perturbation to the matter configuration. Therefore, the radial velocity of the flow vanishes and the fluid particles describe circular orbits.

The Schwarzchild spacetime is described by the metric
\begin{eqnarray}
ds^2=-\left(1-\frac{2M}{r}\right)dt^2 +\left(1-\frac{2M}{r}\right)^{-1} dr^2+r^2 d\Omega^2\,,  
\nonumber \\
\label{metric-1}
\end{eqnarray}
where $d\Omega^2=d\theta^2+ \sin^2\theta d\phi^2$. Since the fluid particles follow circular orbits their four-velocity $u^{\mu}$, subject to the normalization condition $u^{\alpha}u_{\alpha}=-1$,  is given by
\begin{eqnarray}
u^{\mu}=(u^{t},0,0,u^{\phi})\,,
\end{eqnarray}
where $u^{\mu} \equiv u^{\mu}(r,\theta)$, with $\mu=t,\phi$.  
The specific angular momentum $l$ and the angular velocity $\Omega$ are given by
\begin{eqnarray}
l(r,\theta)=-\frac{u_{\phi}}{u_{t}}, \quad  \Omega(r,\theta)=\frac{u^{\phi}}{u^{t}}\,, 
\end{eqnarray}
so that the following relationship holds between both quantities
\begin{eqnarray}
l(r,\theta)=-\frac{g_{\phi \phi}}{g_{tt}}\Omega(r,\theta)= \frac{ r \sin \theta}{(1-\frac{2M}{r})}\Omega(r,\theta)\,.
\end{eqnarray}
In our study we consider the Eckart frame for addressing viscous hydrodynamics which is a common choice of reference frame in relativistic astrophysics~\cite{ROMATSCHKE_2010}.  The energy-momentum tensor of viscous matter in the presence of a magnetic field is given by
\begin{eqnarray}
T^{\mu \nu}= (w+b^2)u^{\mu}u^{\nu}+\left(p+\frac{1}{2}b^2\right)g^{\mu\nu}-b^{\mu}b^{\nu}+\pi^{\mu \nu} \,.
\label{en-mom}
\end{eqnarray}
In this expression, the enthalpy density is given by $w=e +p$, where $p$ is the fluid pressure and $e$ is the total energy, and $\pi^{\mu \nu}$ is the shear viscosity tensor.   The dual of the Faraday tensor relative to an observer with four-velocity $u^{\mu}$ is \cite{anile2005relativistic},
  \begin{eqnarray}
  ^{*}F^{\mu \nu}= b^{\mu}u^{\nu}-b^{\nu}u^{\mu}\,,
  \end{eqnarray}
where $b^{\mu}$ is the magnetic field in that frame, which obeys the relation $b^2=b^{\alpha}b_{\alpha}$ and yields to the conservation law $\nabla_{\nu}\,^{*}F^{\mu \nu}=0$, where $\nabla_{\nu}$ is the covariant derivative. In the fluid frame $b^{\mu}=(0,{\bf B})$ where ${\bf B}$ denotes the three-vector of the magnetic field which satisfies the condition $u^{\alpha}b_{\alpha}=0$.
   Since the magnetic field distribution is purely toroidal, it follows that
   \begin{eqnarray}
   b^{r}=b^{\theta}=0, \qquad b^{\mu}=(b^{t},0,0,b^{\phi})\,.
   \end{eqnarray}
   From the condition $u^{\alpha}b_{\alpha}=0$ we obtain
   \begin{eqnarray}
   b^{t} = lb^{\phi}, \qquad
   b_{t} = -\Omega b_{\phi} \,,
   \end{eqnarray}
   and
   \begin{eqnarray}
   b^2= (1-\Omega l)b^{\phi}b_{\phi}= 2p_{m } \label{pm}\,,
   \end{eqnarray}
   where the magnetic pressure is defined as $p_{m}\equiv b^2/2$. 

As mentioned before we consider a second-order theory of viscous hydrodynamics constructed using the gradient expansion scheme which ensures the causality of propagation speeds in the Eckart frame. In this scheme the shear viscosity tensor is  expressed in terms of a causality-preserving term and additional curvature terms which will help investigate the influence of curvature contributions on our system. As a result, the general form of the shear viscosity tensor can be expressed as~\cite{Lahiri_2020},
\begin{eqnarray}
\pi^{\mu \nu}&=&-2\eta \sigma^{\mu \nu}-\tau_2 ^{<}D(-2\eta \sigma^{\mu\nu})^{>} +\kappa_1 R^{<\mu \nu>} 
 \nonumber \\
 && +\kappa_2 u_{\alpha} u_{\beta} R^{\alpha <\mu \nu >\beta}  \,,
 \label{viscosity}
\end{eqnarray} 
with the definition $D\equiv u^{\alpha}\nabla_{\alpha}$.  Here
$R^{\alpha\beta\gamma\delta}$ and $R^{\alpha\beta}$ are the Riemann tensor and the Ricci tensor, respectively, 
$\eta$ is the shear viscosity coefficient and $\tau_2$, $\kappa_1$ and $\kappa_2$ are the second-order transport coefficients. Moreover, the angular brackets in the previous equation indicate traceless  symmetric  combinations. 
The remaining quantities appearing in Eq.~(\ref{viscosity}) are defined as
\begin{small}
\begin{eqnarray}
\sigma^{\mu \nu}&=&\triangle^{\mu \alpha}\triangle^{\nu \beta}\left(\displaystyle \frac{\nabla_{\alpha}u_{\beta}+\nabla_{\beta}u_{\alpha}}{2}\right)-\displaystyle \frac{1}{3} \triangle^{\mu \nu} \triangle^{\alpha \beta} \nabla_{\alpha}u_{\beta} \,, \nonumber \\
^{<}D\sigma^{\mu\nu}\,^{>}&=&\triangle^{\mu \alpha}\triangle^{\nu \beta}\left(\displaystyle \frac{D\sigma_{\alpha \beta}+D\sigma_{\beta \alpha}}{2}\right)\displaystyle -\frac{1}{3} \triangle^{\mu \nu} \triangle^{\alpha \beta}D\sigma_{\alpha \beta} \,, \nonumber \\ 
R^{<\mu \nu>}&=&\triangle^{\mu \alpha}\triangle^{\nu \beta}\left(\displaystyle \frac{R_{\alpha \beta}+R_{\beta \alpha}}{2}\right)\displaystyle -\frac{1}{3} \triangle^{\mu \nu} \triangle^{\alpha \beta}R_{\alpha \beta} \,, \nonumber \\ 
R^{\alpha <\mu \nu >\beta}&=&\left[\triangle^{\mu \rho} \triangle^{\nu \sigma}\left(\displaystyle \frac{  R^{\alpha}_{\;\rho \sigma \gamma}+ R^{\alpha}_{\; \sigma \rho \gamma}}{2}\right)-\frac{1}{3} \triangle^{\mu \nu} \triangle^{\rho \sigma}  R^{\alpha}_{\;\rho \sigma \gamma}\right]g^{\beta \gamma} \,, \nonumber
\end{eqnarray}
\end{small}
\noindent
where the projection tensor is given by $\triangle^{\mu \nu}=g^{\mu \nu}+u^{\mu}u^{\nu}$.
Using Eq.~(\ref{en-mom}), the momentum conservation equation $\triangle_{\mu \nu} \nabla_{\lambda}T^{\lambda \nu}=0 $ can be written as
\begin{eqnarray}
(e+p)a_{\mu} &+& \triangle_{\mu}^ {\rho}\nabla_{\rho}\,p +\frac{\partial_{\mu}({\cal{L}}b^2)}{2{\cal{L}}} +g_{\mu \rho}\pi^{\rho \nu}a_{\nu} 
\nonumber \\
&+& \triangle_{\mu \gamma} \triangle_{\kappa \tau}\nabla^{\tau}\pi^{\gamma \kappa} = 0   \,,
\label{2}
\end{eqnarray} 
which is the general form of the momentum conservation equation in the presence of a magnetic field.  The four-acceleration is given by $a^{\mu}=u^{\rho}\nabla_{\rho}u^{\mu}=Du^{\mu}$ and ${\cal{L}}\equiv -g_{tt}g_{\phi \phi}$. 

\subsection{Perturbation of the magnetized torus}

Since we consider disks with constant specific angular momentum distributions we take $l(r)\equiv l_0$. We further assume that the internal energy density, $\varepsilon$, is very small and, therefore, the total energy is approximately equal to the rest-mass density i,e. $e=\rho(1+\varepsilon) \approx \rho$. For the Schwarzschild black hole, the term $R^{<\mu \nu>}=0$ and therefore it does not contribute to the shear viscosity tensor. 
We also assume that the shear viscosity is small in the sense that the coefficients $\eta$ and $\kappa_2$ can be considered as perturbations in the disk fluid.
These two coefficients will be assumed to be constant and to act as perturbations with the perturbation parameter $\lambda$ as follows,
\begin{equation}
\eta= \lambda m_1, \qquad \qquad \kappa_2=\lambda m_2\,.  \label{transport}
\end{equation}
The shear viscosity perturbation in the disk fluid generates linear perturbations in the energy density, pressure, and  magnetic field. Up to linear order, we can express each of these quantities as follows:
\begin{eqnarray}
e(r,\theta)&=& e_{(0)}(r,\theta)+\lambda  e_{(1)}(r,\theta) \,, \\
p(r,\theta)&=& p_{(0)}(r,\theta)+\lambda  p_{(1)}(r,\theta) \label{perturb-1} \,, \\
b^{t}(r,\theta)&=&b^{t}_{(0)}(r,\theta) + \lambda b^{t}_{(1)}(r, \theta) \,, \\
b^{\phi}(r,\theta)&=&b^{\phi}_{(0)}(r,\theta) + \lambda b^{\phi}_{(1)}(r,\theta) \,,  \label{perturb-2}
\end{eqnarray}
where, as usual, index $(0)$ denotes background quantities and index $(1)$ quantities at linear perturbation order.
By using Eqs.~(\ref{pm}) and (\ref{perturb-2}) the magnetic pressure at both zeroth order and first order  reads
\begin{eqnarray}
p_{m}^{(0)}&=&\displaystyle \frac{1}{2}(1-\Omega l) \,b^{\phi}_{(0)}b_{(\phi)}^{(0)} \,, \\
p_{m}^{(1)}&=&\displaystyle \frac{1}{2} \left[b^{\phi}_{(0)}\left(l b_{t}^{(1)}+b_{\phi}^{(1)}\right) +b_{\phi}^{(0)}\left(b^{\phi}_{(1)}-\Omega b^{t}_{(1)}\right)\right] \,.
\end{eqnarray} 
Defining the magnetization parameter as $\beta_{m}\equiv {p}/{p_{m}}$, the zeroth-order and first-order changes in this parameter can be written as follows
\begin{eqnarray}
\beta_{m}^{(0)}=  \displaystyle \frac{p_{(0)}}{p_{m}^{(0)}}\,,
\end{eqnarray}
and
\begin{eqnarray}  
 \beta_{m}^{(1)}&=&\displaystyle \frac{p_{(1)}}{p_{m}^{(0)}}-\beta_{m}^{(0)} \, \frac{p_{m}^{(1)}}{p_{m}^{(0)}} \,.
 \label{beta1}
\end{eqnarray}
From the momentum conservation equation (\ref{2}) we see that there are four unknown quantities to be determined, namely, $p_{(1)}, e_{(1)}, b_{t}^{(1)}$ and $b_{\phi}^{(1)}$. 
However, the variables $p_{(1)}$ and $e_{(1)}$ are not independent under the assumption of a barotropic equation of state. Following  \cite{Komissarov:2006, Gimeno_Soler_2017}  we take the same polytropic index $\gamma$ for the equations of state corresponding to both the fluid pressure $p$ and the magnetic pressure $p_{m}$, given by,
\begin{eqnarray}
p = K e^{\gamma}, \label{eos-1}
\end{eqnarray}
and
\begin{eqnarray}
 p_{m} = K_m \mathcal{L}^{\gamma-1} e^{\gamma}. 
 \label{eos-2}
\end{eqnarray}
 Now, expanding up to linear order one can write the equations of state at zeroth order and first order as
\begin{eqnarray}
p_{(0)}&=&K e_{(0)}^{\gamma}, \qquad \qquad p_{(1)}=\gamma K e_{(0)}^{\gamma-1} e_{(1)}\,,\\
p^{(0)}_m&=&K_m \mathcal{L}^{\gamma-1} e_{(0)}^{\gamma}, \quad p^{(1)}_m=\gamma K_m \mathcal{L}^{\gamma-1} e_{(0)}^{\gamma-1} e_{(1)}\,.
\end{eqnarray}
From the above relations, we find that $p_{(0)}$, $p_{(1)}$, $p^{(0)}_m$ and $p^{(1)}_m$ are related by
\begin{eqnarray}
p_{(1)}= p_{(0)} \frac{\gamma e_{(1)}}{e_{(0)}}\,, \label{eq:p1}\\
p^{(1)}_m= p^{(0)}_m \frac{\gamma e_{(1)}}{e_{(0)}}\label{eq:p1m}.
\end{eqnarray}
Using the last two equations we obtain the following condition
\begin{eqnarray} 
\displaystyle\frac{p_{(1)}}{p_{m}^{(1)}} =\frac{p_{(0)}}{p_{m}^{(0)}}\,.
\end{eqnarray} 
Substituting Eqs.(\ref{eq:p1}) and (\ref{eq:p1m}) in Eq.~(\ref{beta1}) leads to 
\begin{eqnarray}
\beta_{m}^{(1)} &=& \frac{\gamma p_{(0)} e_{(1)}}{e_{(0)} p_{m}^{(0)}} - \beta_{m}^{(0)} \frac{\gamma e_{(1)}}{e_{(0)}} \nonumber \\
&=& \beta_{m}^{(0)} \frac{\gamma e_{(1)}}{e_{(0)}} - \beta_{m}^{(0)} \frac{\gamma e_{(1)}}{e_{(0)}} = 0\,,  \label{eq-1}
\end{eqnarray}
which shows that the linear corrections $p_{(1)}$ and $p^{(1)}_m$ do not affect the value of the magnetization parameter in the disk. 

Moreover, from the orthogonality relation $u^{\alpha}b_{\alpha} = 0$ we obtain
\begin{eqnarray}
b^{t}_{(0)}= l b^{\phi}_{(0)},  \label{bphi}\qquad
b^{t}_{(1)}= l b^{\phi}_{(1)}\label{bt}\,,
\end{eqnarray}
which imply that $b^{\phi}_{(1)}$ and  $b^{t}_{(1)}$ are not independent variables. Using the relations, $b^2=(1-l\Omega)b^{\phi}b_{\phi}$ and $b^2=2p_{m}$, the zeroth-order and first-order terms of the magnetic field read
\begin{eqnarray}
b^{\phi}_{(0)}&=&\sqrt{\frac{2\beta_{m}}{p_{(0)}(1-l\Omega)g_{\phi \phi}}} \,,
\\
 b^{\phi}_{(1)} &=& \frac{p_{(1)}}{\beta_{m}} \sqrt{\frac{p_{(0)}}{2 \beta_{m} (1-\Omega l) g_{\phi \phi}}} \,,
 \label{eq-3}
\end{eqnarray}
where we have also used Eq.~(\ref{bphi}).
Hence, the variables $p_m^{(1)}, b^{t}_{(1)}$ and $b^{\phi}_{(1)}$ are all related to $p_{(1)}$. 
The pressure correction $p_{(1)}$ is determined by solving the momentum conservation law given by Eq.~(\ref{2}) with a constant angular momentum distribution $l=l_{0}$. Using Eq.~(\ref{eq-3}) and expanding Eq.~(\ref{2}) up to linear order in the variables $p_{(1)}$, $e_{(1)}$ and $b_{(1)}$, the fluid pressure correction equation can be expressed as
\begin{eqnarray}
(e_{(1)}+p_{(1)})a_{\mu} + \triangle_{\mu}^{\rho} \nabla_{\rho}p_{(1)} \,+\,\frac{\partial_{\mu}\left[\frac{{\cal L}}{\beta_{m}}p_{(1)}\right]}{{\cal L}} \nonumber \\[2mm]
+g_{\mu \rho}\pi^{\rho \nu}a_{\nu} 
\,+\, \triangle_{\mu \gamma} \triangle_{\kappa \tau}\nabla^{\tau}\pi^{\gamma \kappa}&=& 0 \,,
\label{4}
\end{eqnarray}
where $e_{(1)}$ is related to $p_{(1)}$  by Eq.~(\ref{eq:p1}). 
Once $p_{(1)}$ is determined by solving the above equation, we can also determine the impact of the shear viscosity on the magnetic pressure $p_{m}^{(1)}$ through Eq.~(\ref{eq:p1m}). 

Let us now for simplicity take the black hole mass $M=1$ in the rest of our calculations. Both the temporal and azimuthal components of Eq.~(\ref{4}) lead to
\begin{eqnarray}
\frac{2\eta  l_0 r   \left[ r (r-3) \sin ^4 \theta -l_{0}^2 (1-2 /r)^2 (r-3\sin^2 \theta )\right]}{\sin^6 \theta \sqrt{r-2}\left(r^3 + l_{0}^2 (2-r) \csc ^2\theta \right)^{5/2}}=0\,. \nonumber
\end{eqnarray}
For $\eta \neq 0$, the above equation in the equatorial plane reduces to
\begin{equation}
r^3-l_{0}^2(r-2)^2=0\,.
\end{equation}
Correspondingly, the radial and angular components of (\ref{4}) are, respectively,
\begin{widetext}	
	\begin{flalign}
	&\frac{(\tau_2 m_1) l_{0}^2(r-3) }{2 r^2 \sin^6 \theta \left(r^3+l_{0}^2(2-r)  \csc ^2\theta \right)^3}
	\left[r^3  \cos 4 \theta(10 r-21)  +\cos 2 \theta  \left\lbrace 4 r^3 (2r^2 -14r
	+21)-8 l_{0}^2(r-3) (r-2)\right\rbrace \right.  \nonumber \\
	&  \hspace{165pt} \left.    -r^3(2 r-7) (4 r-9) -8 l_{0}^2 (r-2) (r^2- 3r +3)\right]\nonumber \\[5pt]
	& + m_2 \frac{3 r^6+  r^6 \cos 4 \theta +2 r^3 \cos 2 \theta \left\lbrace l_{0}^2(r-2) (5 r-14)-2 r^3\right\rbrace -2 r^3 l_{0}^2(r-2) (5 r-14)
		-4l_{0}^4 (r-2)^3 }{4r^5 \sin^4 \theta \left(r^3 + l_{0}^2 (2-r)\csc^2 \theta \right)^2}\nonumber \\[5pt]
	&+\left(1+\frac{1}{\beta_{m}}\right)\frac{ (r-2 )}{r} \frac{\partial p_{(1)}}{\partial r} + \left[\frac{2(r-1)}{r^2 \beta_{m}}+	\frac{\left\lbrace r^3 - l_{0}^2 (2-r)^2 \csc^2 \theta \right\rbrace \left(\gamma  K +  e_{(0)}^{1-\gamma }\right)}{\gamma  K r^2 \left(r^3 + l_{0}^2 (2-r)\csc^2 \theta \right) } \right] p_{(1)}\quad = \quad 0\,,
	\label{NSE-radial}
	\end{flalign}
	
	\begin{flalign}
	&\tau_2 m_1\frac{ 4 l_{0}^2 \cot \theta  \left\lbrace r^3(4 r-9) \sin^2 \theta +(r-2) \left(2 l_{0}^2 (4 r-9)-r l_{0}^2(r-2) \csc ^2 \theta -2r^4 \right)\right\rbrace }{r^2 \sin^4 \theta \left(r^3+l_{0}^2(2-r) \csc ^2 \theta \right)^3} \nonumber\\[5pt]
	&+ \,m_2 \frac{2 l_{0}^2(r-2) \cot \theta  \left(2r^3 \sin^2 \theta +l_{0}^2(r-2)\right)}{r^3 \sin^4 \theta \left(r^3+l_{0}^2 (2-r)\csc^2 \theta\right)^2} 
	+\left(1+\frac{1}{\beta_{m}}\right) \frac{\partial p_{(1)}}{\partial  \theta}   \nonumber \\[5pt]
	&+\left[ \frac{2 \cot \theta}{\beta_{m}}-\frac{(r-2) l_{0}^2 \cot \theta \left(\gamma  K +  e_{(0)}^{1-\gamma }\right)}{\gamma  K \sin^2 \theta \left(r^3 + l_{0}^2(2-r) \csc^2 \theta \right)} \right] p_{(1)}\quad = \quad 0\,.
	\label{NSE-angular}
	\end{flalign}
\end{widetext}
In the limit $\beta_{m} \rightarrow \infty$, 
Eqs.~(\ref{NSE-radial}) and (\ref{NSE-angular}) reduce to the corresponding equations obtained in~\cite{lahiri2019toy} for a purely hydrodynamical viscous thick disk. 
Substituting $p_{(1)}$ from Eq.~(\ref{NSE-angular}) in Eq.~(\ref{NSE-radial}) we obtain the following equation:
\begin{widetext}
	\begin{eqnarray}
	&2l_{0}^2\cot \theta \left( \tilde{A}+\displaystyle\frac{\tilde{B}k_1}{C}\right)(\tau_2 m_1)+\displaystyle\frac{1}{2}\cot\theta \left(\tilde{f}_1-\frac{\tilde{f}_2 k_1}{C}\right)m_{2}\nonumber \\[2pt]
	+& \sin^6 \theta \left[r^3+l_{0}^2(2-r)\csc^2\theta \right]^3 \displaystyle\frac{(1+\beta_m)}{\beta_{m}}\left[r(r-2)\cot \theta \displaystyle\frac{\partial p_{(1)}}{\partial  r}- \frac{4 k_1}{C}\displaystyle\frac{\partial p_{(1)}}{\partial  \theta} \right] =0 \,, \label{eq:PDE_fin}
	\end{eqnarray} 
	with the definitions
	\begin{flalign*}
	&\tilde{A}=-2(r-3)\left[ r^3(10r-21)\cos 4\theta+\cos 2\theta\left\lbrace 4r^3(2r^2-14r+21)-8l_{0}^2(r-3)(r-2)\right\rbrace \right.\nonumber \\ 
	& \qquad \left. -r^3(2r-7)(4r-9)-8l_{0}^2(r-2)(r^2-3r+3)\right]\,,\nonumber \\[2pt]
	&\tilde{B}= 4 \left[2rl_{0}^2(r-2)^2-\sin^2 \theta \left\lbrace r^3(2r-3)+4l_{0}^2(4r-9)(r-2)-r^3\cos 2\theta (4r-9)\right\rbrace \right]\,, \nonumber \\[2pt]
	& k_1 = \left[-2 \beta_{m} \sin^2 \theta \left(r^3+l_{0}^2(2-r)\csc^2\theta \right)e_{(0)}+K\gamma e_{(0)}^{\gamma}\left\lbrace 8l_{0}^2(1+\beta_m) \right.\right.\nonumber \\[2pt]
	& \qquad \left. \left.+ 2r^2l_{0}^2(\beta_{m}+2)-2r^4-r^3(\beta_m-2)-4l_{0}^2r(3+\beta_{m})+r^3 \cos 2 \theta(2r+\beta_m-2)\right\rbrace  \right]\,,\nonumber \\[3pt]
	& C=(r-2)l_{0}^2\beta_m e_{(0)}+K\gamma e_{(0)}^{\gamma} \left[l_{0}^3(r-2)(2+\beta_m)-2r^3\sin^2 \theta\right]\,, \\[2pt]
	& \tilde{f}_1= \left[10r^6-6r^3l_{0}^2(5r-12)(r-2)+32l_{0}^4(r-3)(r-2)^2+16rl_{0}^6\left(1-\frac{2}{r}\right)^4 \right. \nonumber \\
	& \qquad \left.+\cos 2 \theta \left\lbrace 8(5r-12)r^3l_{0}^2-15r^6-32l_{0}^4(r-3)(r-2)^2 \right\rbrace \right ]\,,\\[2pt]
	& \tilde{f}_2= 4r^3 \left(1-\frac{2}{r}\right)\left[3r^6-2r^3 l_{0}^2 (r-2)+4l_{0}^4(r-2)^2+2r^3 \cos 2\theta(2r^3-(r-2)l_{0}^2)+r^6 \cos 4\theta \right]\,.
	\end{flalign*}
\end{widetext}

We must solve Eq.~(\ref{eq:PDE_fin}) once  the values of the parameters $m_1, \tau_2, l_{0}$ and $\beta_{m}$ are selected and using the appropriate boundary conditions. 
Eq.~(\ref{eq-1}) shows that  $\beta_m^{(1)}=0$. Therefore, the magnetization parameter can be completely expressed in terms of the zeroth-order magnetic pressure and fluid pressure, and is given by  $\beta_{m}(r, \theta)={p_{(0)}}/{p_{m}^{(0)}}$. Using Eqs.~(\ref{eos-1}) and (\ref{eos-2}) we can further express
 \begin{eqnarray}
 \beta_m(r,\theta)=\frac{K}{K_m \mathcal{L}^{\gamma-1}(r,\theta)}\,.
 \end{eqnarray} 
 In addition, we can define the magnetization parameter at the center of the disk as $\beta_{m,c} \equiv \beta_{m}(r_c, \pi/2)$
 and write it as
 \begin{equation}
 \beta_{m,c}=\frac{K}{K_m \mathcal{L}^{\gamma-1}(r_c,\pi/2)}\,.
 \end{equation}
 Then, the magnetization parameter can be  expressed as
 \begin{equation}
 \beta_{m}(r,\theta)=\beta_{m,c} \left(\frac{\mathcal{L}(r_c, \pi/2)}{\mathcal{L}(r, \theta)}\right)^{\gamma-1}\,,
 \end{equation}
 which, for the Schwarzschild metric, reads
 \begin{equation}
 \beta_{m}(r,\theta)=\beta_{m,c} \left(\frac{r_c(r_c-2)}{r(r-2)\sin^2 \theta}\right)^{\gamma-1}\,,
 \end{equation} 
 where $r_c$, $\beta_{m,c}$ and $K$ are constant parameters. Let us compute $r_c$ for a given angular momentum $l_0$. This can be determined by finding the extrema of the effective (gravitational plus centrifugal) potential $W$, as the center of the disk is located at a minimum of the potential (see, e.g.~\cite{Font:2002} for details).
In the Schwarzschild geometry, the total potential $W(r,\theta)$ for constant angular momentum distributions is be defined as,
\begin{equation}
    W(r,\theta) = \frac{1}{2} \ln \frac{r^2(r-2)\sin^2\theta}{r^3\sin^2\theta-l_0^2(r-2)}\,.
\end{equation}
At the equatorial plane, taking $\partial_{r}W = 0$ leads, after some algebra, to
 \begin{equation}
 r^3-l_0^2(r-2)^2=0\,.
 \end{equation}
 The largest root of the above equation corresponds to the disk center, $r_c$.
 In the absence of dissipative terms, the relativistic momentum conservation equation, with our choices of equation of state, can be expressed as follows \cite{Gimeno_Soler_2017}
 \begin{eqnarray}
 W-W_{\mathrm{s}}+\frac{\gamma}{\gamma-1}\left(\frac{p_{(0)}}{e_{(0)}}+\frac{p_m^{(0)}}{e_{(0)}}\right)=0\,,
 \end{eqnarray}
 which can further be rewritten as
 \begin{equation}
 W-W_{\mathrm{s}}+\frac{\gamma K e_{(0)}^{\gamma-1}}{\gamma-1}\left(1+\frac{1}{\beta_m(r,\theta)}\right)\,,
 \end{equation}
where $W_{\mathrm{s}}$ is the potential at the surface of the disk, i.e.~the surface for which $p_{(0)} = p_m^{(0)} = e_{(0)}= 0$. From the above expression, the zeroth-order energy density can be obtained and it reads as
 \begin{equation}
e_{(0)}=\left(\frac{1}{K}\right)^{\frac{1}{\gamma-1}}\left(\frac{\gamma(1+\beta_{m}(r,\theta))}{(1-\gamma)\beta_{m}(r,\theta)(W-W_{\rm s})}\right)^{\frac{1}{1-\gamma}}\,,
\end{equation}
and the zeroth-order pressure, in terms of $\beta_{m,c}, r_{c}$ and $W_{\rm s}$,  becomes
\begin{small}
\begin{equation}
p_{(0)}=K^{\frac{1}{\gamma-1}}\left(\frac{\gamma\left(1+\beta_{m,c}\left(\frac{r_c(r_c-2)}{r(r-2)\sin^2\theta}\right)^{\gamma-1}\right)}{(1-\gamma)\beta_{m,c}\left(\frac{r_c(r_c-2)}{r(r-2)\sin^2\theta}\right)^{\gamma-1}(W-W_{\rm s})}\right)^{\frac{\gamma}{1-\gamma}}\,,
\end{equation}
\end{small}
which corresponds to the fluid pressure of the magnetized ideal fluid.
From this equation it follows that for the term inside the parenthesis to be positive, we require that $W-W_{\mathrm{s}}<0$. On the contrary, if $W-W_{\mathrm{s}} > 0$, the pressure (and the
energy density) should vanish, which indicates regions outside the disk.

\begin{table*}[h!]
\caption{Location of $r_{\mathrm{cusp}}$ and the magnitudes of pressure $p_{(0),\mathrm{cusp}}$ at $r_{\mathrm{cusp}}$ in an ideal fluid magnetized disk with two different choices of magnetization parameter $\beta_{m,c}$. }   
\begin{tabular}{c | c c | c c | c c}
\hline\hline  
  & \multicolumn{2}{c}{$W_{\mathrm{s}} = -0.039$} & \multicolumn{2}{c}{$W_{\mathrm{s}} = -0.040$} & \multicolumn{2}{c}{$W_{\mathrm{s}} = -0.041$}\\ 
\hline
$\beta_{\mathrm{m,c}}$ & $r_{\mathrm{cusp}}$ & $p_{(0), \mathrm{cusp}}$ & $r_{\mathrm{cusp}}$ & $p_{(0), \mathrm{cusp}}$ & $r_{\mathrm{cusp}}$ & $p_{(0), \mathrm{cusp}}$ \\
\hline        
$10^3$ & $4.576$ & $1.041 \times 10^{-4}$ & $4.576$ & $3.556 \times 10^{-5}$ & $4.576$ & $3.700 \times 10^{-6}$\\ 
$10^{-3}$ & $4.644$ & $1.229 \times 10^{-6}$ & $4.617$ & $4.252 \times 10^{-7}$ & $4.591$ & $4.480 \times 10^{-8}$\\ 
\hline
\end{tabular} \label{table-I}
\end{table*}

\section{Methodology}
\label{methodology}
\subsection{Formalism}
\begin{figure}[t]
\includegraphics[width=0.4\textwidth]{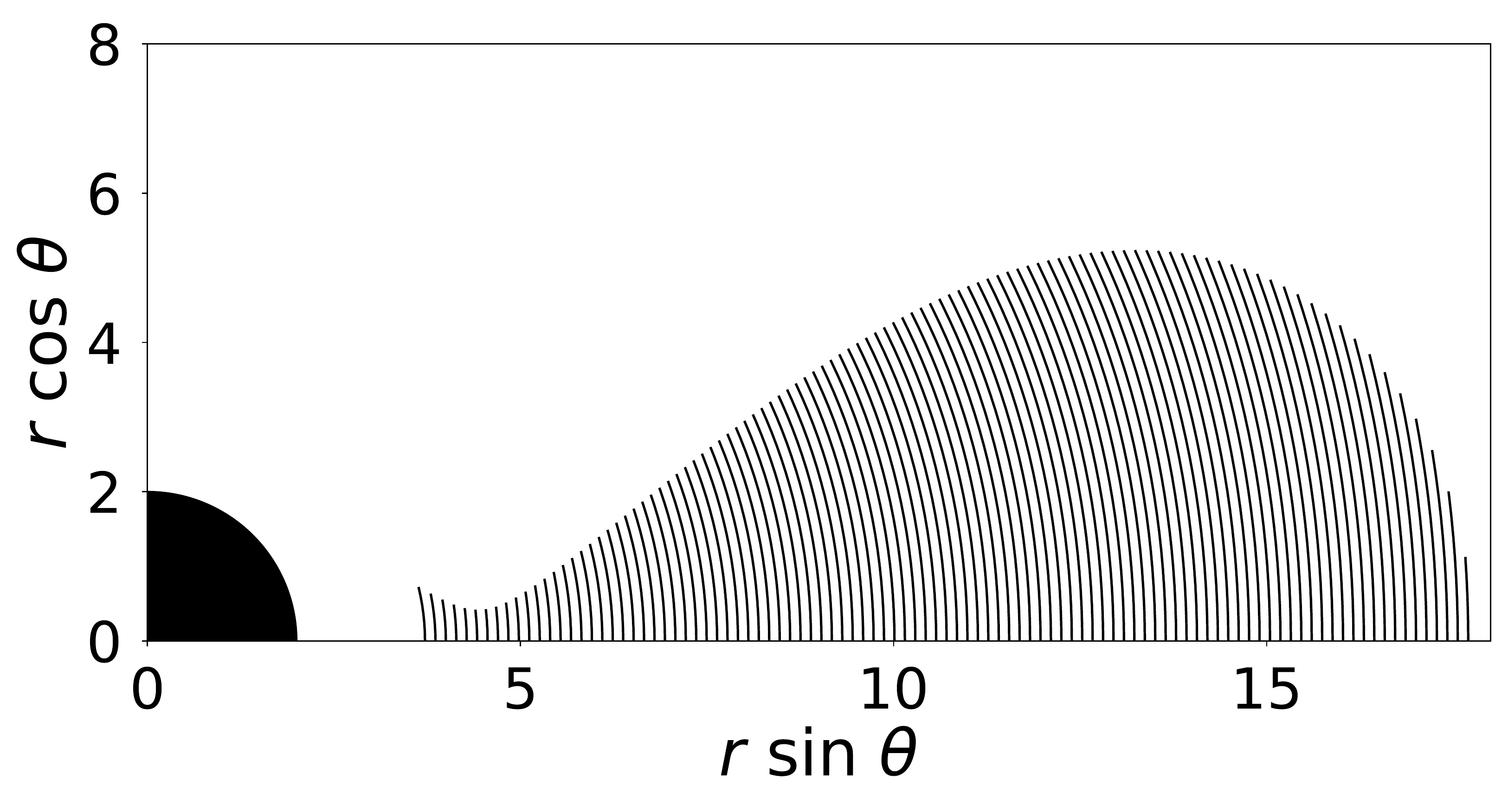}
\caption{Characteristic curves computed for $W_{\mathrm{s}} = -0.039$ (i.e.~the solutions of Eq.~\eqref{eq:characteristic_curves_final}.). For visualization purposes we only show a sample of 101 curves instead of the complete set of 703 curves we have computed. The black circle represents the black hole.}
\label{characteristic_curves}
\end{figure}

We solve Eq.~\eqref{eq:PDE_fin} with the domain of definition set by the conditions $W(r, \theta) \leq W_{\mathrm{s}}$, $r_{\mathrm{in}} \leq r \leq r_{\mathrm{out}}$ where $r_{\mathrm{in}}$ and $r_{\mathrm{out}}$ are the inner and the outer boundary of the disk at the equatorial plane. As in this work we are considering disks slightly overflowing their Roche lobe (i.e.~ $W_{\mathrm{s}} \gtrsim W(r_{\mathrm{cusp}}, \pi/2)$ where $r_{\mathrm{cusp}}$ corresponds to the location of the self-crossing point of the critical equipotential surface) it is important to note that the disks do not possess an inner edge (i.e.~the outermost equipotential surface is attached to the event horizon of the black hole) and thus our choice of $r_{\mathrm{in}}$ is arbitrary. Here, we choose the value of $r_{\mathrm{in}}$ such that $r_{\mathrm{in}} \lesssim r_{\mathrm{cusp}}$ so we can study the cusp region, and exclude the region closest to the black hole, as it is irrelevant for our study (the reason will become clear in Section~\ref{results}). In addition, 
we exclude the funnel region along the symmetry axis ($\theta=0$) by further restricting our domain by only considering the region containing equipotential surfaces that cross the equatorial plane at least once.
As our system has axisymmetry and reflection symmetry with respect to the equatorial plane, we can further restrict our domain to $0 < \theta < \pi/2$.

\begin{figure*}[t]\label{full_disks}
\includegraphics[scale=0.27]{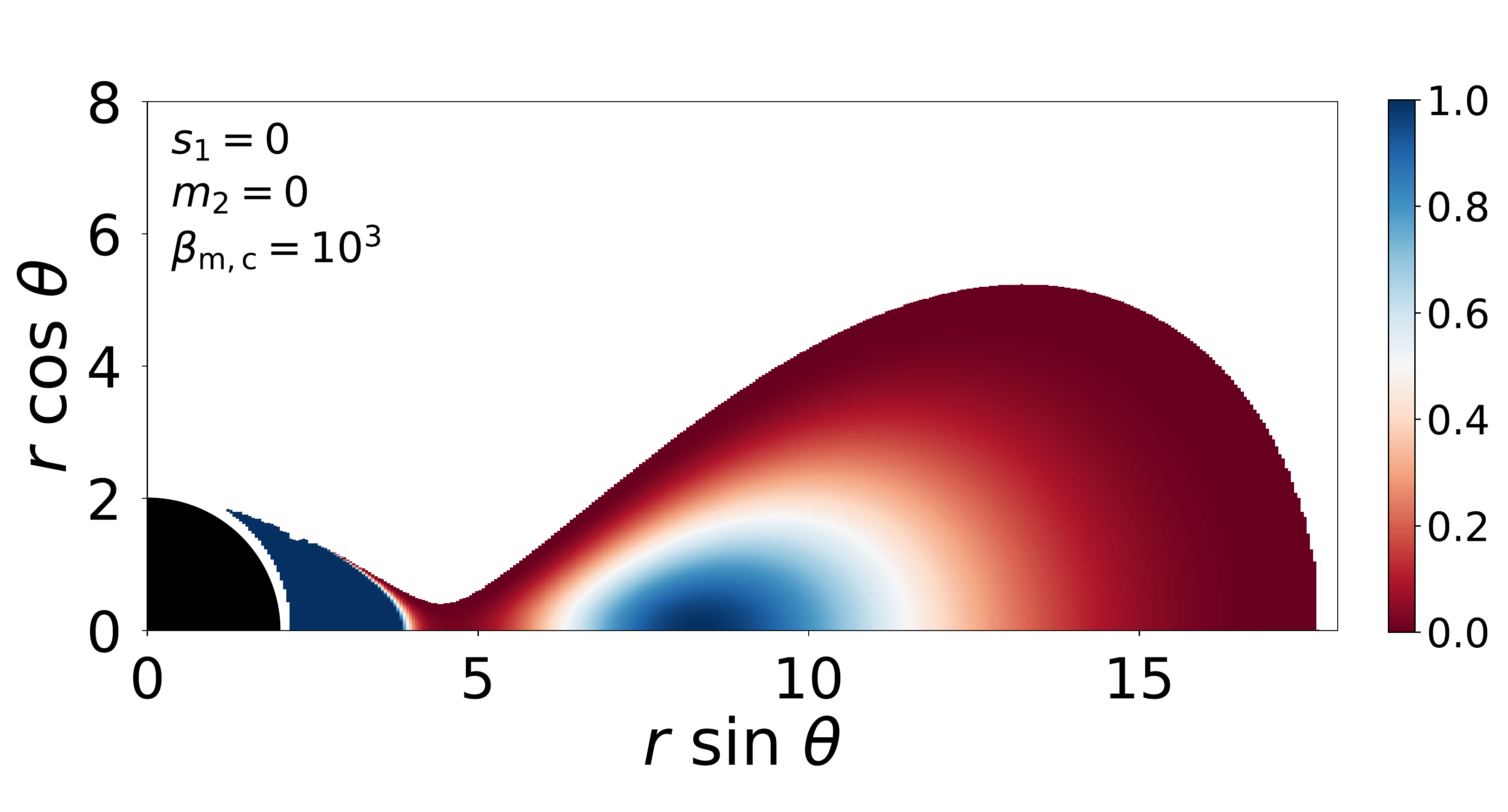}
\hspace{-0.3cm}
\includegraphics[scale=0.27]{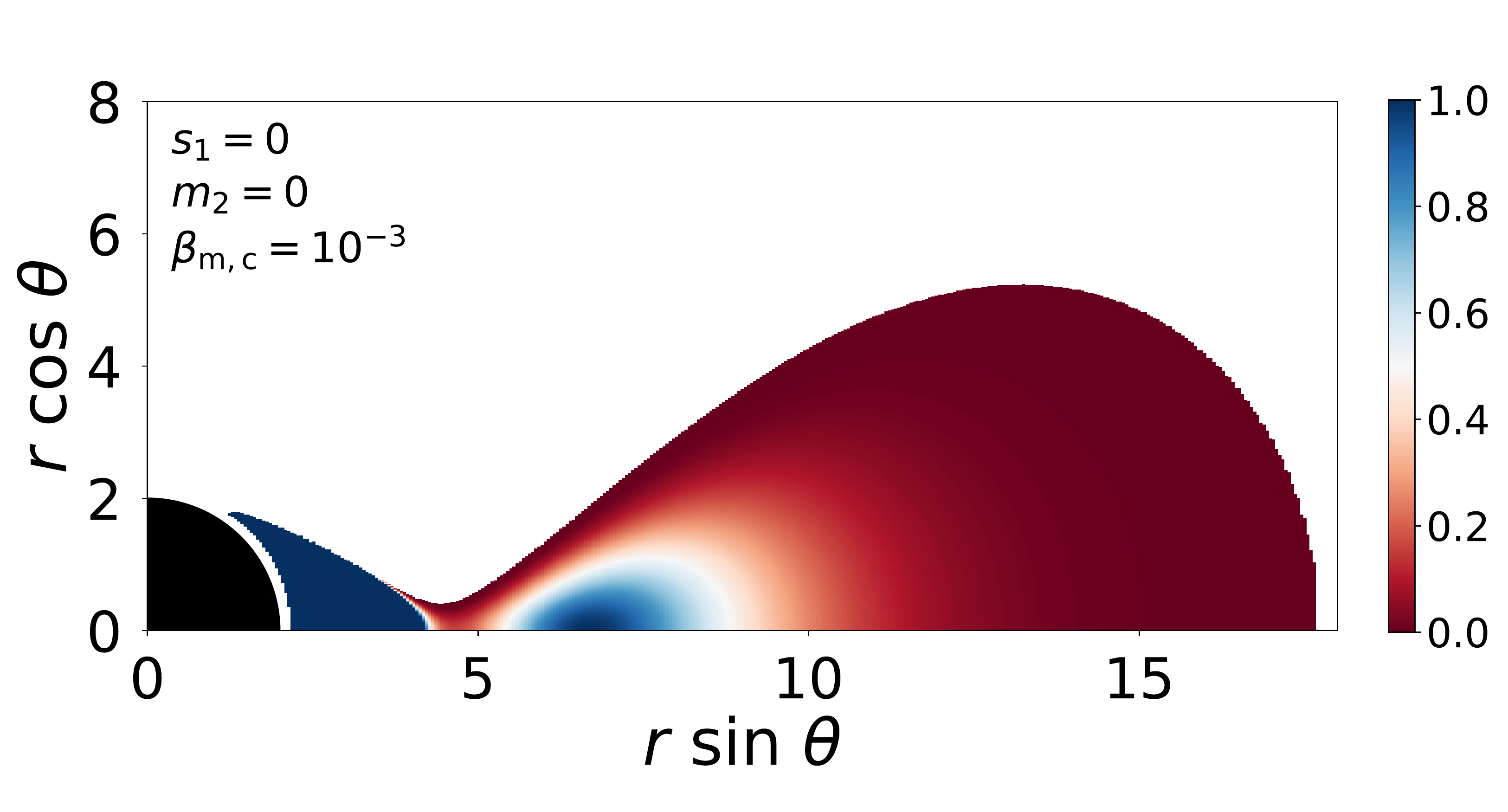}
\\
\vspace{-0.2cm}
\includegraphics[scale=0.27]{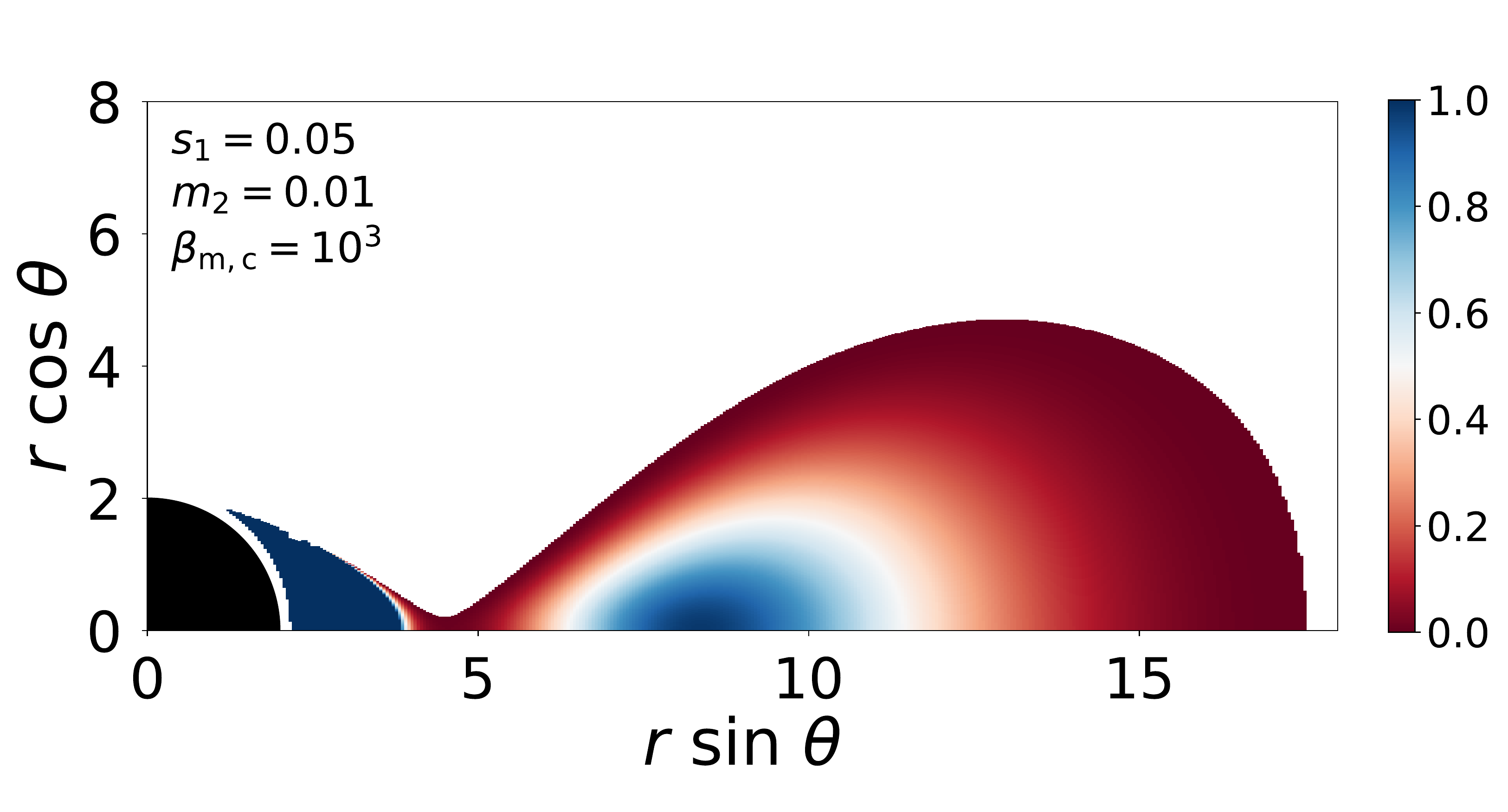}
\hspace{-0.3cm}
\includegraphics[scale=0.27]{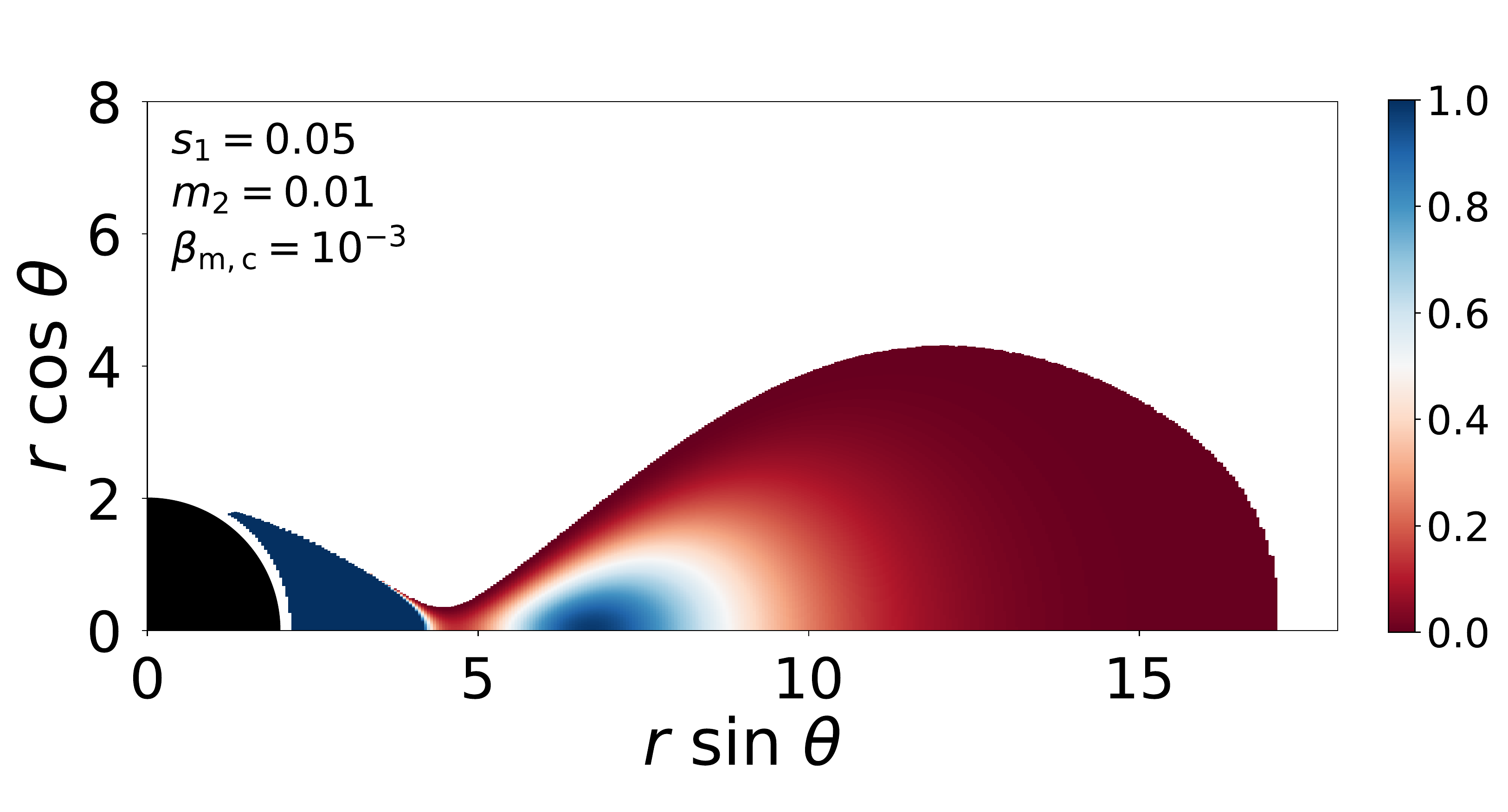}
\caption{Comparison of the full disk structure for $W_{\mathrm{s}} = -0.039$. The top row shows inviscid solutions and the bottom row shows viscous solutions for $s_1 = 0.05$ and $m_2 = 0.01$. The left panels correspond to non-magnetized disks ($\beta_{\mathrm{m,c}} = 10^3$) and the right panels to highly-magnetized disks ($\beta_{\mathrm{m,c}} = 10^{-3}$).  In all figures the colour gradient refers to the distributions of the total fluid pressure. Furthermore,  in the four cases the pressure has been normalized to the corresponding value of $p_{\mathrm{(0), max}}$. The morphology of the disks remains essentially the same for all cases, the only noticeable difference being a small decrease in size at the very low pressure region for the viscous cases. Note that, for visualization purposes we have extended our original domain of integration ($r_{\mathrm{in}} = 2.1$ instead of the original $r_{\mathrm{in}} = 3.7$) so that it is apparent that the inner region of the disk is attached to the event horizon of the black hole.}
\label{full-disk}
\end{figure*}

Eq.~\eqref{eq:PDE_fin} can be rewritten in a more compact form as
\begin{equation} \label{eq:partial_polar_compact}
\vec{\alpha}(r, \theta) \cdot \vec{\nabla}_{(r, \theta)} p_{(1)} - \tilde{c}(r,\theta) = 0,
\end{equation}
with the following definitions
\begin{small}
\begin{equation}\label{eq:polar_coeffs}
  \begin{array}{lr}
    \alpha_r(r,\theta) = \sin^6 \theta \left[r^3+l_{0}^2(2-r)\csc^2\theta \right]^3 \displaystyle\frac{(1+\beta_m)}{\beta_{m}} \left(r(r-2)\cot \theta \right)\,, 
    \\
    \alpha_{\theta}(r,\theta) = r\sin^6 \theta \left[r^3+l_{0}^2(2-r)\csc^2\theta \right]^3 \displaystyle\frac{(1+\beta_m)}{\beta_{m}} \left( - \frac{4 k_1}{C} \right) \,,
    \\
    \tilde{c}(r,\theta) = -2l_{0}^2\cot \theta \left( \tilde{A}+\displaystyle\frac{\tilde{B}k_1}{C}\right)(\tau_2 m_1)-\displaystyle\frac{m_{2}}{2}\cot\theta \left(\tilde{f}_1-\frac{\tilde{f}_2 k_1}{C}\right)\,.
  \end{array}
\end{equation}
\end{small}
Close examination of the coefficients in Eq.~\eqref{eq:polar_coeffs} reveals that, at the equatorial plane ($\theta = \pi/2$), Eq.~\eqref{eq:partial_polar_compact} is simply
\begin{equation}\label{eq:eq_ode}
\frac{\partial p_{(1)}}{\partial \theta} = 0.
\end{equation}
Eq.~\eqref{eq:eq_ode} has two relevant consequences for our solution. The first one is that surfaces of constant $p_{\mathrm{(1)}}$ are orthogonal to the equatorial plane (a consequence of the reflection symmetry of the problem). The second one is that one cannot extract information of the distribution of $p_{\mathrm{(1)}}$ at the equatorial plane directly from Eq.~\eqref{eq:partial_polar_compact} at $\theta = \pi/2$. To know the values of $p_{\mathrm{(1)}}$ at the equatorial plane we must look for the solution $p_{\mathrm{(1)}}(r, \theta)$ when $\theta \to \pi/2$ i.e.~a point that belongs to the domain of the $\theta$ coordinate. Thus, to maximize the accuracy of the solution is convenient to solve Eq.~\eqref{eq:partial_polar_compact} in Cartesian coordinates, as the distance between the last point of our domain and the equatorial plane will remain the same. Then, we can rewrite this equation as
\begin{equation}
\vec{\alpha}'(x, y) \cdot \vec{\nabla}_{(x,y)} p_{(1)} - c'(x, y) = 0,
\end{equation}
in which we used the change of coordinates defined by $x = r \sin \theta$, $y = r \cos \theta$,
and the new expressions for the coefficients
\begin{equation}
\begin{array}{lr}
\alpha_x'(x(r, \theta),y(r, \theta)) = \alpha_r(r,\theta) \sin \theta + \alpha_{\theta}(r,\theta) \cos \theta
\,,
\\
\alpha_y'(x(r, \theta),y(r, \theta)) = \alpha_r(r,\theta) \cos \theta - \alpha_{\theta}(r,\theta) \sin \theta
\,,
\\
c'(x(r, \theta),y(r, \theta)) = \tilde{c}(r, \theta)\,,
\end{array}
\end{equation}
where $\alpha_x'$ and $\alpha_y'$ are the $x$ and $y$ components of the vector of coefficients $\vec{\alpha}'(x, y)$.
Taking into account that $\alpha_y'(x, y) \neq 0$ in our domain, we can redefine all the coefficients as 
\begin{equation}
\begin{array}{lr}
a(x,y) = \alpha_x'(x, y)/\alpha_y'(x, y)\,,
\\
b(x,y) = 1\,,
\\
c(x,y) = c'(x,y)/\alpha_y'(x,y)\,.
\end{array}
\end{equation}
Therefore, the final form of the partial differential equation (PDE) we want to solve reads
\begin{equation}\label{eq:eq_cartesian}
a(x,y) \frac{\partial p_{(1)}}{\partial x} + \frac{\partial p_{(1)}}{\partial y} - c(x, y) = 0\,.
\end{equation}
To solve Eq.~\eqref{eq:eq_cartesian} we use the so-called method of characteristics, in which we can reduce a PDE to a set of ordinary differential equations (ODEs), one for each initial value defined at the boundary of the domain. 
The final form of the characteristic equations is
\begin{eqnarray}
\label{eq_characterstics_cart_final}
    \displaystyle\frac{d x}{d t} &=& a(x,y)\,,\label{eq:characterstics_x}
    \\
    \displaystyle\frac{d y}{d t} &=& 1 \,, \label{eq:characterstics_y}
    \\
    \displaystyle\frac{d p_{\mathrm{(1)}}}{d t} &=& c(x,y)\,.\label{eq:characterstics_z}
\end{eqnarray}
To solve this system, we start from a point $(x_0, y_0)$ in the boundary of the domain (i.e.~$\{(x_0,y_0) \; / \, W(x_0,y_0) = W_{\mathrm{s}}\}$). Then, we can integrate the system of ODEs as follows: first, the solution of Eq.~\eqref{eq:characterstics_y} is trivially $y(t) = t + y_0$. Using this result, we can rewrite Eq.~\eqref{eq:characterstics_x} as
\begin{equation}\label{eq:characteristic_curves_final}
\frac{d x}{d y} = a(x,y)\,.
\end{equation}
We can integrate numerically this equation starting from the selected point $(x_0,y_0)$. The solution of this equation ($x(y)$) will give us a characteristic curve of the problem, i.e.~a curve along which the solution of our PDE coincides with the solution of the ODE. To finish the procedure, we take Eq.~\eqref{eq:characterstics_z} and rewrite it in the same way as the previous one.
\begin{eqnarray}
\label{eq:p1_diff}
\frac{d p_{\mathrm{(1)}}}{d y} = c(x(y),y)\,.
\end{eqnarray}
Then, we can integrate $p_{\mathrm{(1)}}$
\begin{eqnarray}
\label{eq:p1_y}
p_{\mathrm{(1)}}(y) = \int^y_{y_0} c(x(y),y) dy + p_{\mathrm{(1)}_0}\,,
\end{eqnarray}
where we have used that $p_{\mathrm{(1)}}(x_0, y_0) = p_{\mathrm{(1)}_0}$.
It is easy to see that we can recover $p_{\mathrm{(1)}}(x,y)$ by using both Eq.~\eqref{eq:p1_y} and the expression for the characteristic curve $x(y)$.
Repeating this three-step procedure over a sufficiently large and well-chosen sample of initial points will give us a mapping of the domain and hence, the solution of the PDE for the whole domain.

\begin{table*}[t]
\caption{Values of $\Delta p_{\mathrm{cusp}}$ and $\Delta r_{\mathrm{cusp}}$ corresponding to different choices of $s_1$, $m_2$ and $\beta_{m,c}$. The considered values of $W_{\mathrm{s}}$ are respectively $-0.039$, $-0.040$ and $-0.041$. Bold-faced values of $\Delta p_{\mathrm{cusp}}$ are employed to indicate the regions when $p_{(1)} \gtrsim p_{(0)}$, such that $p_{(1)} $ cannot be treated as a perturbation for a given value of $W_{\mathrm{s}}$ and $\beta_{m,c}$.}   
\begin{tabular}{c c c | c c | c c | c c}
\hline\hline       
 &  &  & \multicolumn{2}{c}{$W_{\mathrm{s}} = -0.039$} & \multicolumn{2}{c}{$W_{\mathrm{s}} = -0.040$} & \multicolumn{2}{c}{$W_{\mathrm{s}} = -0.041$}\\ 
\hline
$s_1$ & $m_2$ & $\beta_{\mathrm{m,c}}$ & $\Delta r_{\mathrm{cusp}}$ & $\Delta p_{\mathrm{cusp}}$ &  $\Delta r_{\mathrm{cusp}}$ & $\Delta p_{\mathrm{cusp}}$ & $\Delta r_{\mathrm{cusp}}$ & $\Delta p_{\mathrm{cusp}}$ \\
\hline        
$0.001$ & $0$ & $10^3$ & $-9.71 \times 10^{-5}$ & $-6.66 \times 10^{-3}$ & $-1.10 \times 10^{-4}$ & $-1.20 \times 10^{-2}$ & $-1.05 \times 10^{-4}$ & $-4.40 \times 10^{-2}$\\ 
$0.001$ & $0$ & $10^{-3}$ & $-2.45 \times 10^{-5}$ & $-1.43 \times 10^{-3}$ & $-2.95 \times 10^{-5}$ & $-2.61 \times 10^{-3}$ & $-4.38 \times 10^{-5}$ & $-9.71 \times 10^{-3}$\\ 
$0.001$ & $0.001$ & $10^{3}$ & $-3.27 \times 10^{-4}$ & $-3.64 \times 10^{-2}$ & $-3.95 \times 10^{-4}$ & $-6.55 \times 10^{-2}$ & $-2.33 \times 10^{-4}$ & $-0.241$\\ 
$0.001$ & $0.001$ &  $10^{-3}$ &  $-8.55 \times 10^{-5}$ & $-7.99 \times 10^{-3}$ & $-1.10 \times 10^{-4}$ & $-1.45 \times 10^{-2}$ & $-1.68 \times 10^{-4}$ & $-5.35 \times 10^{-2}$\\
$0.001$ & $0.005$ & $10^{3}$ & $-1.35 \times 10^{-3}$ & $-0.156$ & $-1.97 \times 10^{-3}$ & $-0.281$ & $-7.03 \times 10^{-4}$ & \bf{-1.03} \\ 
$0.001$ & $0.005$ & $10^{-3}$ & $-3.27 \times 10^{-4}$ & $-3.43 \times 10^{-2}$ & $-4.72 \times 10^{-4}$ & $-6.20 \times 10^{-2}$ & $-6.97 \times 10^{-4}$ & $-0.229$\\
$0.001$ & $0.01$ & $10^{3}$ & $-2.65 \times 10^{-3}$ & $-0.306$ & $-3.62 \times 10^{-3}$ & $-0.552$ & $-1.32 \times 10^{-2}$ & \bf{-1.92} \\ 
$0.001$ & $0.01$ & $10^{-3}$ & $-6.19 \times 10^{-4}$ & $-6.71 \times 10^{-2}$ & $-9.97 \times 10^{-4}$ & $-0.122$ & $-1.35 \times 10^{-3}$ & $-0.449$\\
$0.001$ & $0.05$ & $10^{3}$ & $-2.48 \times 10^{-2}$ & \bf{-1.48} & $-4.72 \times 10^{-2}$ & \bf{-2.03} & $-7.06 \times 10^{-2}$ & \bf{-2.33} \\ 
$0.001$ & $0.05$ & $10^{-3}$ & $-2.58 \times 10^{-3}$ & $-0.331$ & $-3.74 \times 10^{-3}$ & $-0.602$ & $-1.66 \times 10^{-2}$ & \bf{-2.02} \\
\hline  
$0.005$ & $0$ & $10^{3}$ & $-5.00 \times 10^{-4}$ & $-3.34 \times 10^{-2}$ & $-5.99 \times 10^{-4}$ & $-6.00 \times 10^{-2}$ & $-5.10 \times 10^{-4}$ & $-0.220$\\  
$0.005$ & $0$ & $10^{-3}$ & $-1.22 \times 10^{-4}$ & $-7.16 \times 10^{-3}$ & $-1.51 \times 10^{-4}$ & $-1.31 \times 10^{-2}$ & $-2.23 \times 10^{-4}$ & $-4.86 \times 10^{-2}$\\ 
$0.005$ & $0.001$ & $10^{3}$ & $-7.46 \times 10^{-4}$ & $-6.32 \times 10^{-2}$ & $-9.51 \times 10^{-4}$ & $-0.114$ & $-6.29 \times 10^{-4}$ & $-0.417$ \\ 
$0.005$ & $0.001$ & $10^{-3}$ & $-1.83 \times 10^{-4}$ & $-1.37 \times 10^{-2}$ & $-2.36 \times 10^{-4}$ & $-2.49 \times 10^{-2}$ & $-3.51 \times 10^{-4}$ & $-9.25 \times 10^{-2}$\\
$0.005$ & $0.005$ & $10^{3}$ & $-1.80 \times 10^{-3}$ & $-0.183$ & $-2.56 \times 10^{-3}$ & $-0.330$ & $-4.66 \times 10^{-3}$ & \bf{-1.20}\\ 
$0.005$ & $0.005$ & $10^{-3}$ & $-4.23 \times 10^{-4}$ & $-4.00 \times 10^{-2}$ & $-6.15 \times 10^{-4}$ & $-7.25 \times 10^{-2}$ & $-8.86 \times 10^{-4}$ & $-0.268$\\
$0.005$ & $0.01$ & $10^{3}$ & $-3.03 \times 10^{-3}$ & $-0.333$ & $-3.98 \times 10^{-3}$ & $-0.601$ & $-1.71 \times 10^{-2}$ & \bf{-2.01}\\ 
$0.005$ & $0.01$ & $10^{-3}$ & $-7.12 \times 10^{-4}$ & $-7.22 \times 10^{-2} $ & $-1.15 \times 10^{-3}$ & $-0.132$ & $-1.51 \times 10^{-3}$ & $-0.488$\\ 
$0.005$ & $0.05$ & $10^{3}$ & $-2.52 \times 10^{-2}$ & \bf{-1.50} & $-4.77 \times 10^{-2}$ & \bf{-2.06} & $-7.12 \times 10^{-2}$ & \bf{-2.34}\\ 
$0.005$ & $0.05$ & $10^{-3}$ & $-2.66 \times 10^{-3}$ & $-0.337$ & $-3.79 \times 10^{-3}$ & $-0.613$ & $-1.75 \times 10^{-2}$ & \bf{-2.04}\\ 
\hline
$0.01$ & $0$ & $10^{3}$& $-1.03 \times 10^{-3}$ & $-6.69 \times 10^{-2}$ & $-1.30 \times 10^{-3}$ & $-0.120$ & $-1.03 \times 10^{-3}$ & $-0.441$ \\ 
$0.01$ & $0$ & $10^{-3}$ & $-2.44 \times 10^{-4}$ & $-1.43 \times 10^{-2}$ & $-3.10 \times 10^{-4}$ & $-2.61 \times 10^{-2}$ & $-4.53 \times 10^{-4}$ & $-9.73 \times 10^{-2}$ \\
$0.01$ & $0.001$ & $10^{3}$ & $-1.29 \times 10^{-3}$ & $-9.68 \times 10^{-2}$ & $-1.70 \times 10^{-3}$ & $-0.174$ & $-1.17 \times 10^{-3}$ & $-0.638$\\  
$0.01$ & $0.001$ & $10^{-3}$ & $-3.04 \times 10^{-4}$ & $-2.09 \times 10^{-2}$ & $-4.01 \times 10^{-4}$ & $-3.80 \times 10^{-2}$ & $-5.86 \times 10^{-4}$ & $-0.141$\\ 
$0.01$ & $0.005$ & $10^{3}$ & $-2.33 \times 10^{-3}$ & $-0.217$ & $-3.18 \times 10^{-3}$ & $-0.391$ & $-8.24 \times 10^{-3}$ & \bf{-1.41}\\ 
$0.01$ & $0.005$ & $10^{-3}$ & $-5.41 \times 10^{-4}$ & $-4.72 \times 10^{-2}$ & $-7.98 \times 10^{-4}$ & $-8.57 \times 10^{-2}$ & $-1.12 \times 10^{-3}$ & $-0.317$\\ 
$0.01$ & $0.01$ & $10^{3}$ & $-3.47 \times 10^{-3}$ & $-0.367$ & $-4.43 \times 10^{-3}$ & $-0.662$ & $-2.18 \times 10^{-2}$ & \bf{-2.07}\\  
$0.01$ & $0.01$ & $10^{-3}$ & $-8.26 \times 10^{-4}$ & $-8.01 \times 10^{-2}$ & $-1.34 \times 10^{-3}$ & $-0.145$ & $-1.71 \times 10^{-3}$ & $-0.538$\\
$0.01$ & $0.05$ & $10^{3}$ & $-2.56 \times 10^{-2}$ & \bf{-1.53} & $-4.85 \times 10^{-2}$ & \bf{-2.09} & $-7.22 \times 10^{-2}$ & \bf{-2.33}\\ 
$0.01$ & $0.05$ & $10^{-3}$ & $-2.77 \times 10^{-3}$ & $-0.344$ & $-3.85 \times 10^{-3}$ & $-0.626$ & $-1.86 \times 10^{-2}$ & \bf{-2.05}\\
\hline
$0.05$ & $0$ & $10^{3}$ & $-4.92 \times 10^{-3}$ & $-0.340$ & $-6.10 \times 10^{-3}$ & $-0.613$ & $-2.58 \times 10^{-2}$ & \bf{-1.87}\\  
$0.05$ & $0$ & $10^{-3}$ & $-1.18 \times 10^{-3}$ & $-7.19 \times 10^{-2}$ & $-1.73 \times 10^{-3}$ & $-0.132$ & $-2.15 \times 10^{-3}$ & $-0.491$\\ 
$0.05$ & $0.001$ & $10^{3}$ & $-5.13\times 10^{-3}$ & $-0.370$ & $-6.44 \times 10^{-3}$ & $-0.668$ & $-2.89 \times 10^{-2}$ & \bf{-1.91}\\ 
$0.05$ & $0.001$ & $10^{-3}$ & $-1.23\times 10^{-3}$ & $-7.85 \times 10^{-2}$ & $-1.83 \times 10^{-3}$ & $-0.144$ & $-2.24 \times 10^{-3}$ & $-0.535$\\
$0.05$ & $0.005$ & $10^{3}$ & $-6.12 \times 10^{-3}$ & $-0.491$ & $-9.70 \times 10^{-3}$ & $-0.887$ & $-4.04 \times 10^{-2}$ & \bf{-1.95}\\  
$0.05$ & $0.005$ & $10^{-3}$ & $-1.44 \times 10^{-3}$ & $-0.105$ & $-2.22 \times 10^{-3}$ & $-0.192$ & $-2.56 \times 10^{-3}$ & $-0.711$\\ 
$0.05$ & $0.01$ & $10^{3}$ & $-9.44 \times 10^{-3}$ & $-0.645$& $-1.54 \times 10^{-2}$ & \bf{-1.15} & $-4.62 \times 10^{-2}$ & \bf{-2.16} \\  
$0.05$ & $0.01$ & $10^{-3}$ & $-1.69 \times 10^{-3}$ & $-0.138$ & $-2.64 \times 10^{-3}$ & $-0.252$ & $-2.90 \times 10^{-3}$ & $-0.931$\\ 
$0.05$ & $0.05$ & $10^{3}$ & $-4.12 \times 10^{-2}$ & \bf{-1.68} & $-6.28 \times 10^{-2}$ & \bf{-2.01} & $-8.08 \times 10^{-2}$ & \bf{-2.20}\\ 
$0.05$ & $0.05$ & $10^{-3}$ & $-3.79 \times 10^{-3}$ & $-0.403$ & $-4.28 \times 10^{-3}$ & $-0.733$ & $-2.61 \times 10^{-2}$ & \bf{-2.13}\\ 
\hline
\end{tabular} \label{table-2}
\end{table*}

\begin{figure*}[t]
\includegraphics[scale=0.27]{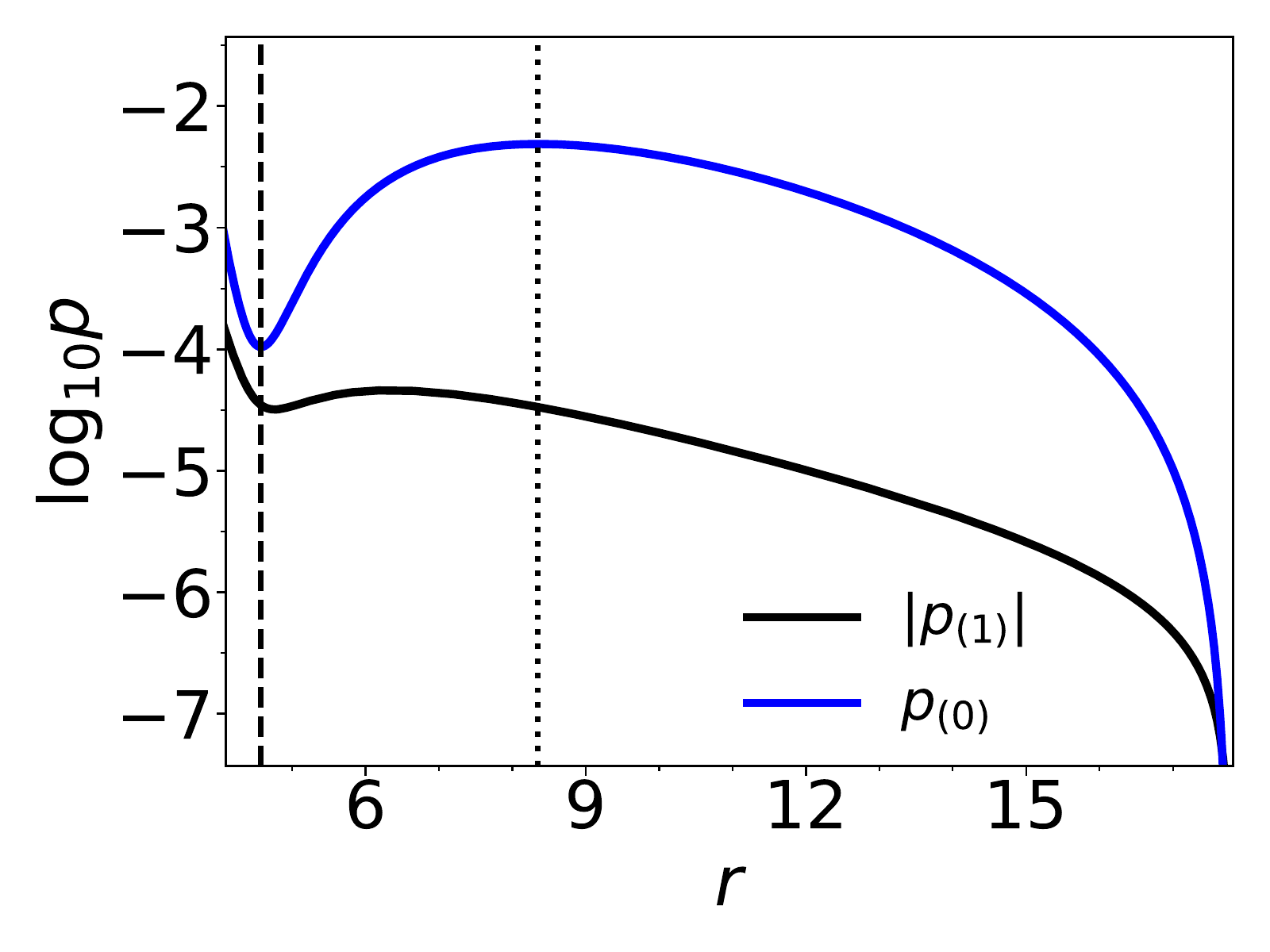}
\hspace{-0.3cm}
\includegraphics[scale=0.27]{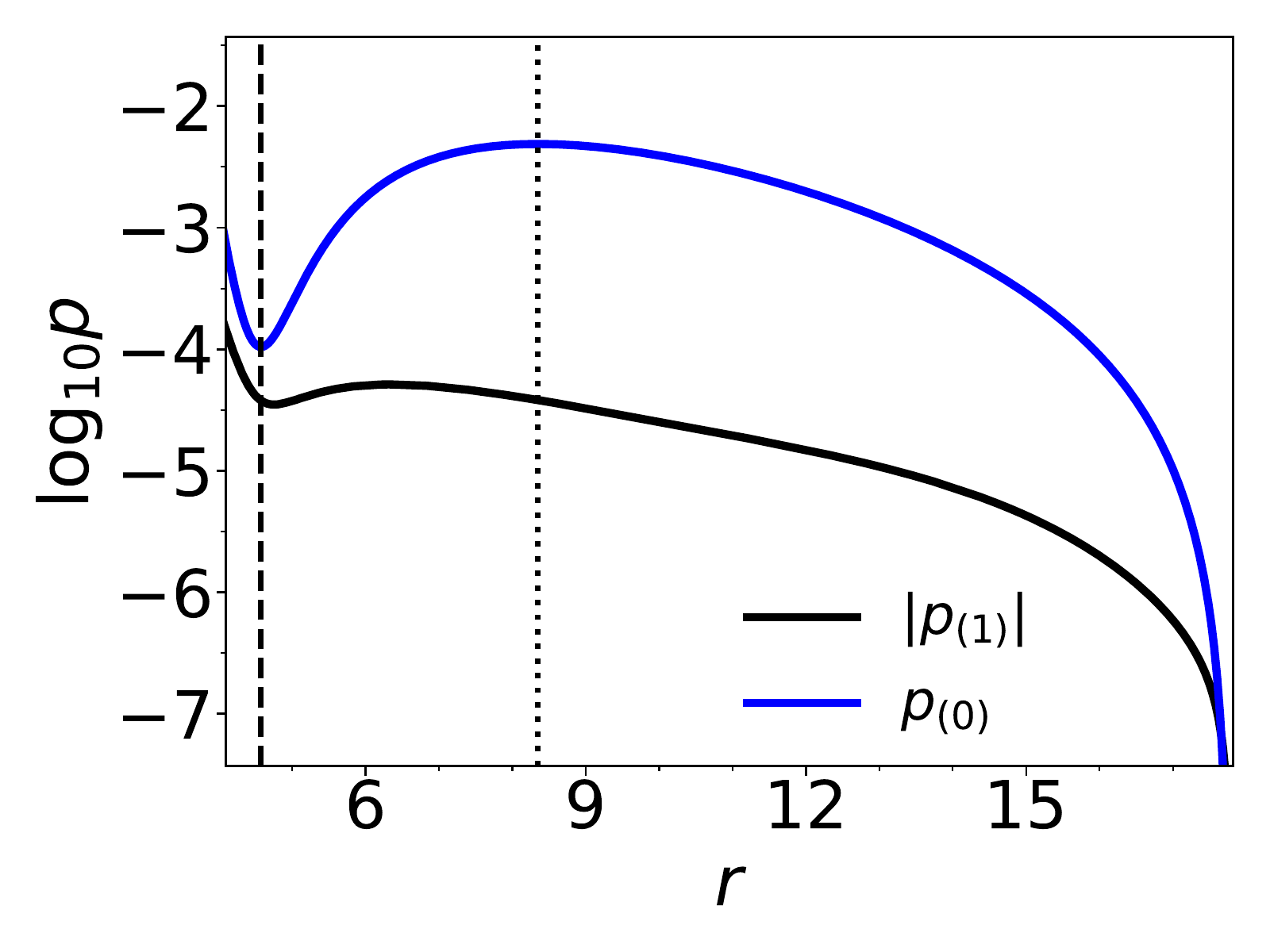}
\hspace{-0.2cm}
\includegraphics[scale=0.27]{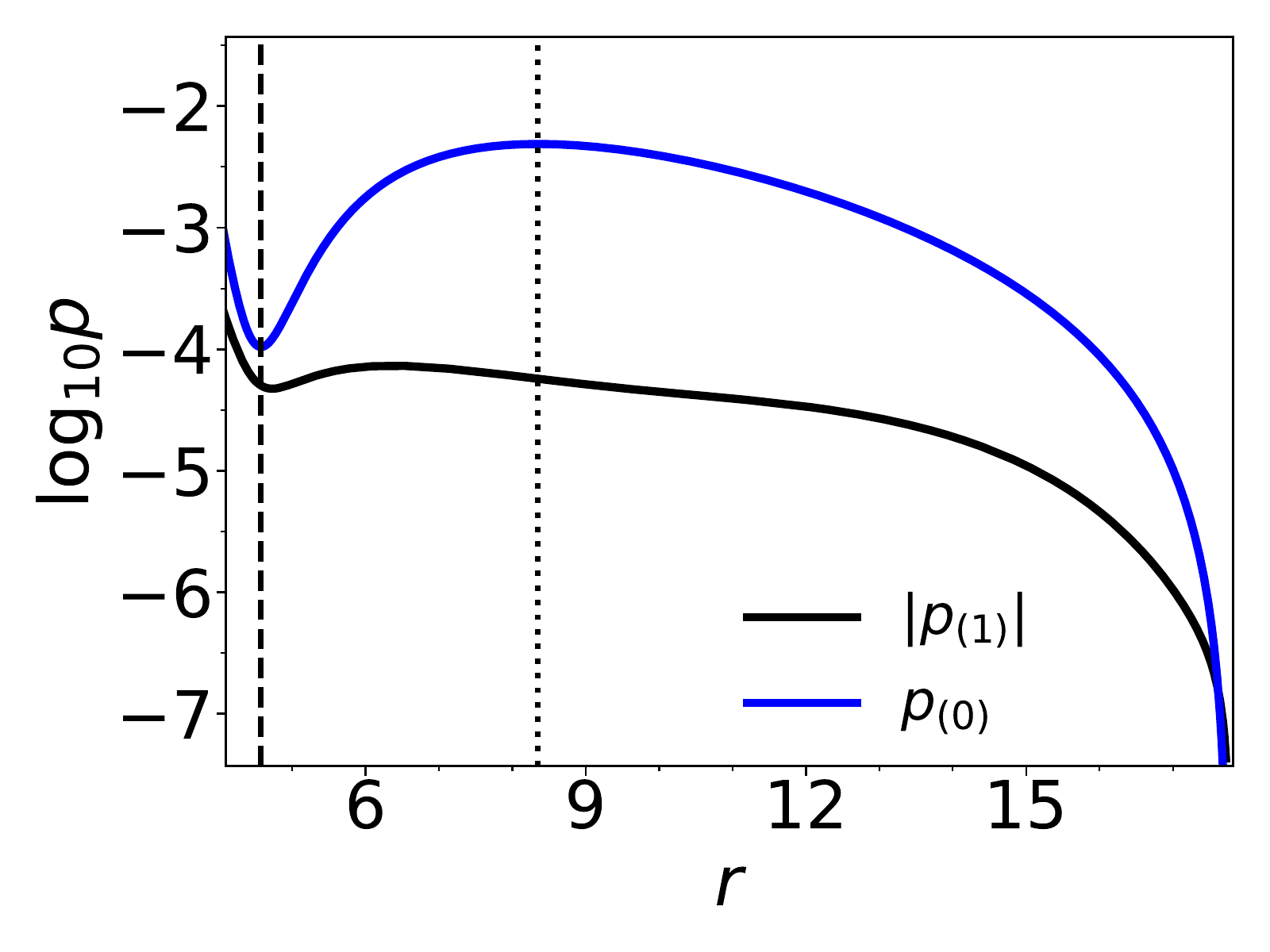}
\hspace{-0.2cm}
\includegraphics[scale=0.27]{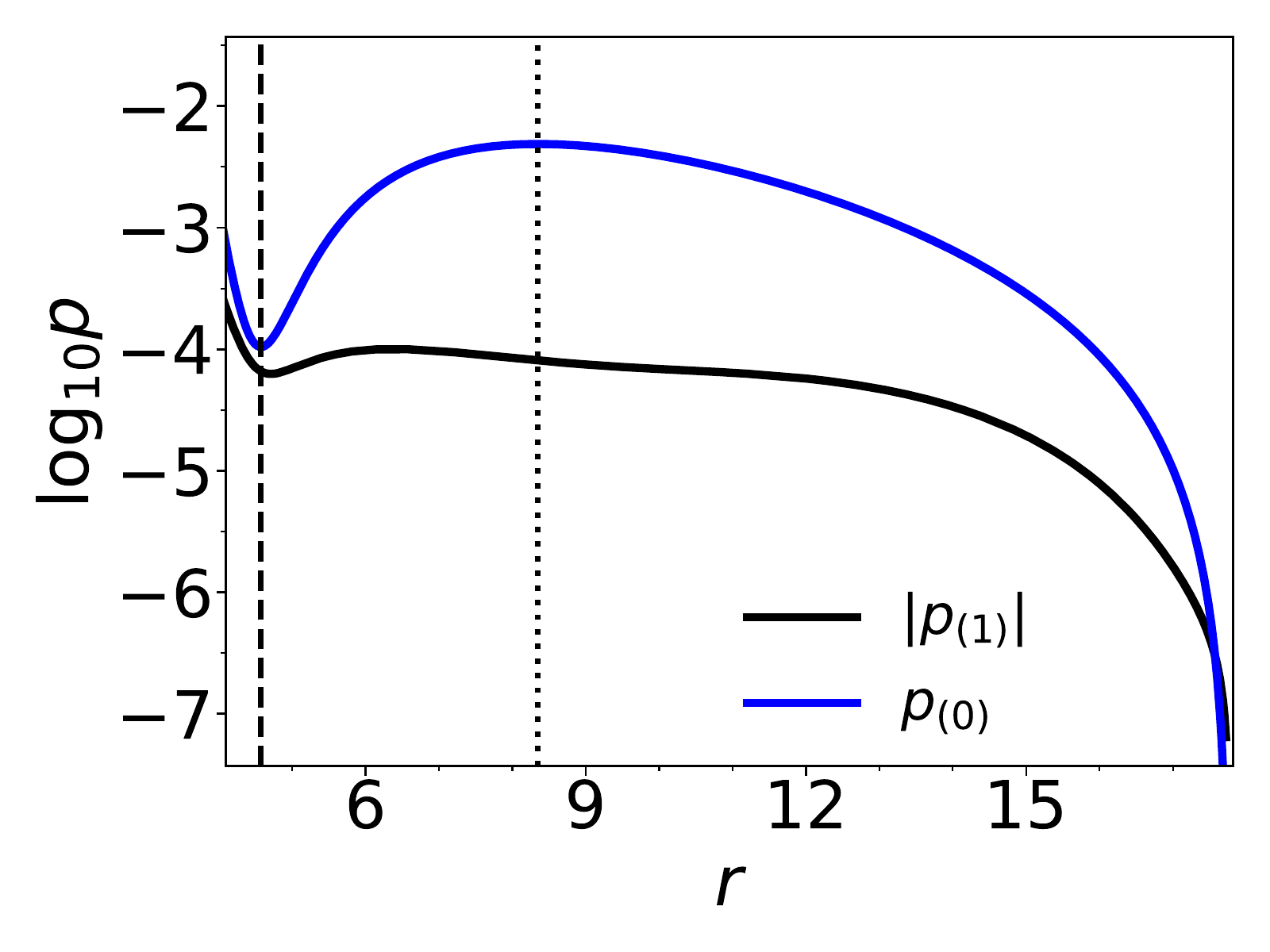}
\\
\includegraphics[scale=0.27]{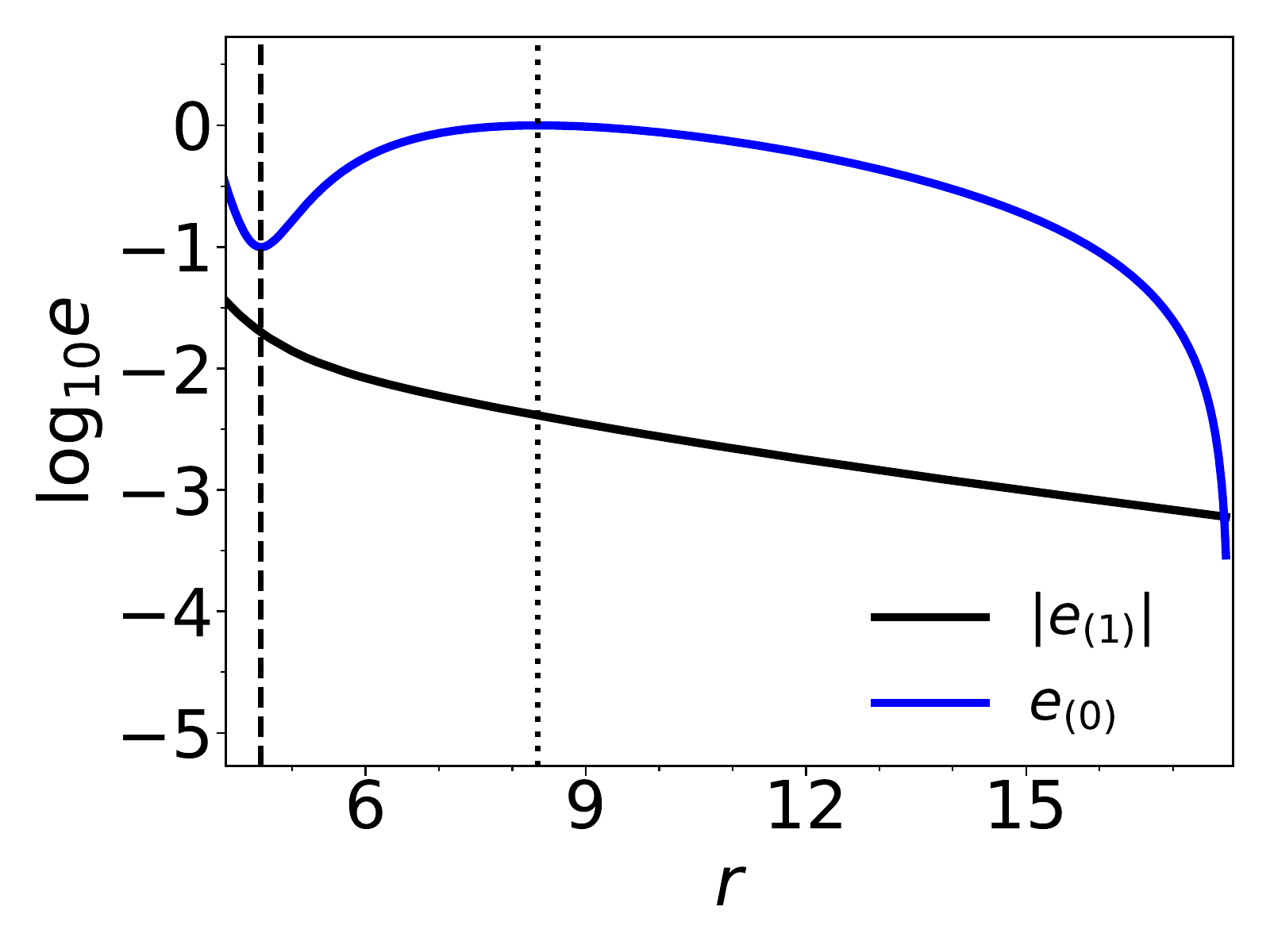}
\hspace{-0.3cm}
\includegraphics[scale=0.27]{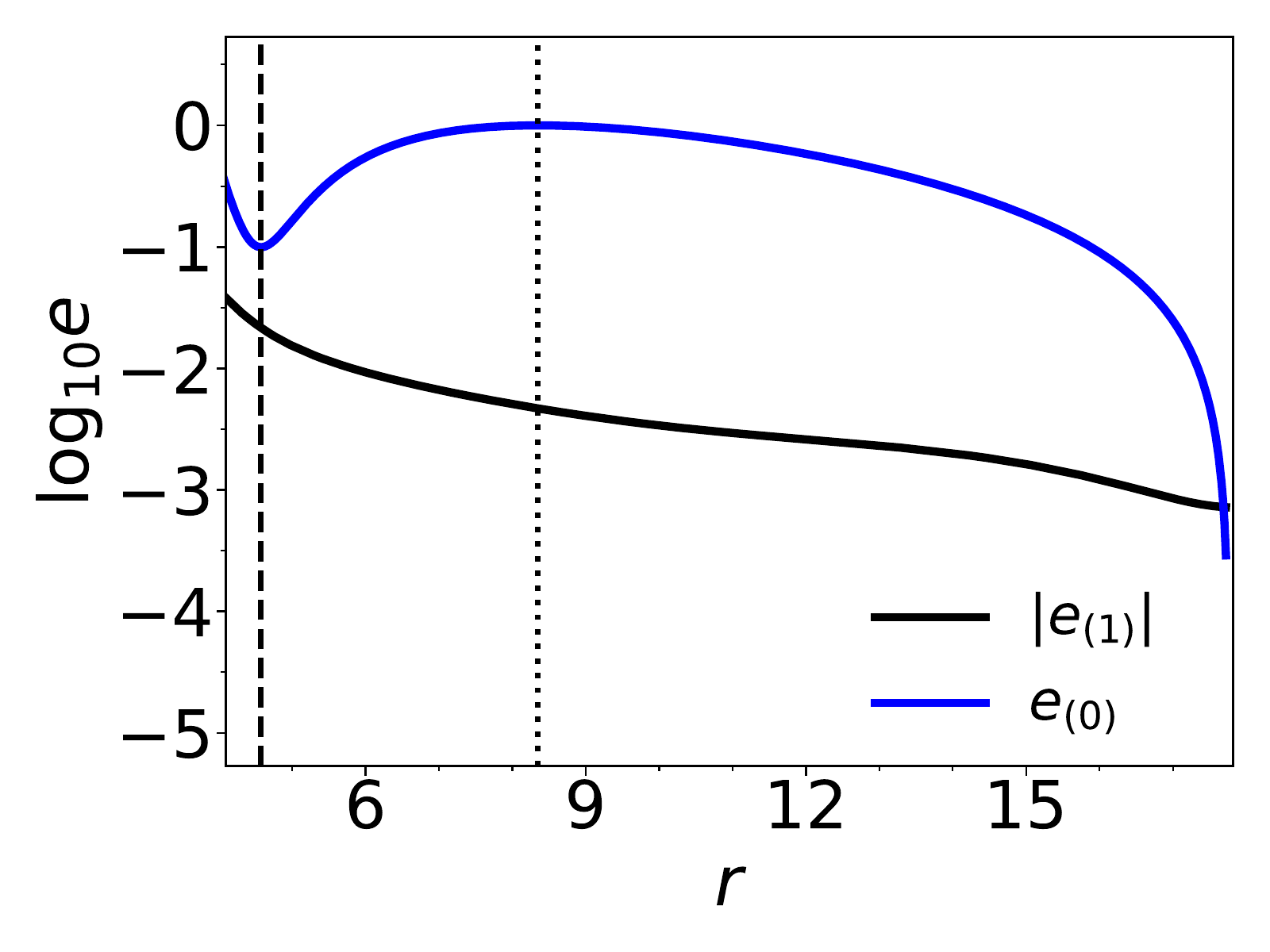}
\hspace{-0.2cm}
\includegraphics[scale=0.27]{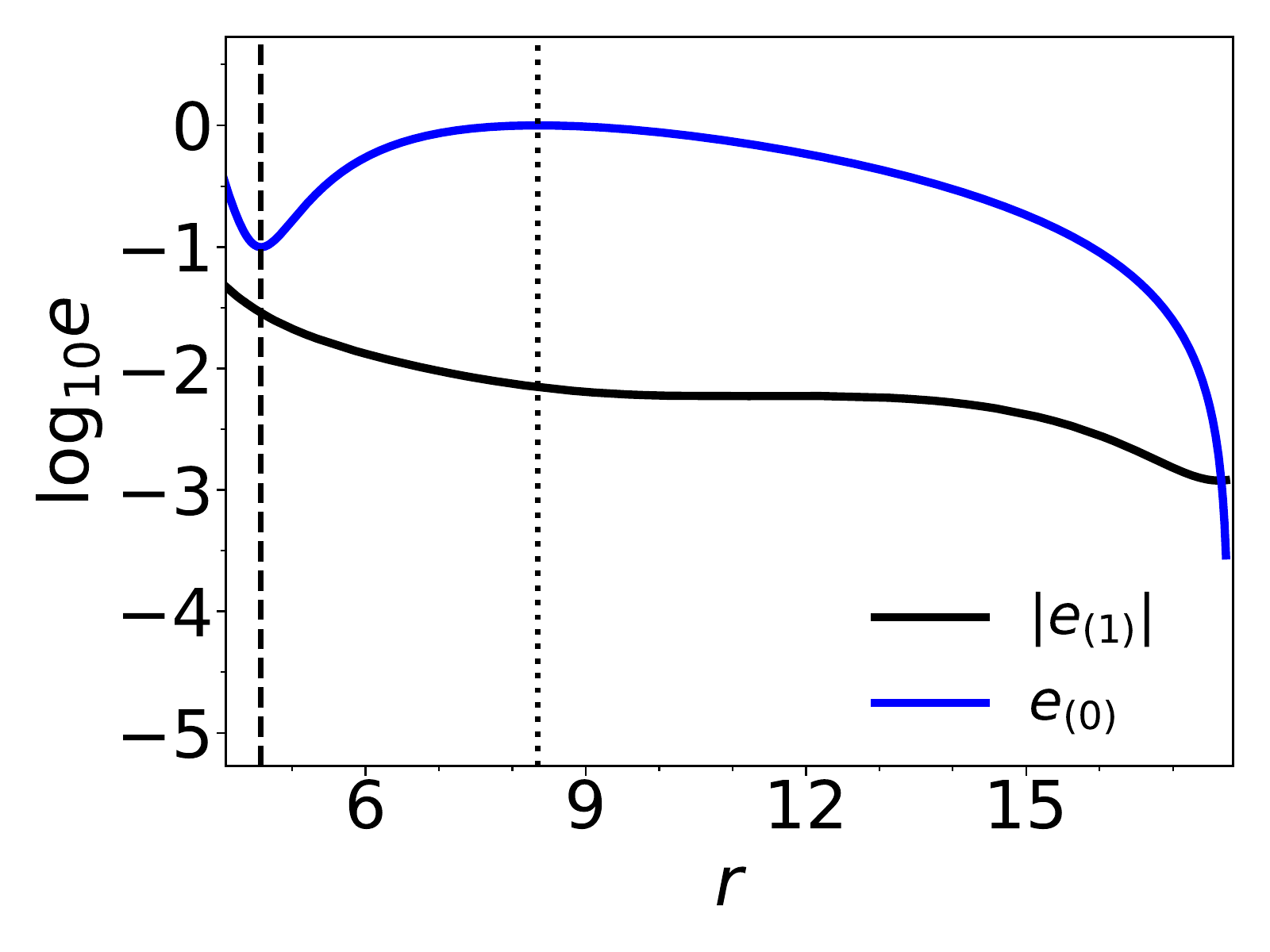}
\hspace{-0.2cm}
\includegraphics[scale=0.27]{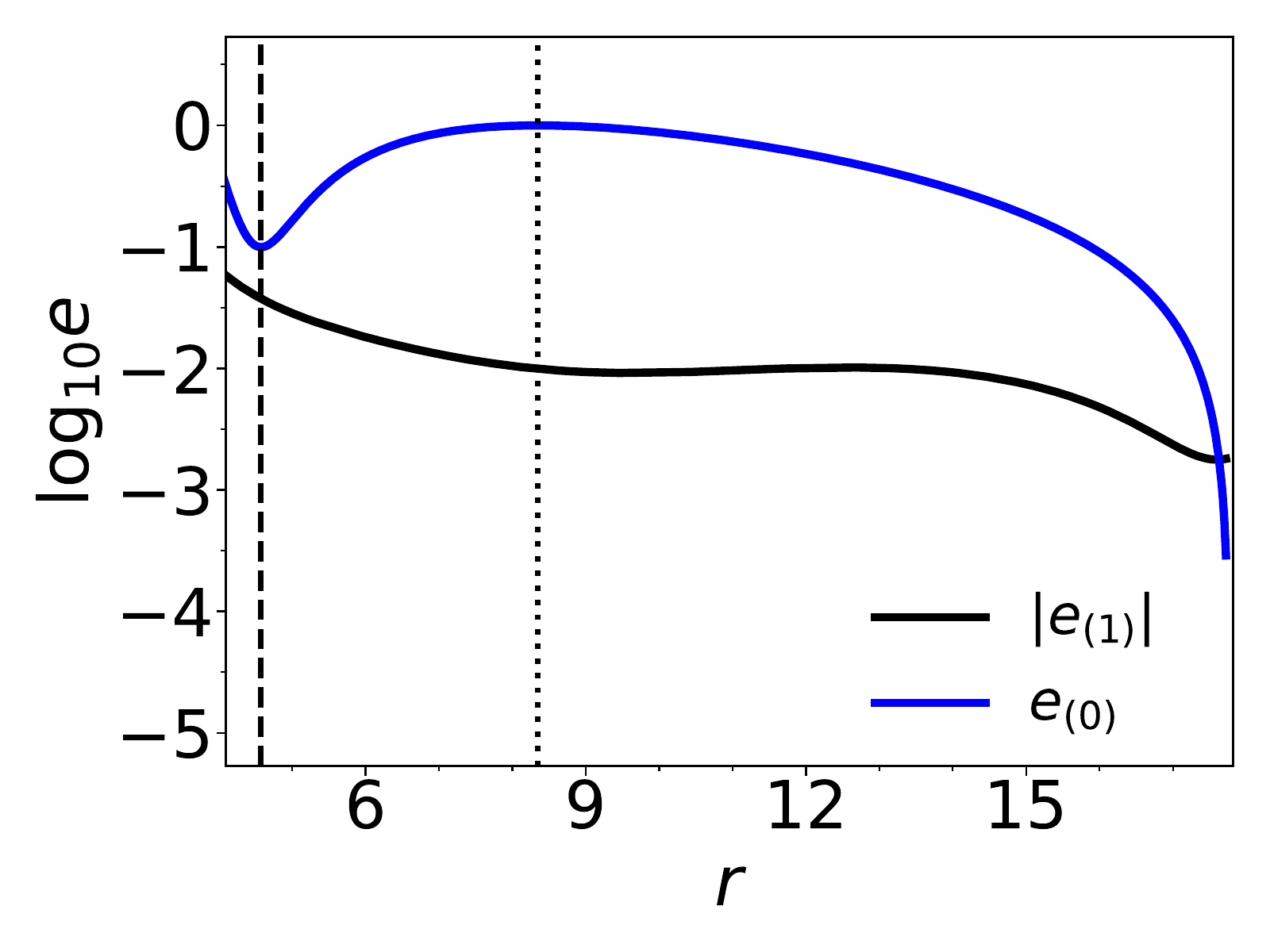}
\caption{Radial plots of $\log p_{(0)}$ and $\log |p_{(1)}|$ (top row) and $\log e_{(0)}$ and $\log |e_{(1)}|$ (bottom row) at the equatorial plane for $W_{\mathrm{s}} = -0.039$, $s_1 = 0.05$, $\beta_{\mathrm{m,c}} = 10^3$ and $m_2 = (0, 0.001, 0.005, 0.01)$. Each column corresponds to an increasing value of $m_2$. The vertical dashed line represents the location of the self-crossing pressure isocontour $r_\mathrm{cusp}$ and the vertical dotted line represents the location of the maximum of the pressure $r_{\mathrm{max}}$, which coincides to the center of the disk $r_{\mathrm{c}}$ for non-magnetized disks.}
\label{radial_plots_beta_3}
\end{figure*}

\begin{figure*}
\centering
\includegraphics[scale=0.27]{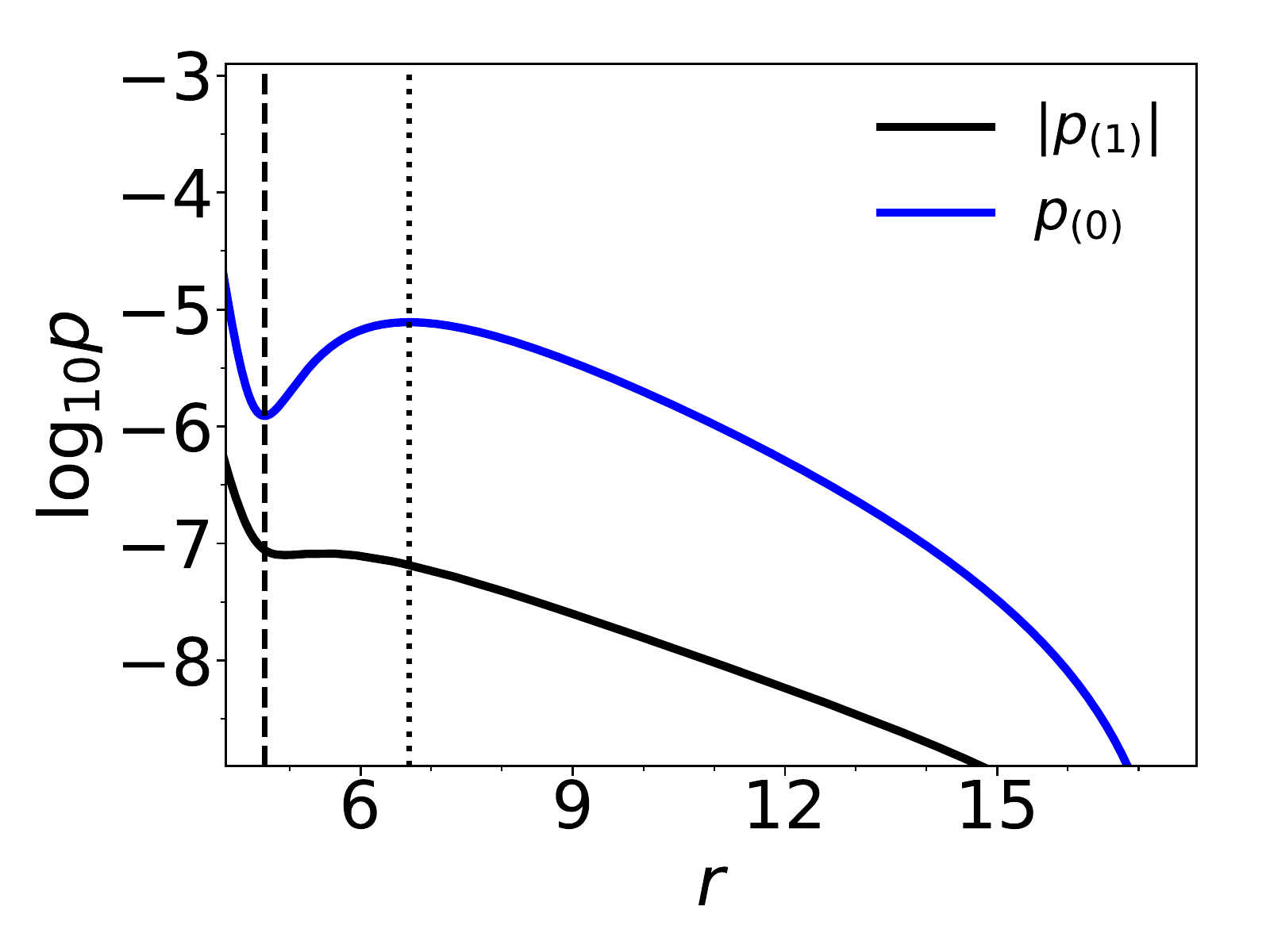}
\hspace{-0.3cm}
\includegraphics[scale=0.27]{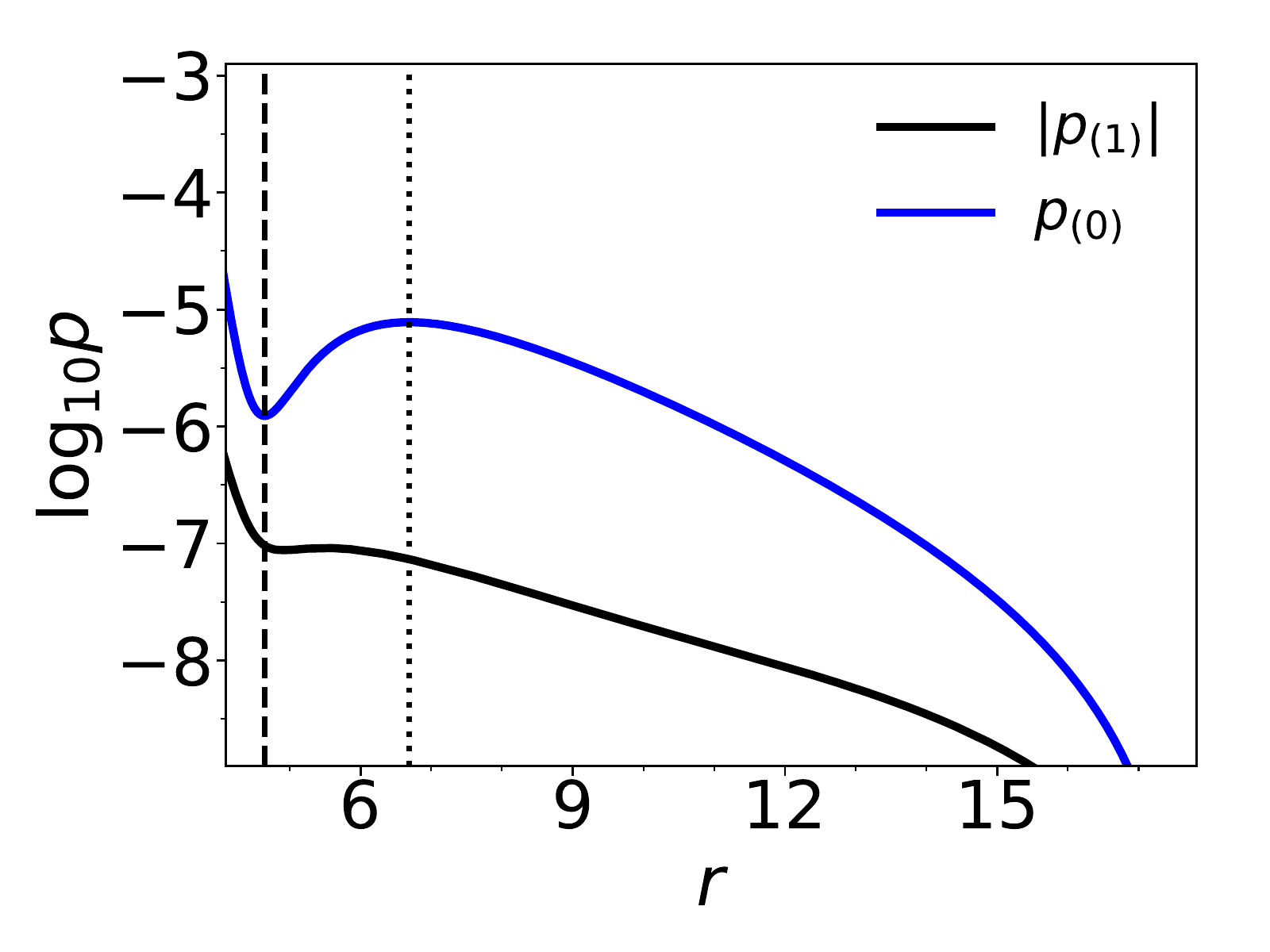}
\hspace{-0.2cm}
\includegraphics[scale=0.27]{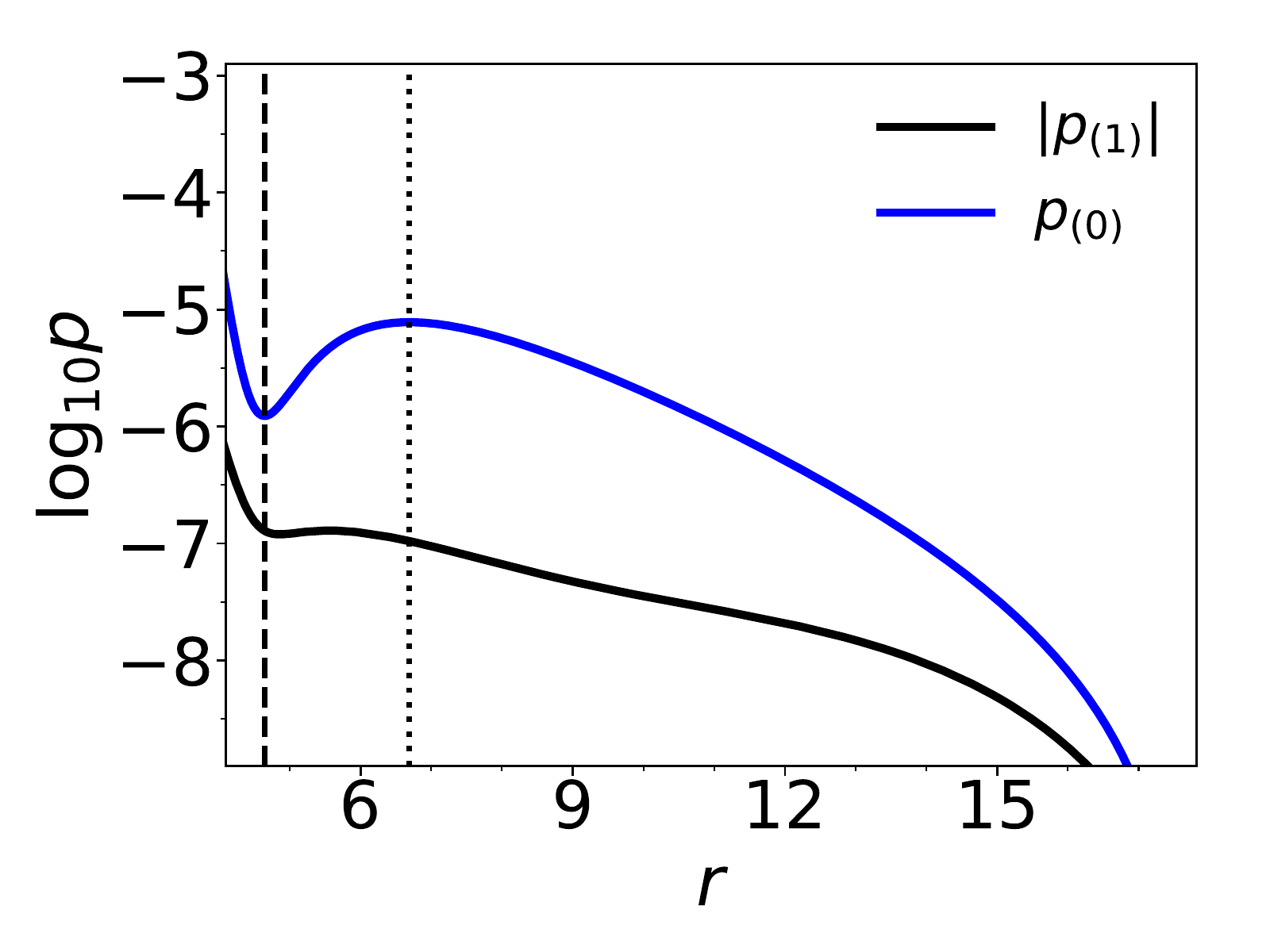}
\hspace{-0.2cm}
\includegraphics[scale=0.27]{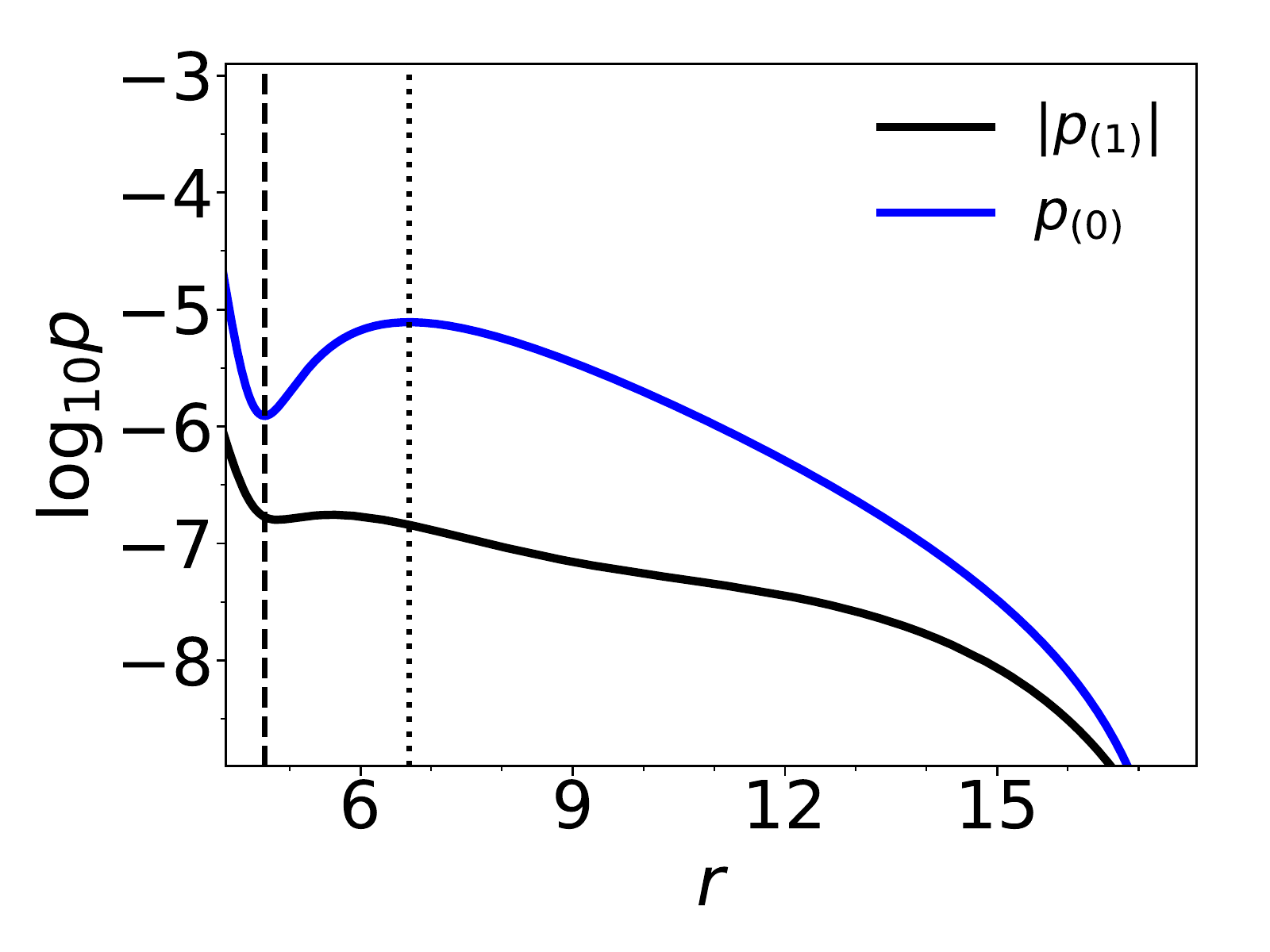}
\\
\includegraphics[scale=0.27]{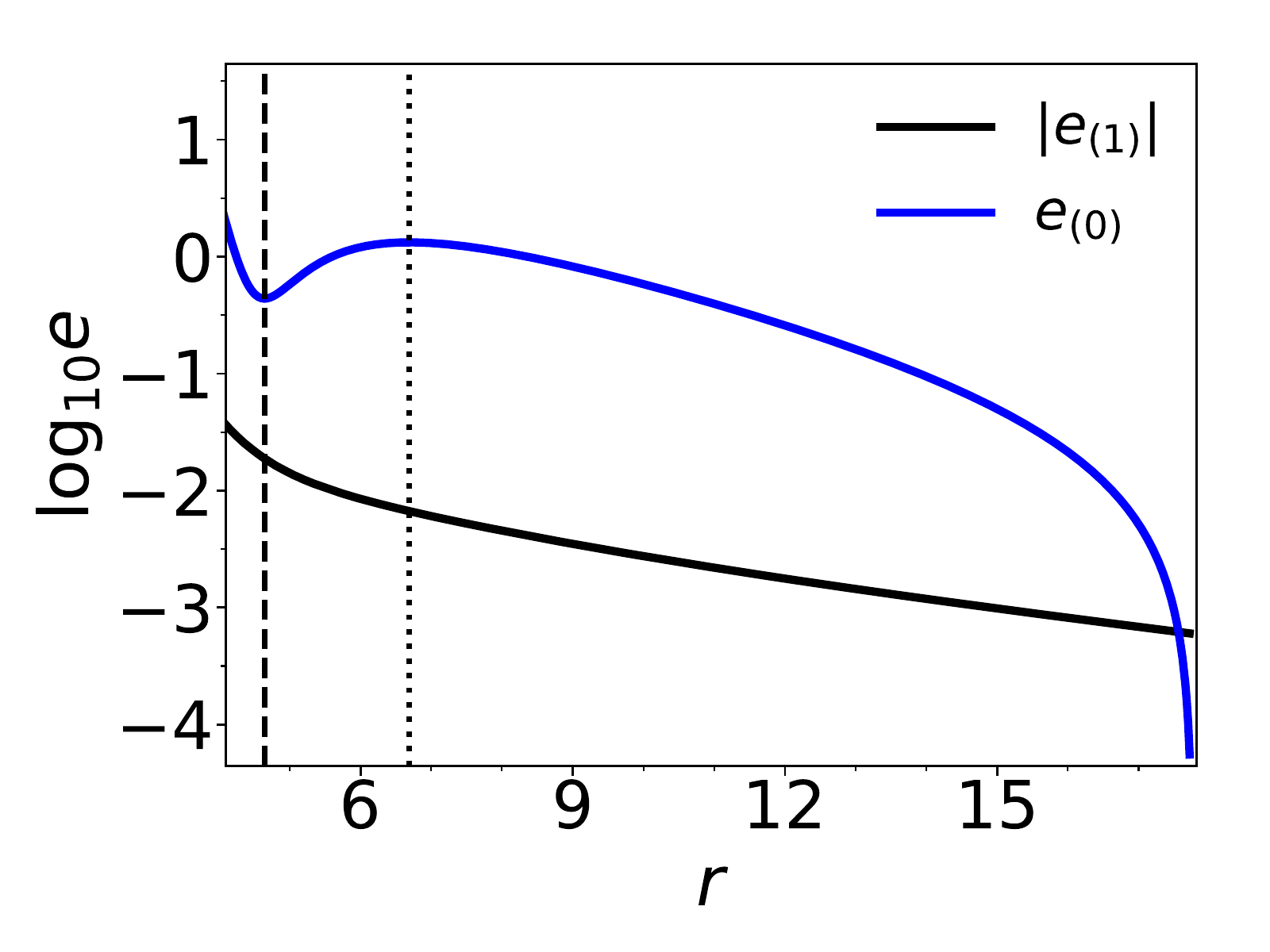}
\hspace{-0.3cm}
\includegraphics[scale=0.27]{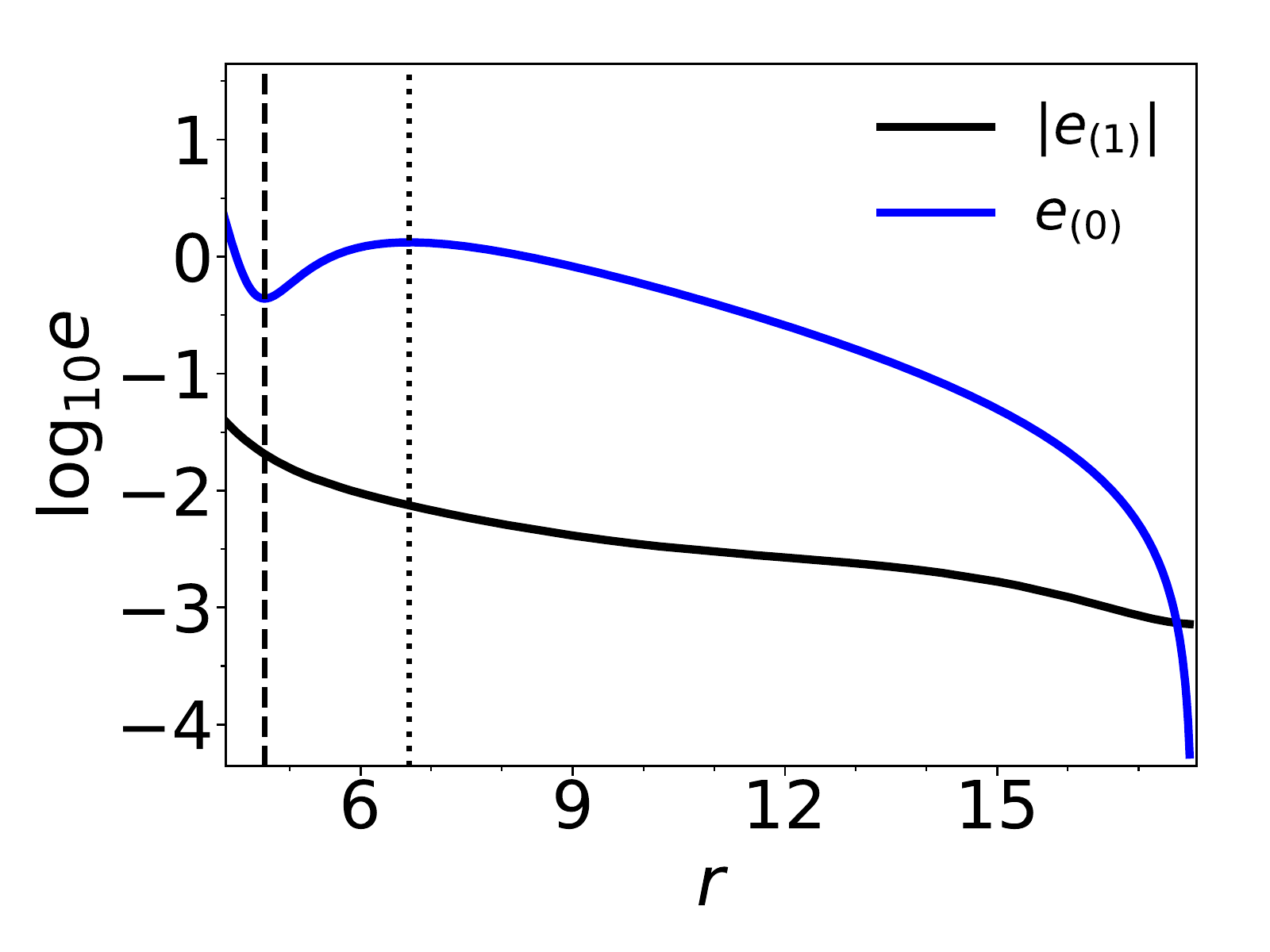}
\hspace{-0.2cm}
\includegraphics[scale=0.27]{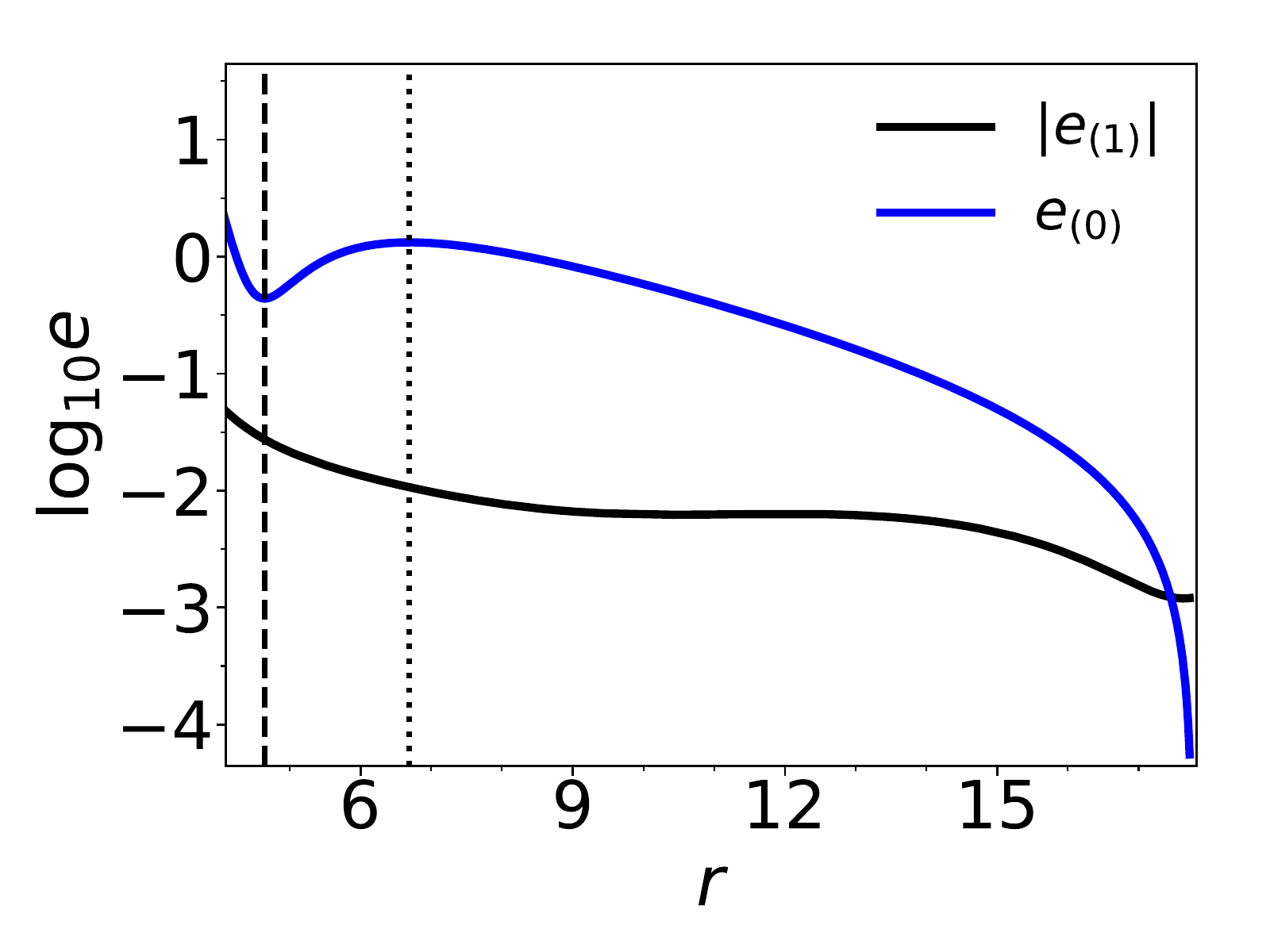}
\hspace{-0.2cm}
\includegraphics[scale=0.27]{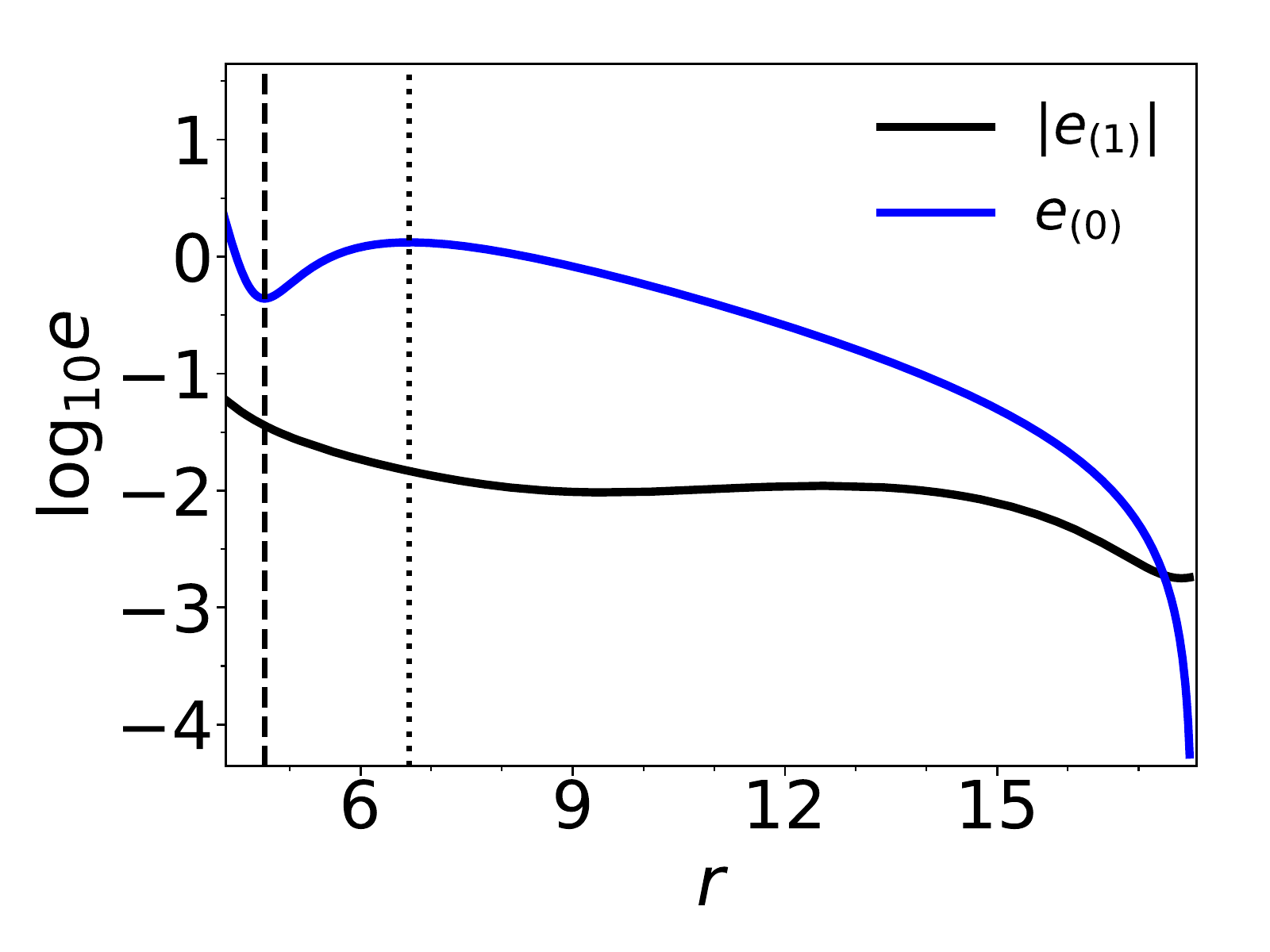}
\caption{Same as Fig.~\ref{radial_plots_beta_3} but for $\beta_{\mathrm{m,c}} = 10^{-3}$.}
\label{radial_plots_beta_4}
\end{figure*}

\subsection{Numerical implementation}

The numerical implementation of the procedure we have just described is as follows: First, we start by defining $N_x$ equally spaced points in the open interval (at the equatorial plane) $(r_{\mathrm{in}}, r_{\mathrm{out}})$,  where $r_{\mathrm{in}} = 3.7$ and $r_{\mathrm{out}}$ is the only solution of the equation $W(r, \pi/2) - W_{\mathrm{s}} = 0$ and its value is $r_{\mathrm{out}} = 17.76$. In this work we fix $N_x = 703$ which corresponds to a distance between points $\Delta x = 0.02$. Starting from this set of points we integrate Eq.~\eqref{eq:characteristic_curves_final} using the fourth-order Runge-Kutta method with $W(x,y) > W_{\mathrm{s}}$ as the terminating condition of the integration and an integration step $h = 10^{-3}$.
As a result of the previous step we obtain a set of points belonging to the boundary of the domain and a set of characteristic curves $\{x(y)_i \; / \; i = [1,N_x] \}$ that start at the boundary and end at the equatorial plane. An example of the distribution of the characteristic curves for the case $W_{\mathrm{s}}=-0.039$ is depicted in Fig.~\ref{characteristic_curves}.
Now, we can integrate Eq.~\eqref{eq:p1_y} along the characteristics, starting from the boundary $(x_0,y_0)_i$. To do this we use the same fourth-order Runge-Kutta solver as before (which in this case reduces to Simpson's rule) and the initial condition $p_{\mathrm{(0)}_0} = 0$.

\begin{figure*}[t]
\includegraphics[scale=0.17]{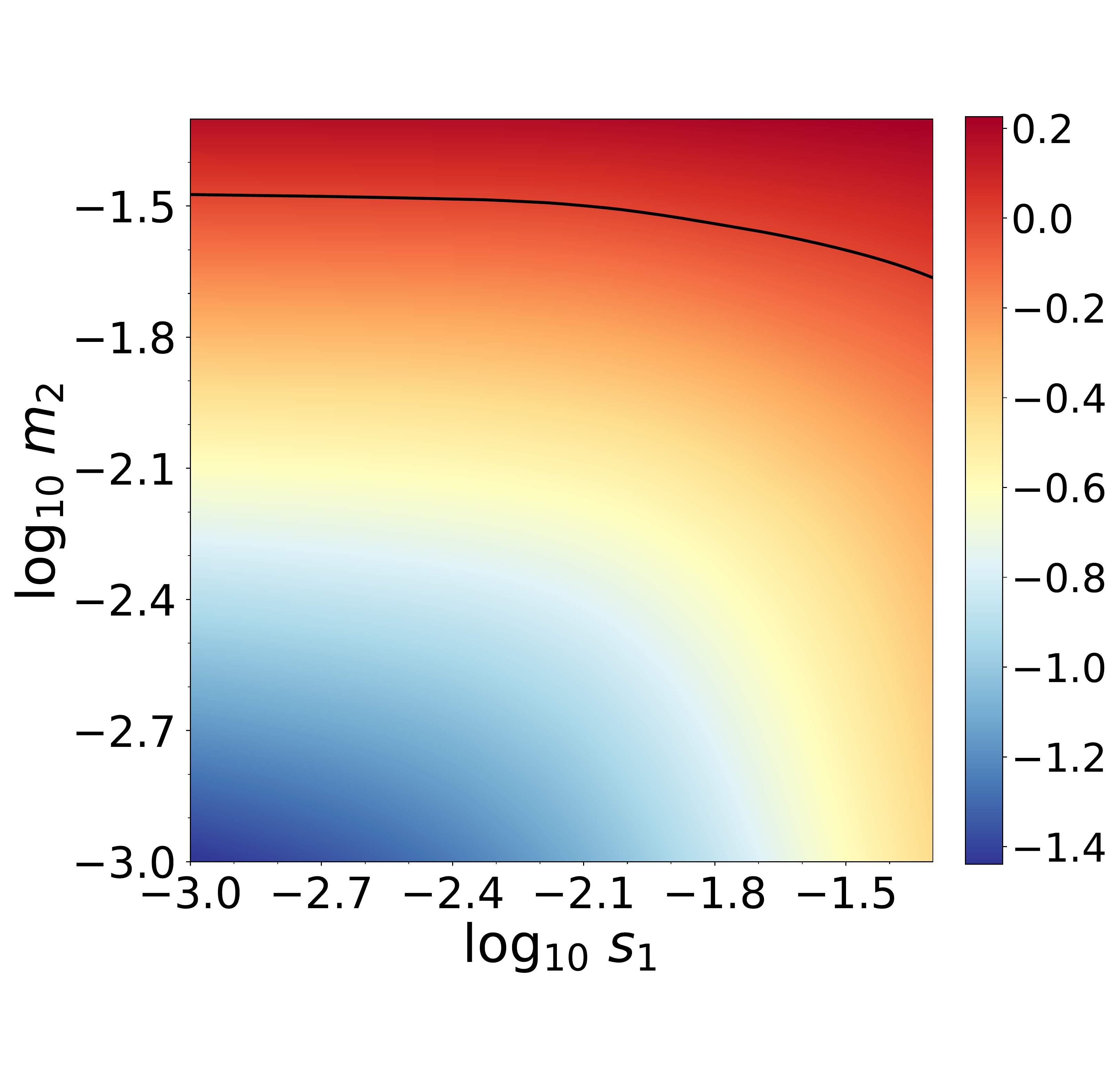}
\hspace{-0.3cm}
\includegraphics[scale=0.17]{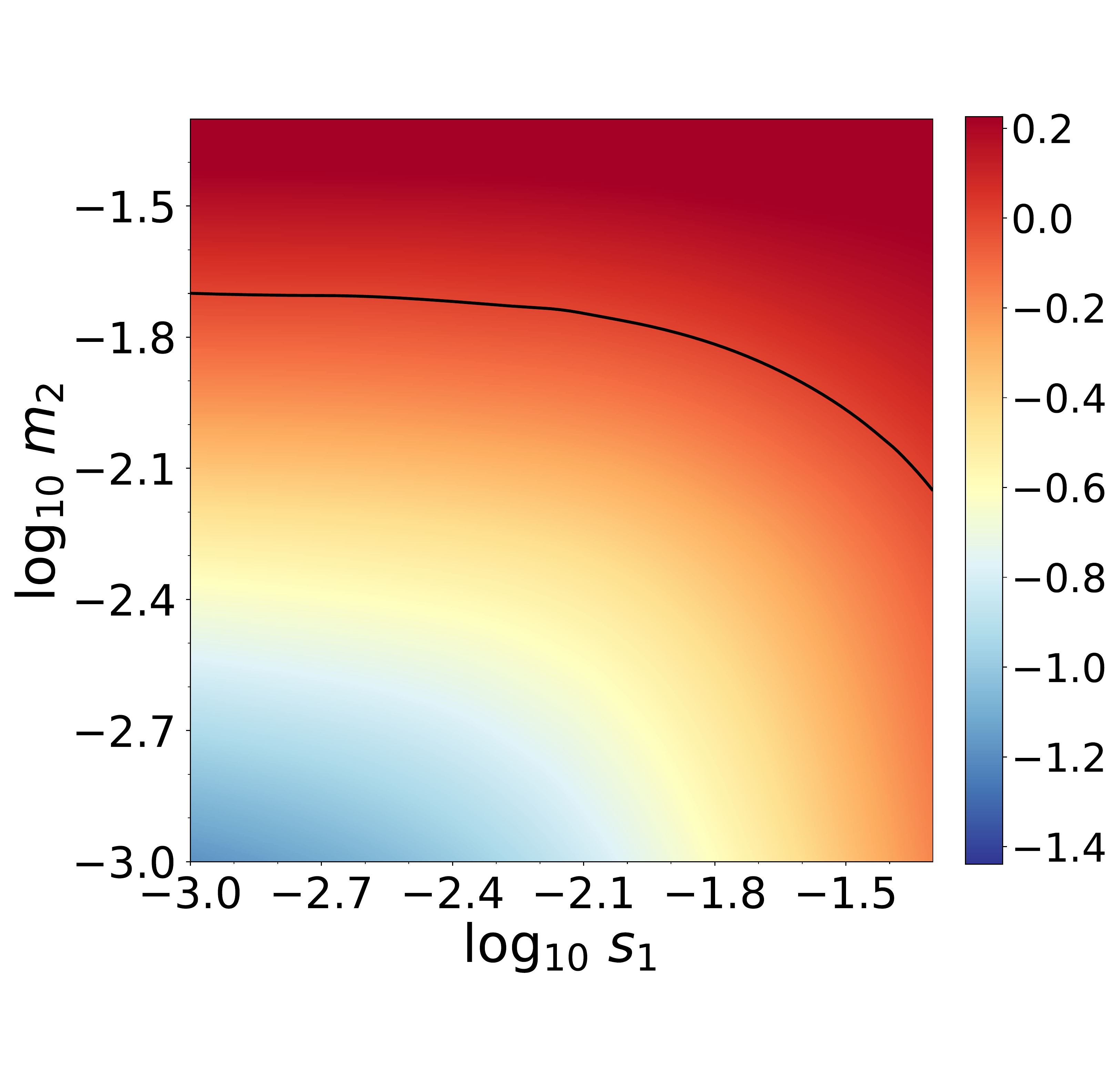}
\hspace{-0.2cm}
\includegraphics[scale=0.17]{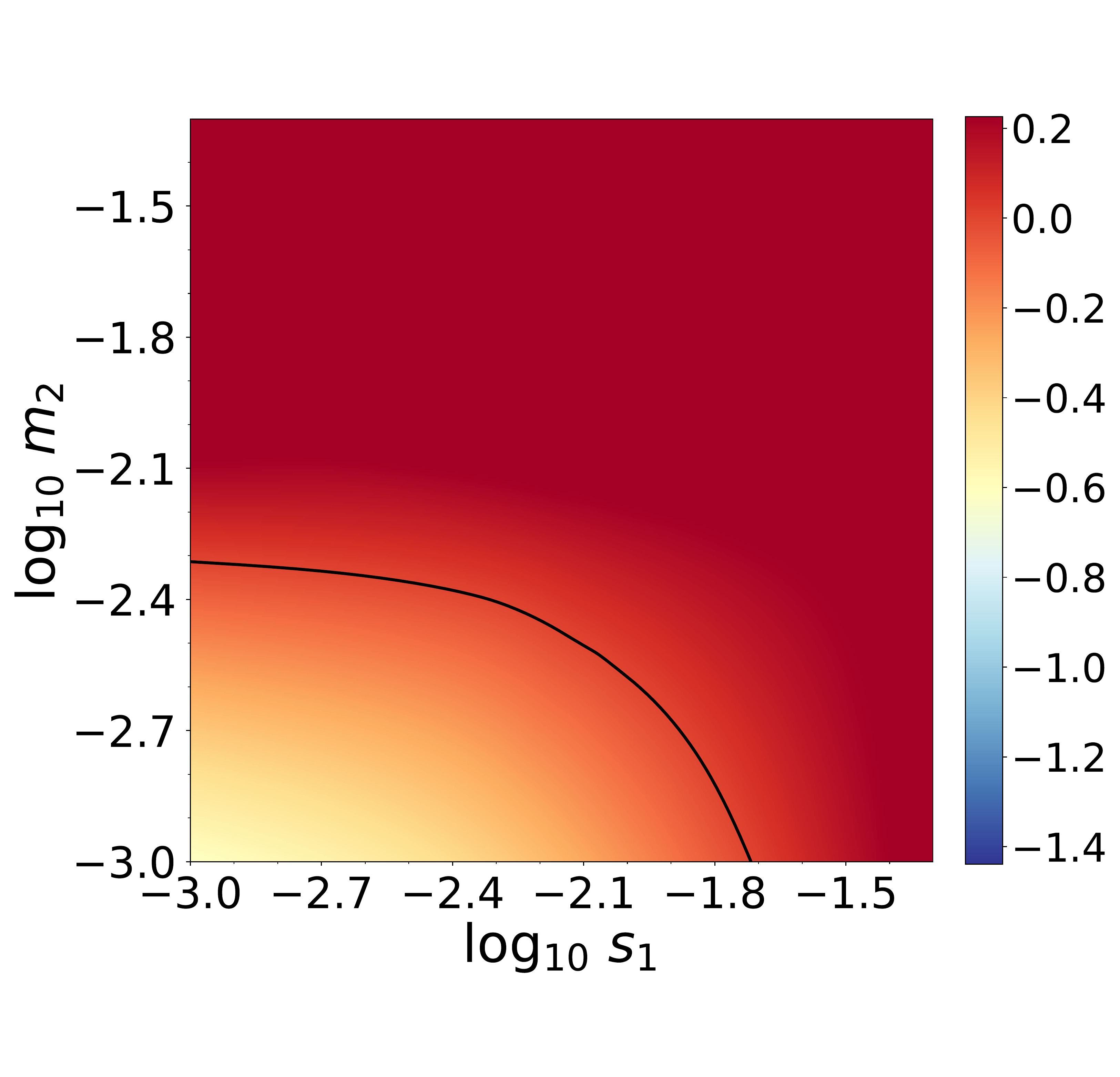}
\\
\vspace{-0.5cm}
\includegraphics[scale=0.17]{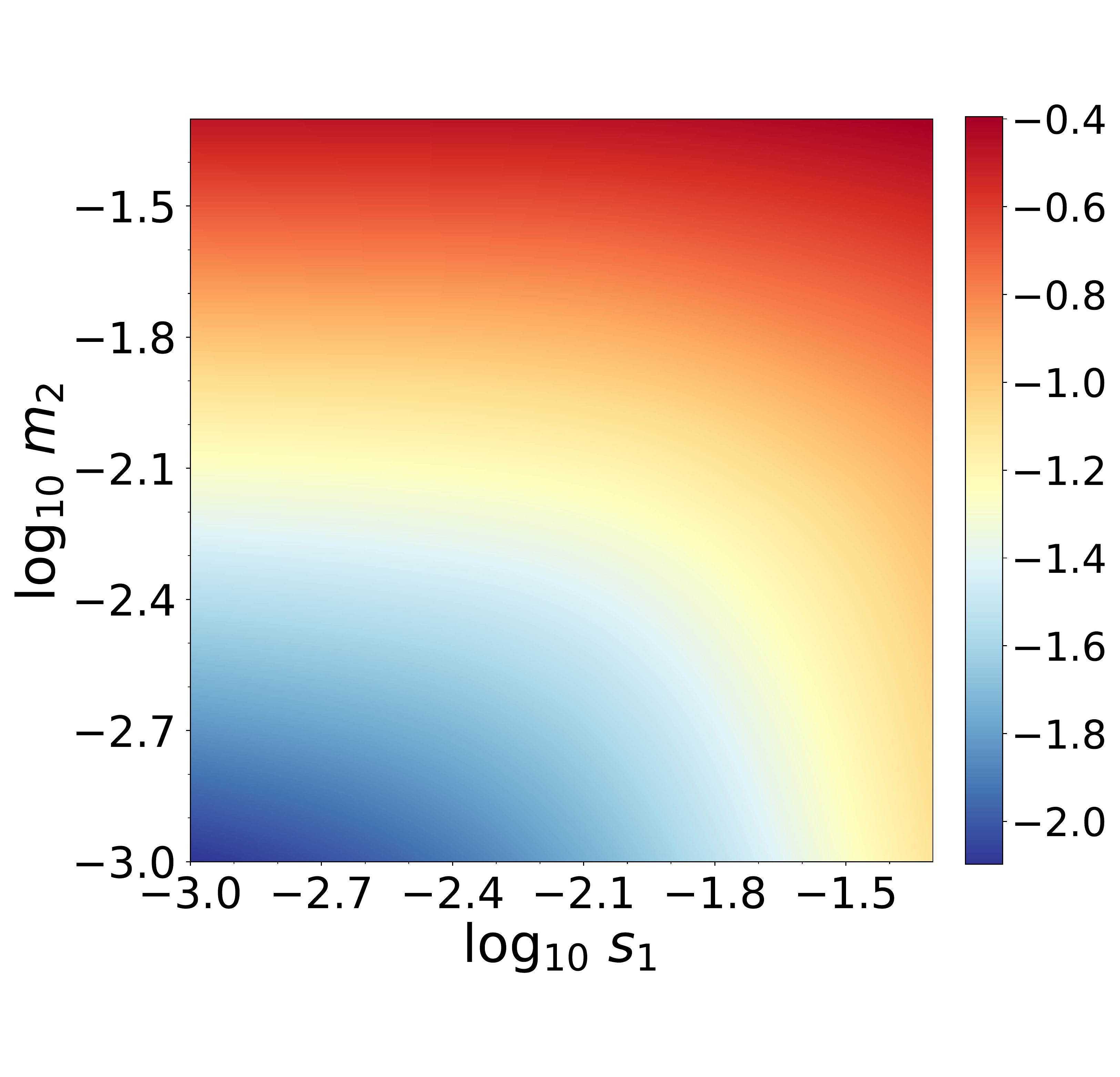}
\hspace{-0.3cm}
\includegraphics[scale=0.17]{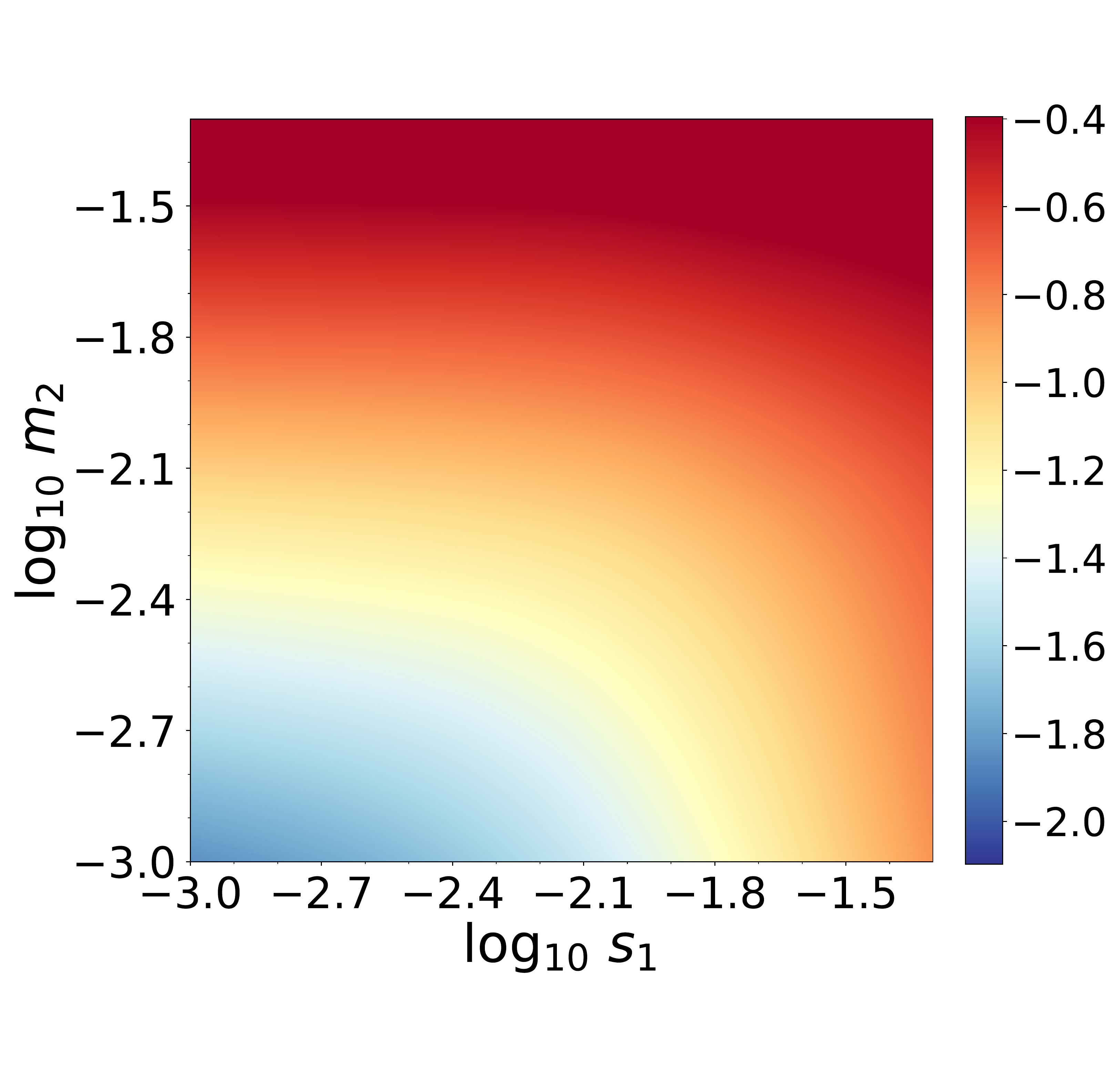}
\hspace{-0.2cm}
\includegraphics[scale=0.17]{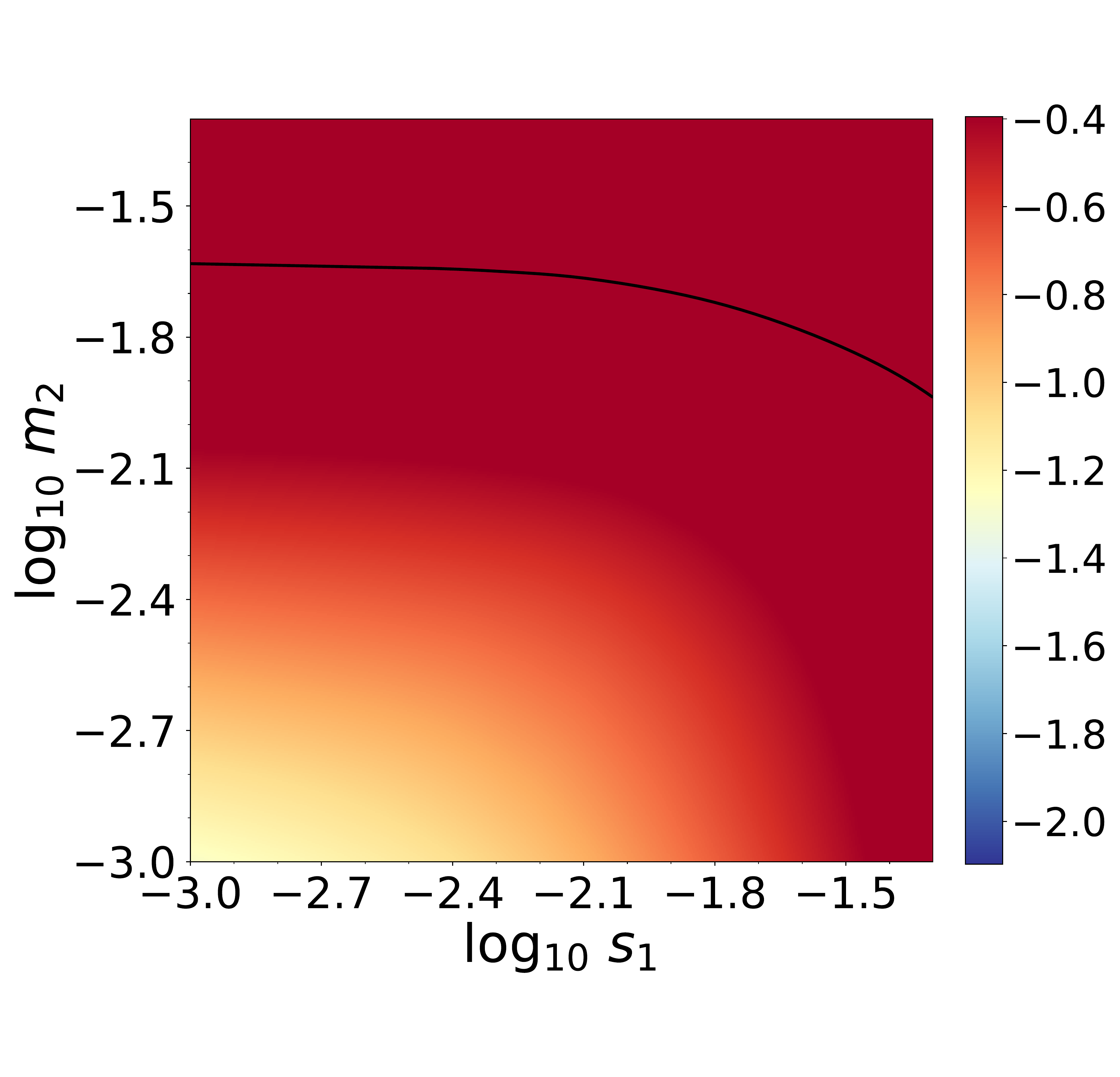}
\caption{Two dimensional plots for $\log_{10} |\Delta p_{\mathrm{cusp}}|$. The first row corresponds to the models with $\beta_{\mathrm{m,c}} = 10^3$ and the second row correspond to the models with $\beta_{\mathrm{m,c}} = 10^{-3}$. The columns correspond, from left to right, to the three different values of $W_{\mathrm{s}}$ we have considered, namely $-0.039$, $-0.040$ and $-0.041$. The black contour appearing in some of the plots corresponds to $\log_{10} |\Delta p_{\mathrm{cusp}}|=0$. }
\label{deltapcusp_2D}
\end{figure*}	 

\section{Results}
\label{results}

The primary motivation of this paper is to determine possible changes in the morphology of geometrically thick magnetized disks in the presence of shear viscosity as compared to the inviscid case. We use a simple setup where stationary viscous disks with constant angular momentum distributions are built around a Schwarzschild black hole. The shear viscosity is assumed to only induce perturbative effects on the fluid so that the fluid in the disk can still move in circular orbits. The analysis of isopressure and isodensity surfaces of our constrained system provides evidences showing that the shear viscous and curvature effects in the stationary disk models are only tractable using the causal approach.

Stationary magnetized tori are constructed for a set of values of the parameters  $\tau_2$, $m_{1}$, $m_{2}$ and the magnetization parameter at the center of the disk, $\beta_{m,c}$ (Note that, to build the solutions, we have to fix the polytropic exponent $\gamma$ and the value of the zeroth-order correction to the energy density at the center $e_{\mathrm{(0)}, c} = e_{\mathrm{(0)}}(r_{\mathrm{c}}, \pi/2)$. In particular, we have chosen $\gamma = 5/3$ and $e_{\mathrm{(0)}, c} = 1$). For convenience, we define a new parameter $s_1=\tau_2 \, m_1$ and set $\tau_2=1$ without loss of generality. We consider two values of the magnetization parameter at the center of the disk, namely $\beta_{m,c}=10^{3}$ (low magnetization, almost a purely hydrodynamical model) and  $\beta_{m,c}=10^{-3}$ (high magnetization) which are sufficient to bring out the effects of a toroidal magnetic field on the viscous disk.

The corrections to the pressure $p_{(1)}$ and to the energy density $e_{(1)}$ for a given choice of parameters are determined by solving Eq.~(\ref{eq:PDE_fin}) numerically, using the method of characteristics as described in the previous section. Our results reveal that the effects of the shear viscosity are particularly noticeable only fairly close to the cusp of the disks. The large-scale morphology of the torus remains essentially unaltered irrespective of the values of the parameters $s_1$ and $m_2$. This can be immediately concluded from figure~\ref{full-disk} which displays the distribution of the pressure in the entire domain for a set of illustrative stationary models. Note that the physical solution is attached to the black hole, even though in the figure there is a gap between the disk and the event horizon. This is due to the fact that Eq.~\eqref{eq:eq_cartesian} is singular at the event horizon, so the solution cannot be extended to it. Figures \ref{radial_plots_beta_3} and \ref{radial_plots_beta_4} display  radial plots at the equatorial plane showing the zeroth-order and first-order corrections of the pressure and of the energy density, corresponding to the low and high value of the magnetization parameter, respectively. We note that, contrary to purely hydrodynamical disks, for magnetized tori the location of the center of the disk $r_{\mathrm{c}}$ does not exactly coincide with the location of the maximum of the pressure but it is slightly shifted towards the black hole~\cite{Gimeno_Soler_2017}. This can be observed for the highly magnetized case in figure \ref{radial_plots_beta_4}.  For both low and high values of $\beta_{m,c}$ the corrections $p_{(1)}$ and $e_{(1)}$ near the cusp remain small in comparison to their respective equilibrium values $p_{(0)}$ and $e_{(0)}$.
As one moves away from the cusp and approaches the outer edge of the disk, the difference between $p_{(0)}$ and $p_{(1)}$ diminishes. This trend is most prominent for low magnetized disk as shown in figure \ref{radial_plots_beta_3}. In addition, by increasing the value of $m_2$, i,e.~the curvature effects (while keeping $m_1$ fixed), the difference between $p_{(0)}$ and $p_{(1)}$ also decreases near the cusp, until a value is reached for which  
${p_{(1)}}/{p_{(0)}} \sim {\cal O}(1) $ and 
${e_{(1)}} / {e_{(0)}} \sim {\cal O}(1)$ and neither $m_1$ nor $m_2$ can further be increased. Under these conditions we are no longer in the regime of  validity of near-equilibrium hydrodynamics where gradients are small. Since we are not addressing the non-equilibrium sector, our analysis can set an upper limit on the contributions of curvature and shear viscosity on  stationary solutions of magnetized viscous disks before far-from-equilibrium effects set in. 

\begin{figure*}[t]
\includegraphics[scale=0.14]{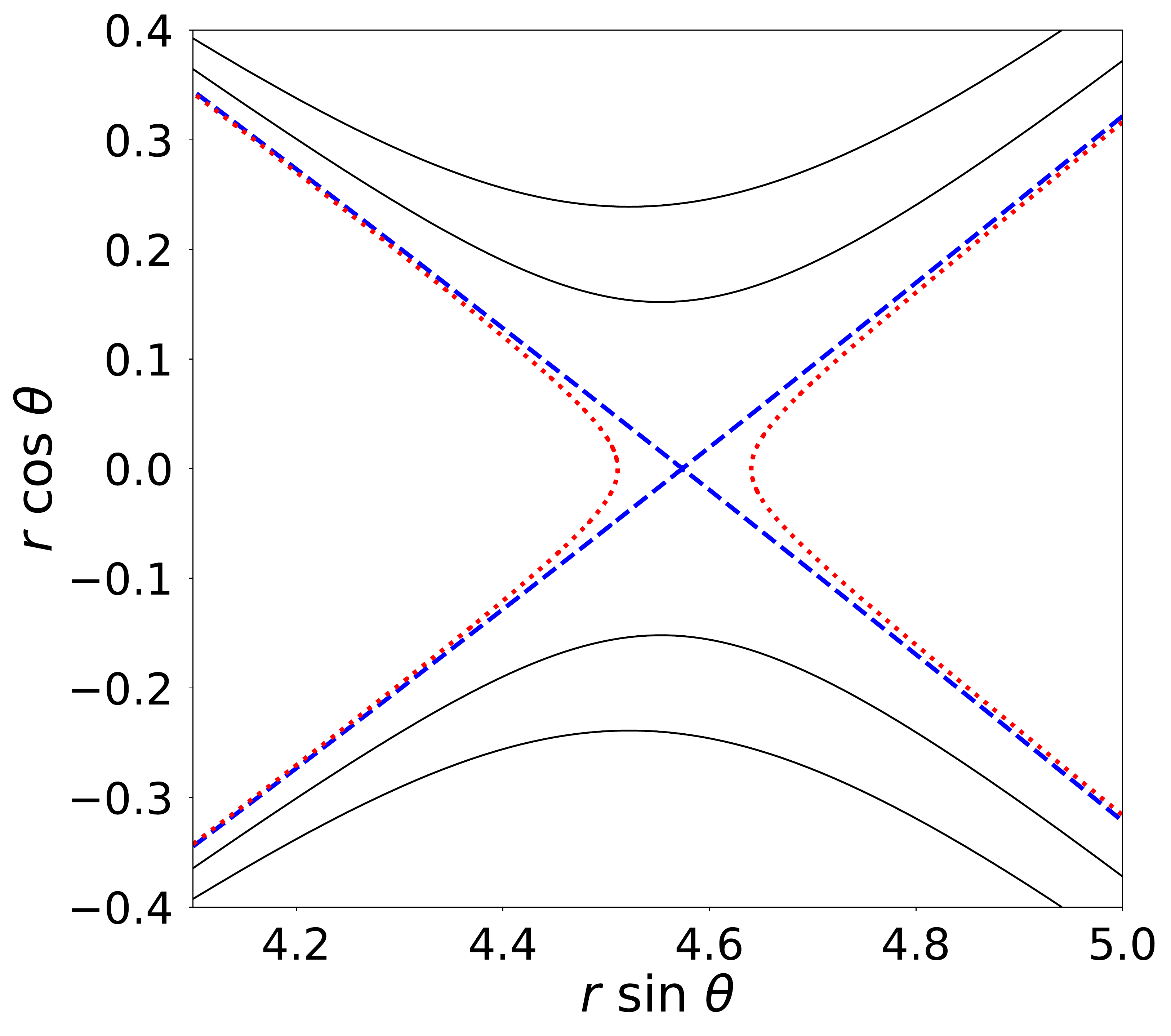}
\hspace{-0.3cm}
\includegraphics[scale=0.14]{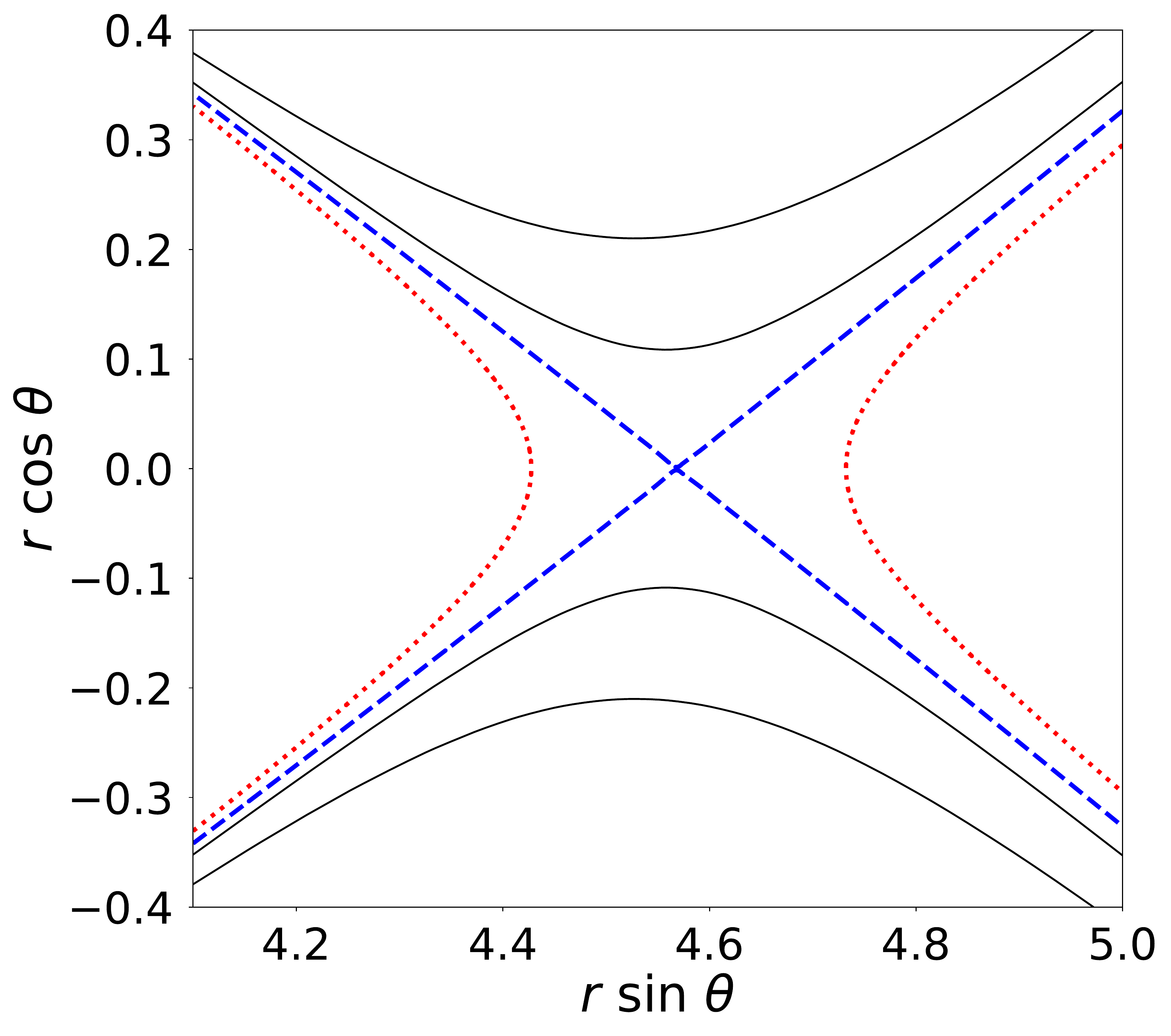}
\hspace{-0.2cm}
\includegraphics[scale=0.14]{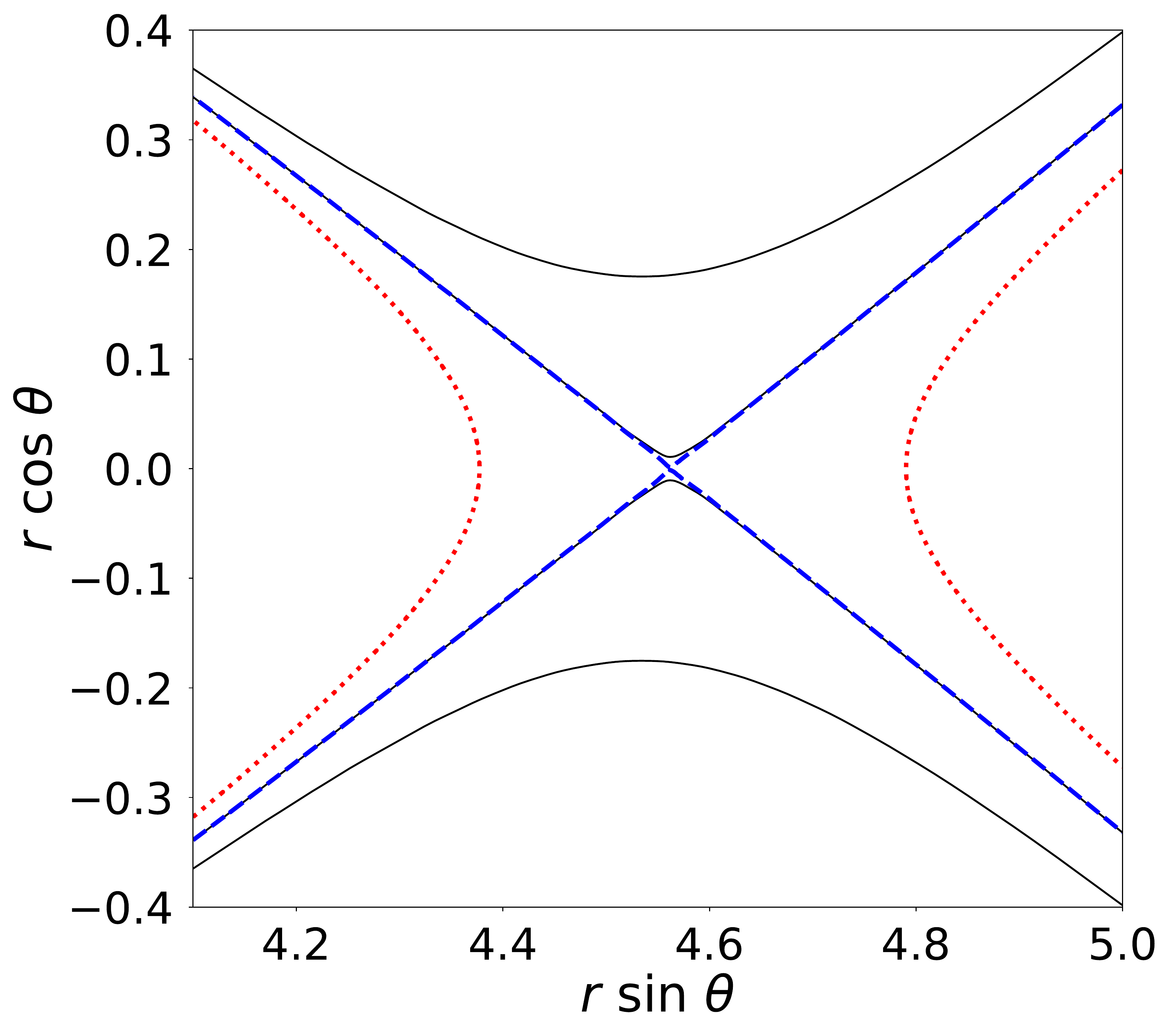}
\\
\includegraphics[scale=0.14]{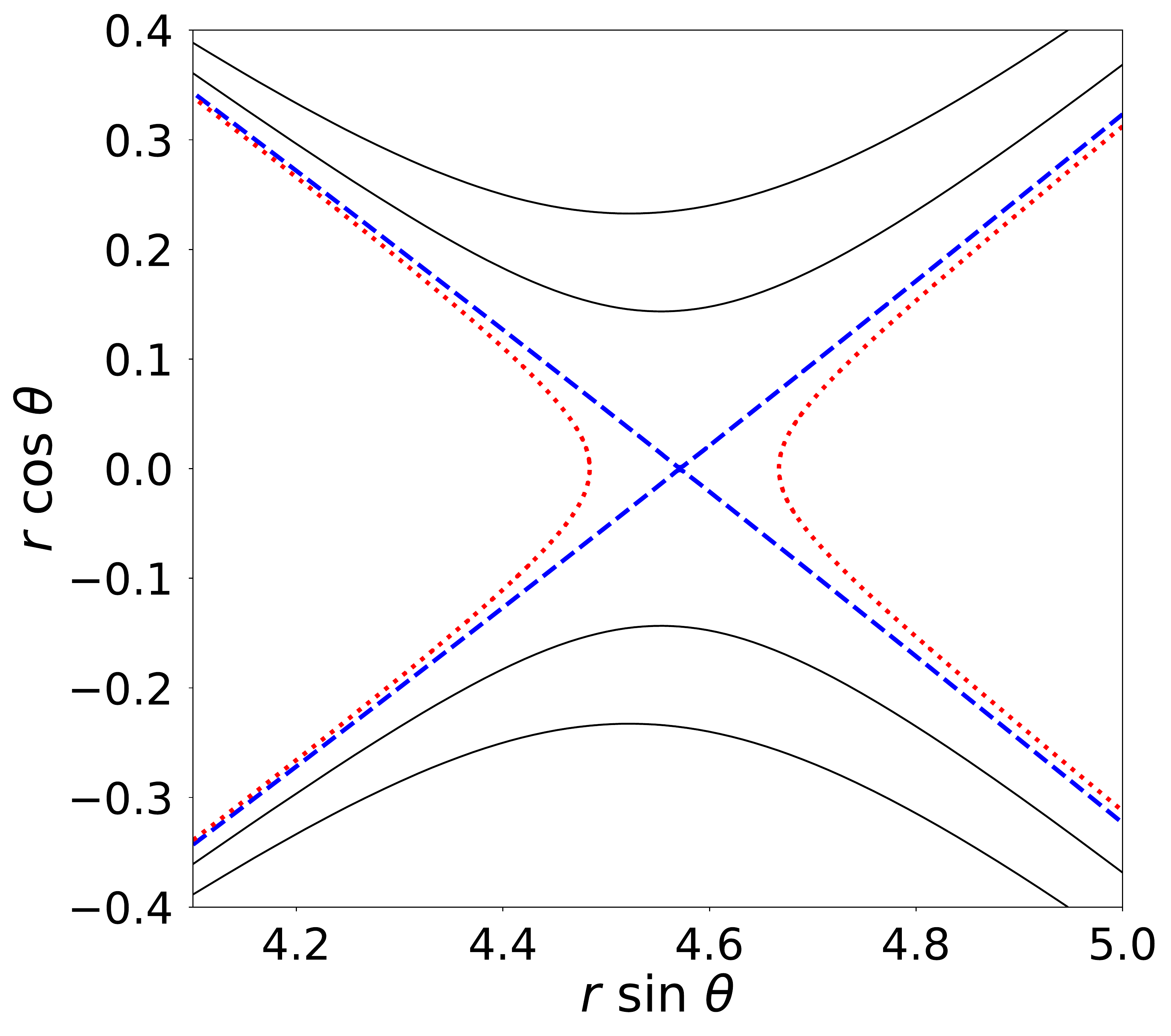}
\hspace{-0.3cm}
\includegraphics[scale=0.14]{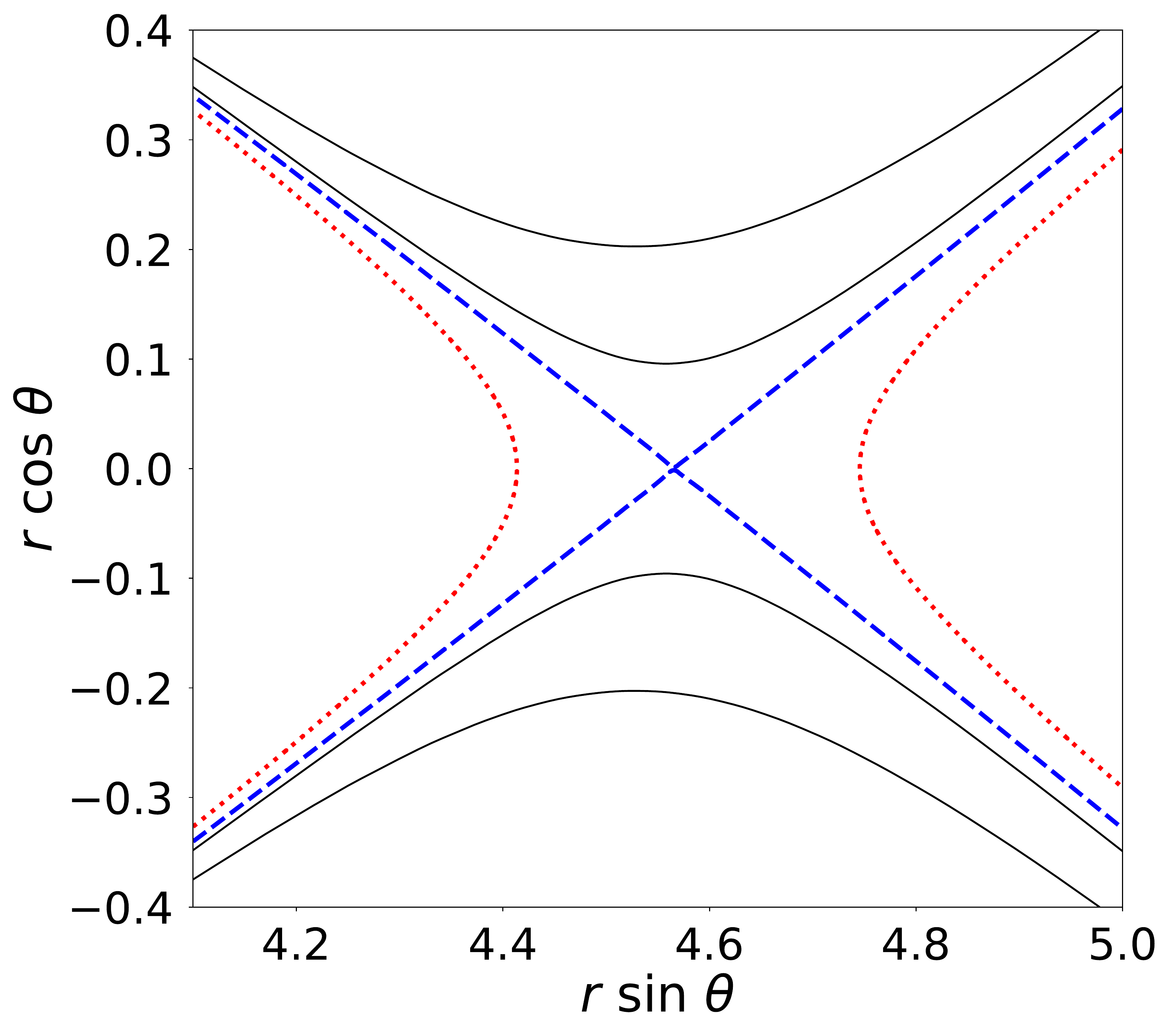}
\hspace{-0.2cm}
\includegraphics[scale=0.14]{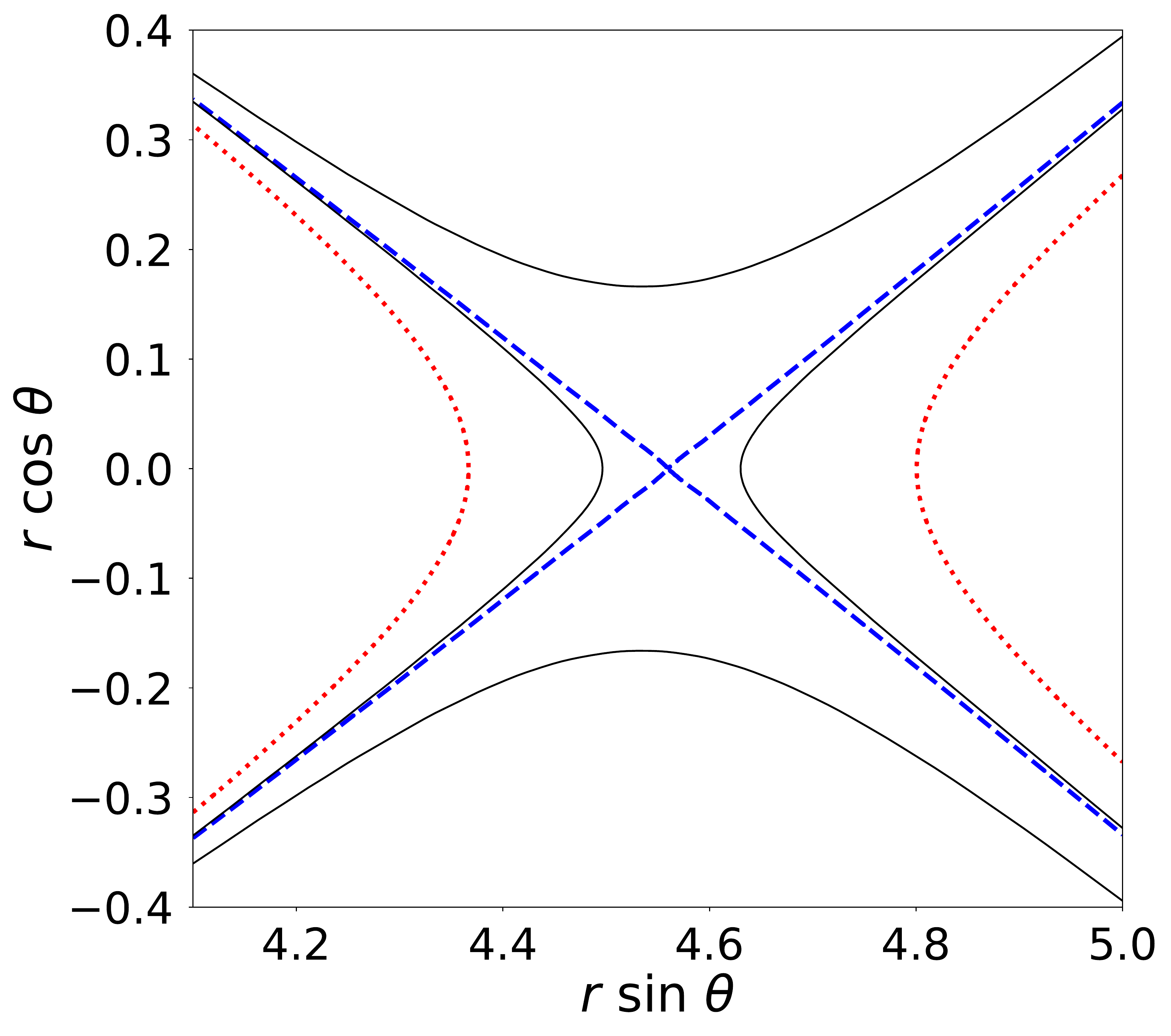}
\\
\includegraphics[scale=0.14]{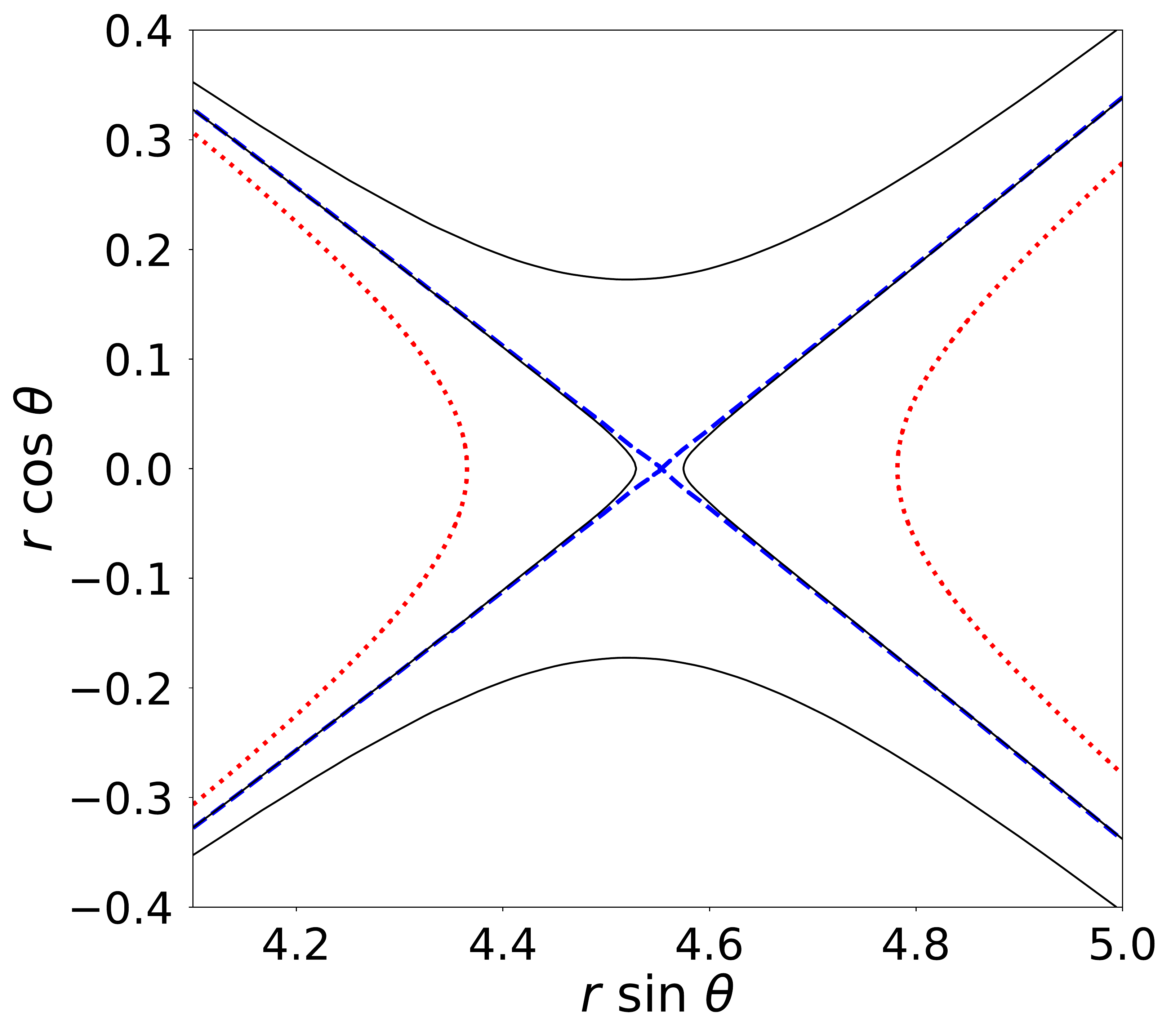}
\hspace{-0.3cm}
\includegraphics[scale=0.14]{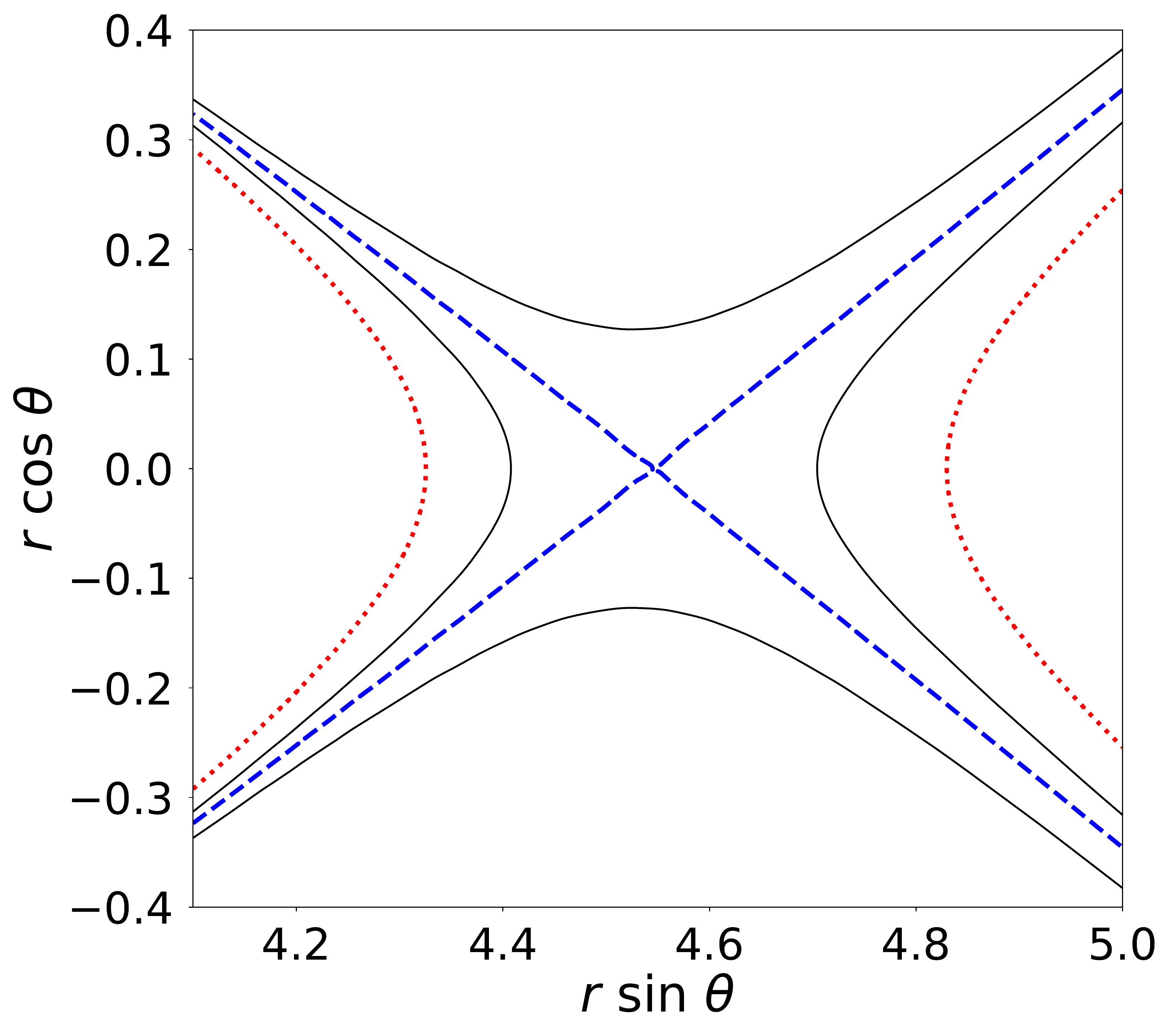}
\hspace{-0.2cm}
\includegraphics[scale=0.14]{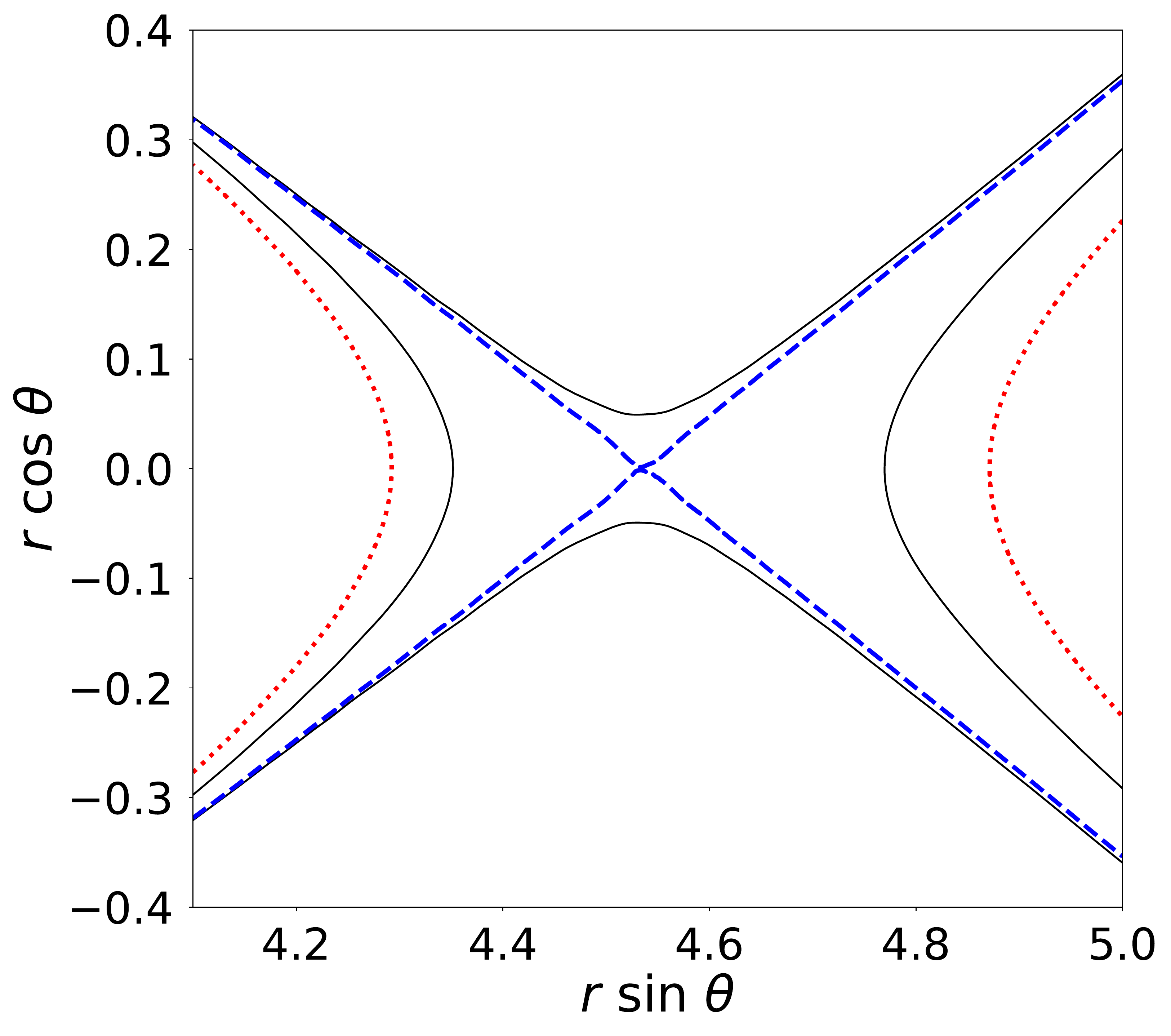}
\caption{Isocontours of $p_{(\mathrm{t})}=p_{(0)}+p_{(1)}$ in the cusp region for $W_{\mathrm{s}} = -0.039$ and $\beta_{\mathrm{m,c}} = 10^3$. From top to bottom the rows correspond to $m_2 = (0, 0.005, 0.01)$. From left to right  the columns correspond to $s_1 = (0.005, 0.01, 0.05)$. Red isocontours correspond to cusp-generating constant pressure surfaces without viscosity and blue isocontours depict newly-formed self-intersecting constant pressure surfaces when viscosity and curvature effects are present. The two black isocontours correspond to the values $p_{\mathrm{t}} = 2p_{\mathrm{0}, \mathrm{cusp}}/3$ and $p_{\mathrm{t}} = p_{\mathrm{0}, \mathrm{cusp}}/3$.}
\label{isocontours_0039_3}
\end{figure*}

\begin{figure*}[t]
\includegraphics[scale=0.14]{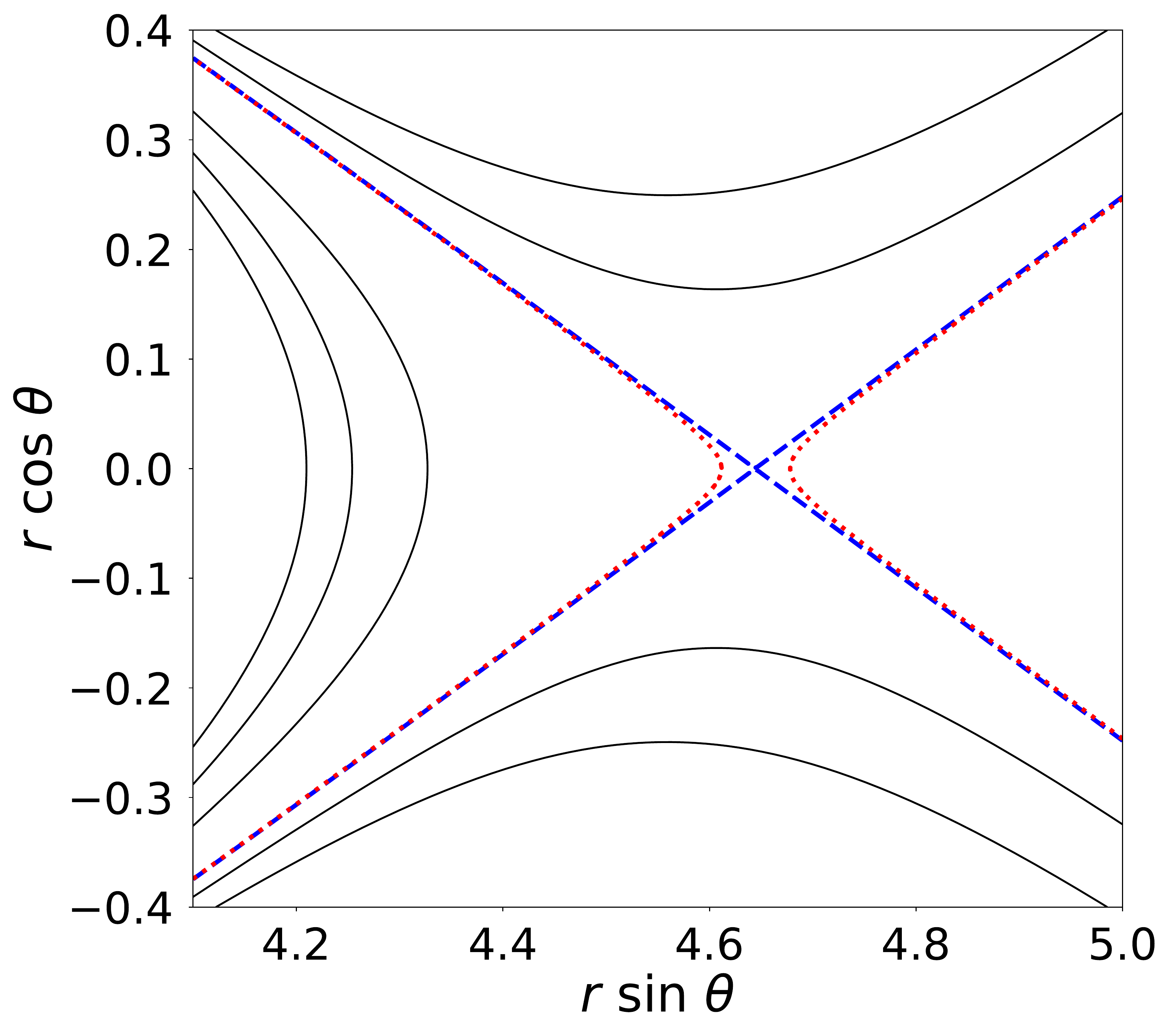}
\hspace{-0.3cm}
\includegraphics[scale=0.14]{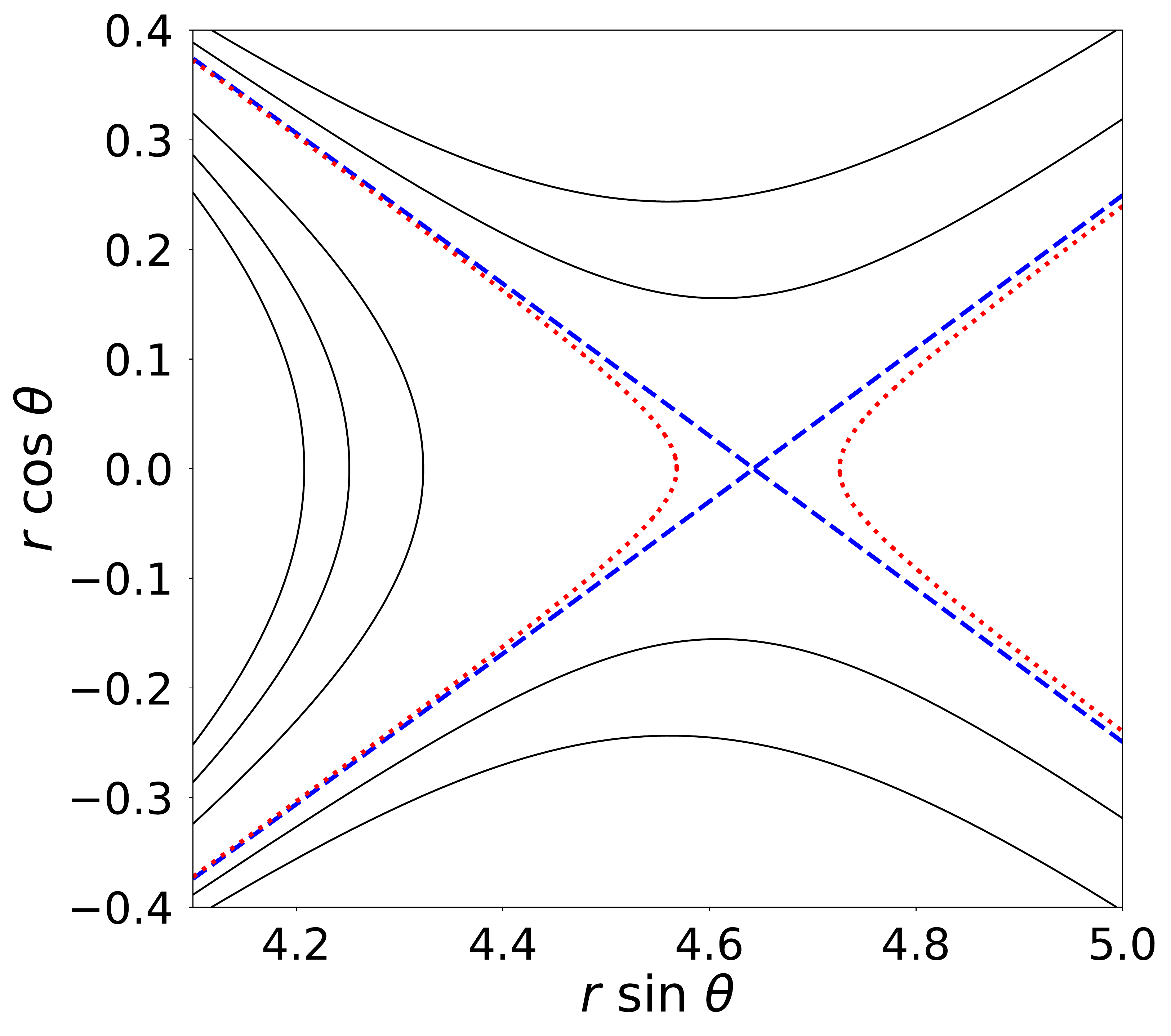}
\hspace{-0.2cm}
\includegraphics[scale=0.14]{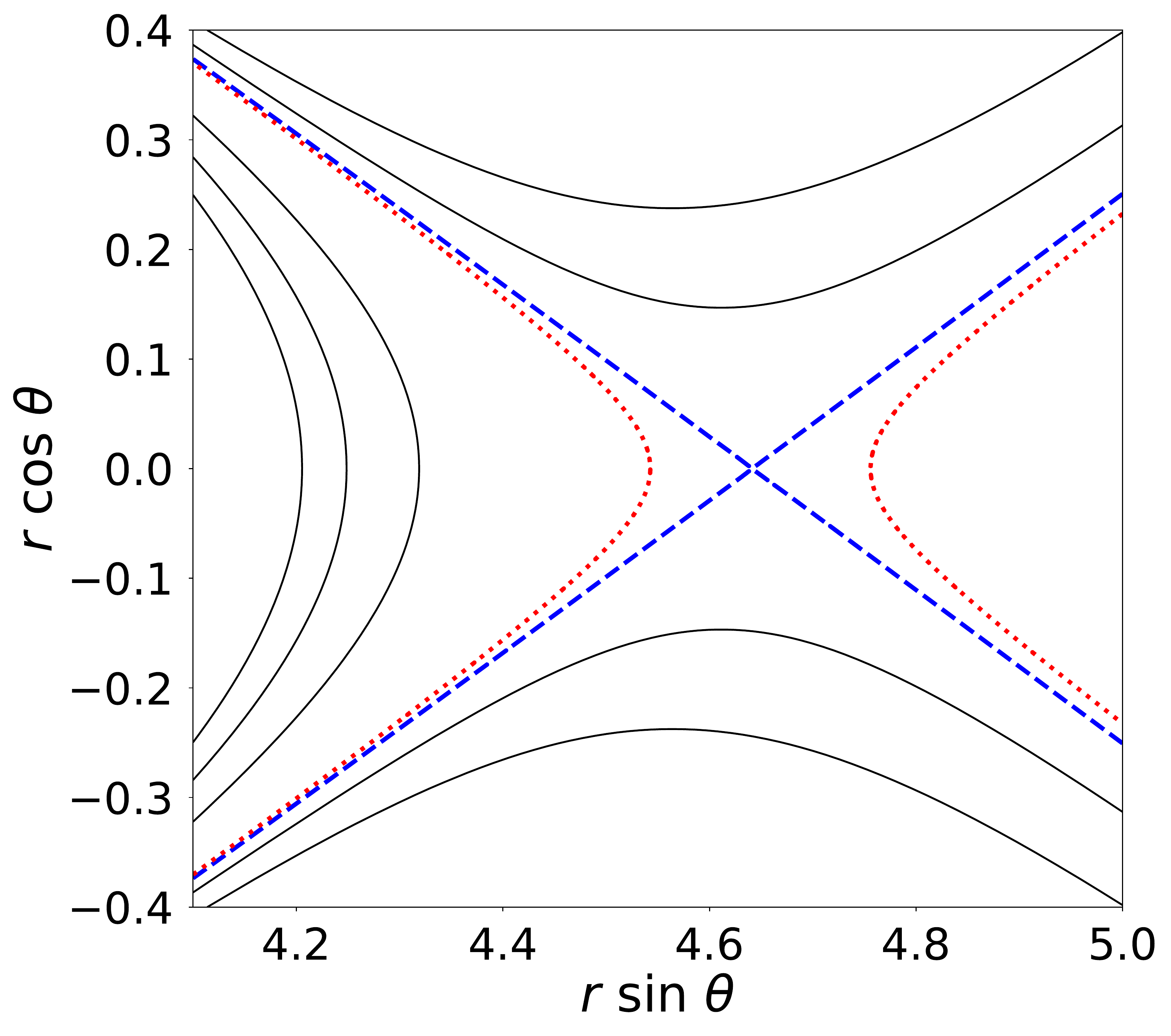}
\\
\includegraphics[scale=0.14]{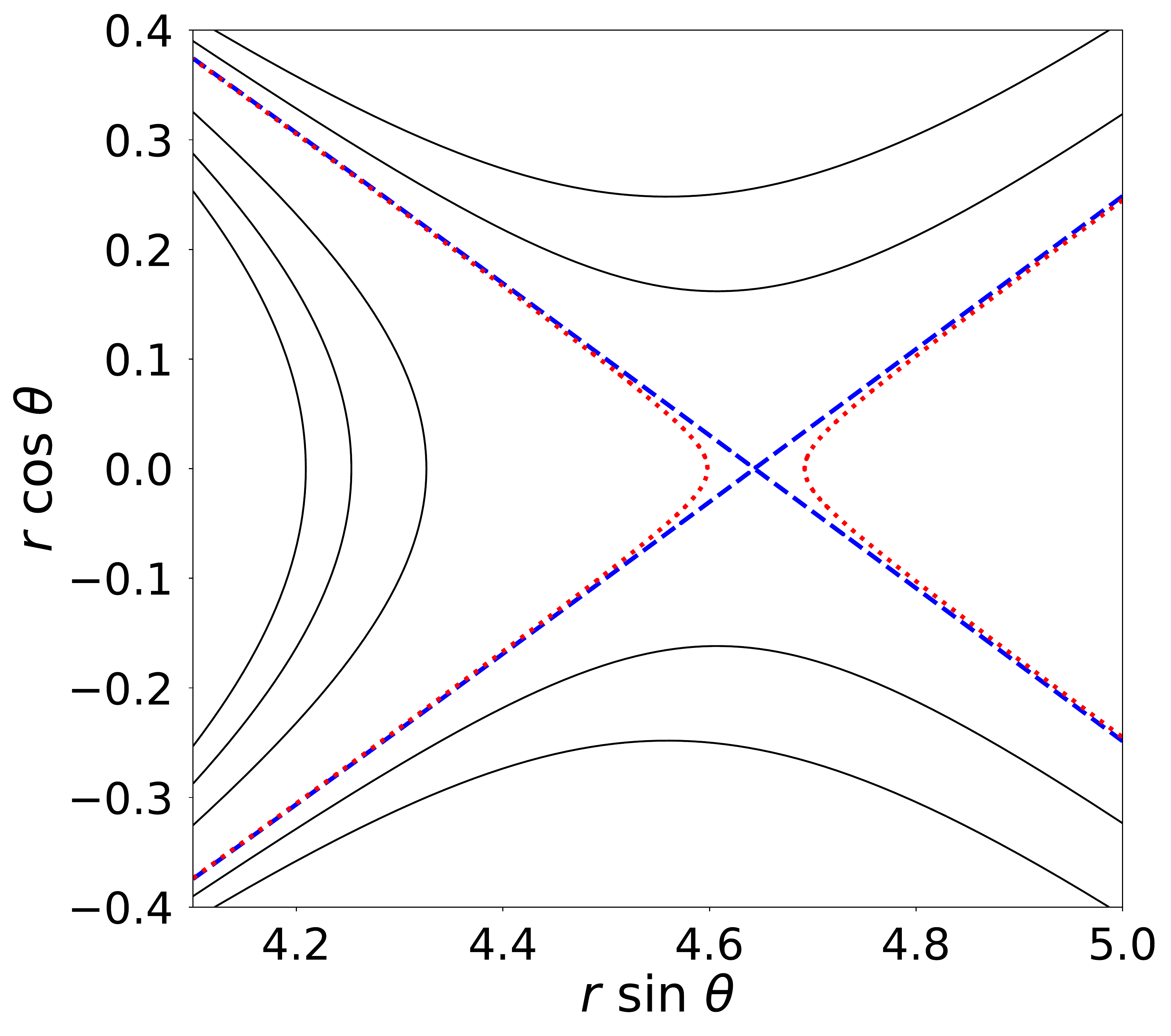}
\hspace{-0.3cm}
\includegraphics[scale=0.14]{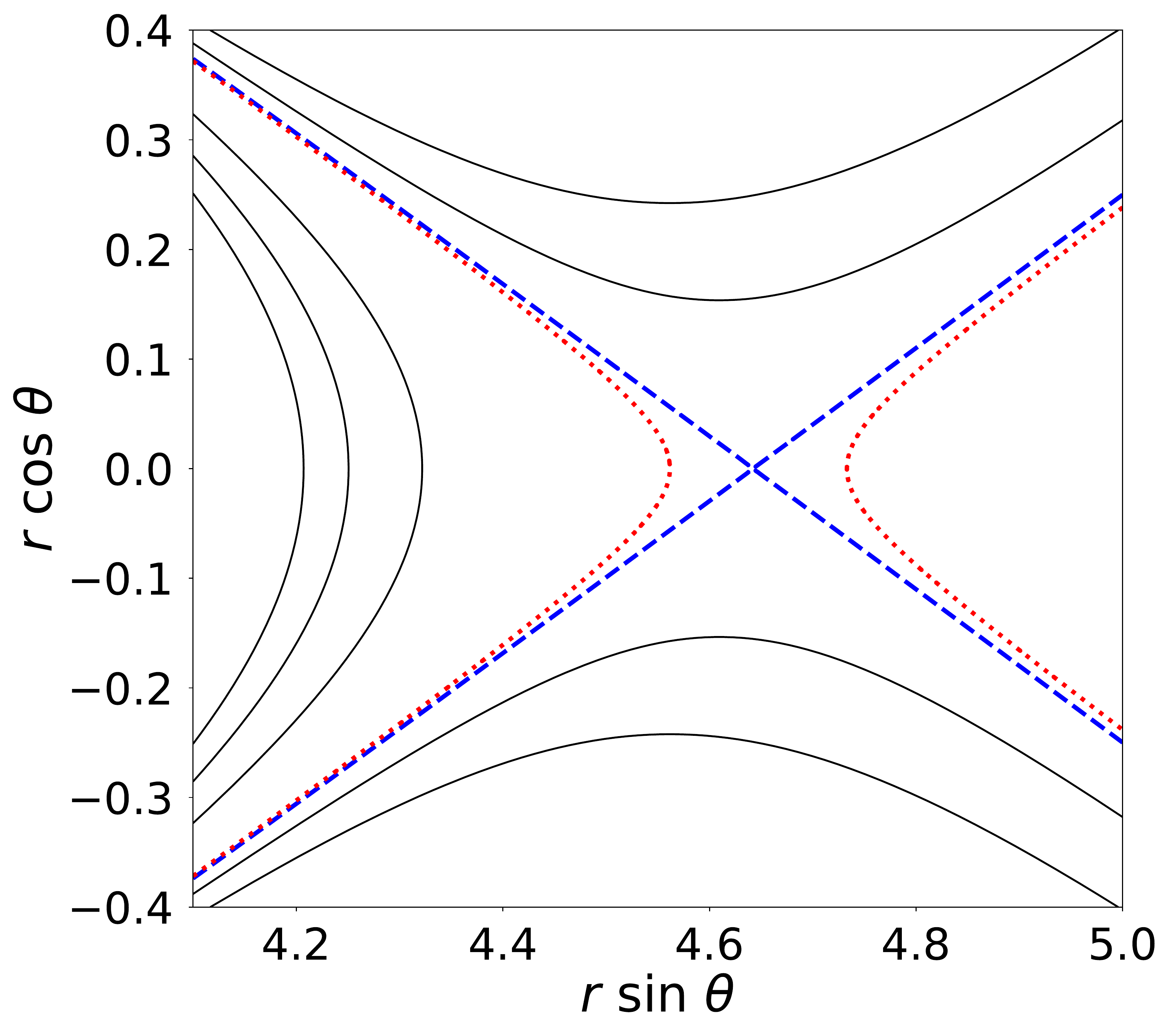}
\hspace{-0.2cm}
\includegraphics[scale=0.14]{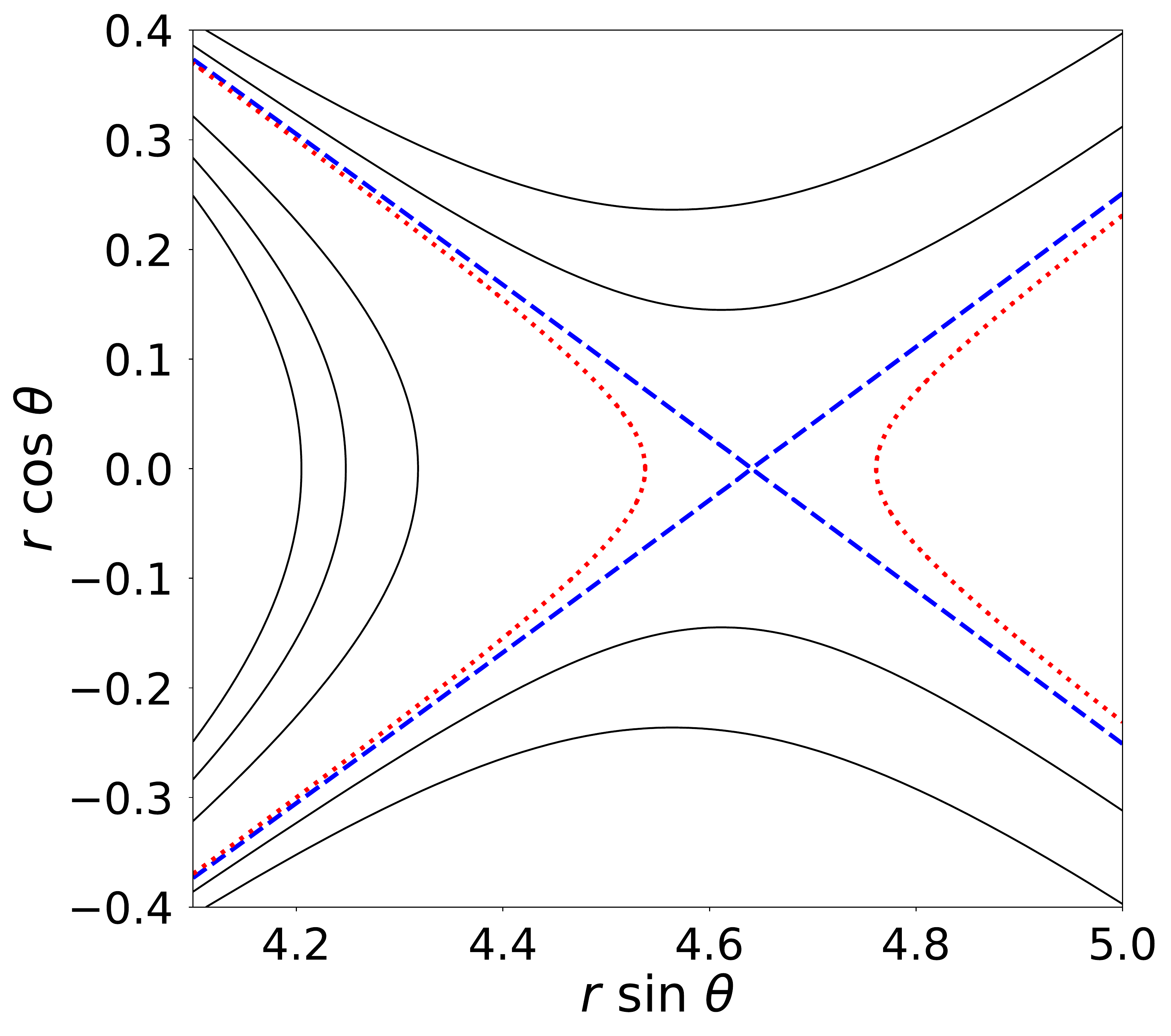}
\\
\includegraphics[scale=0.14]{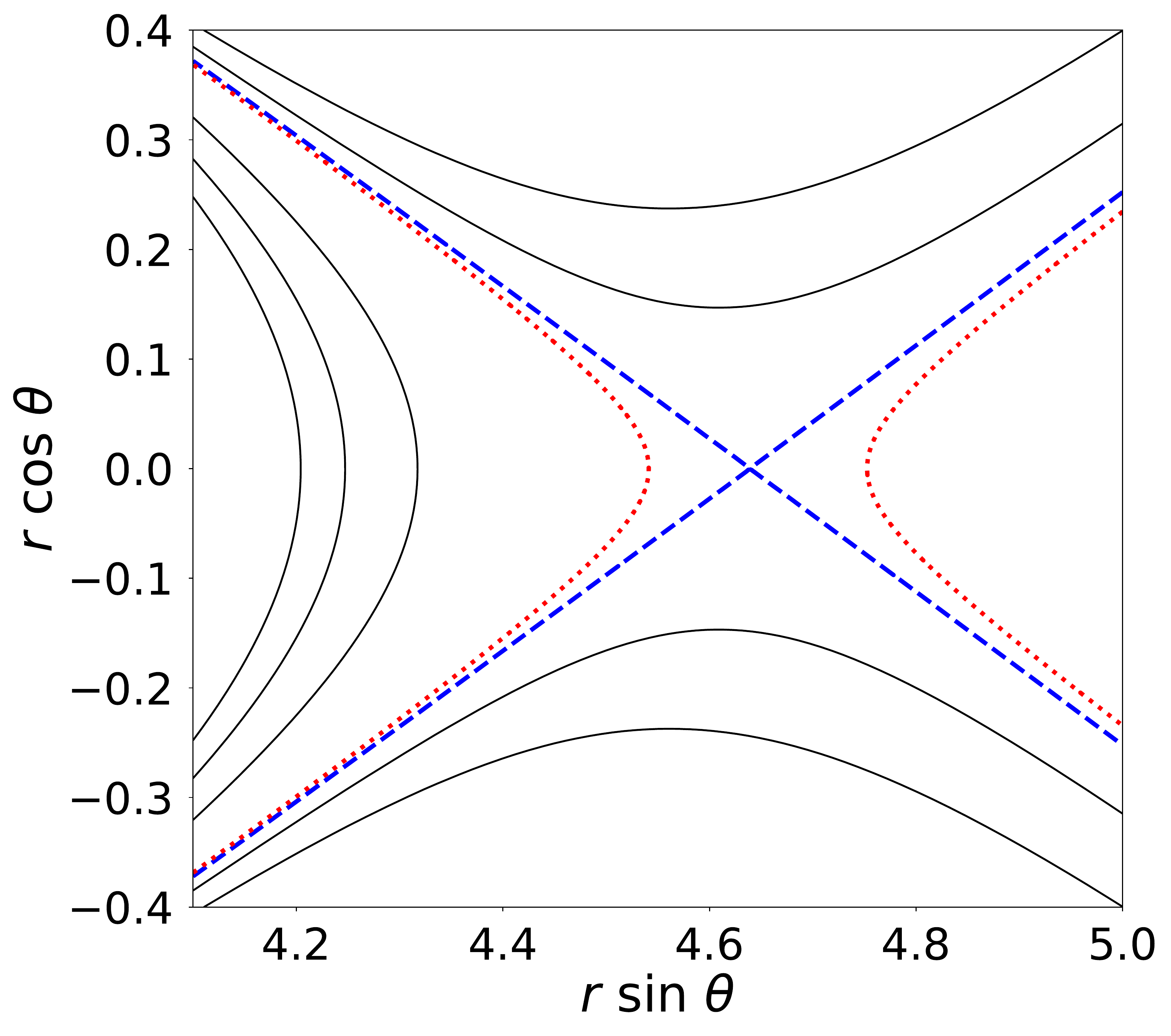}
\hspace{-0.3cm}
\includegraphics[scale=0.14]{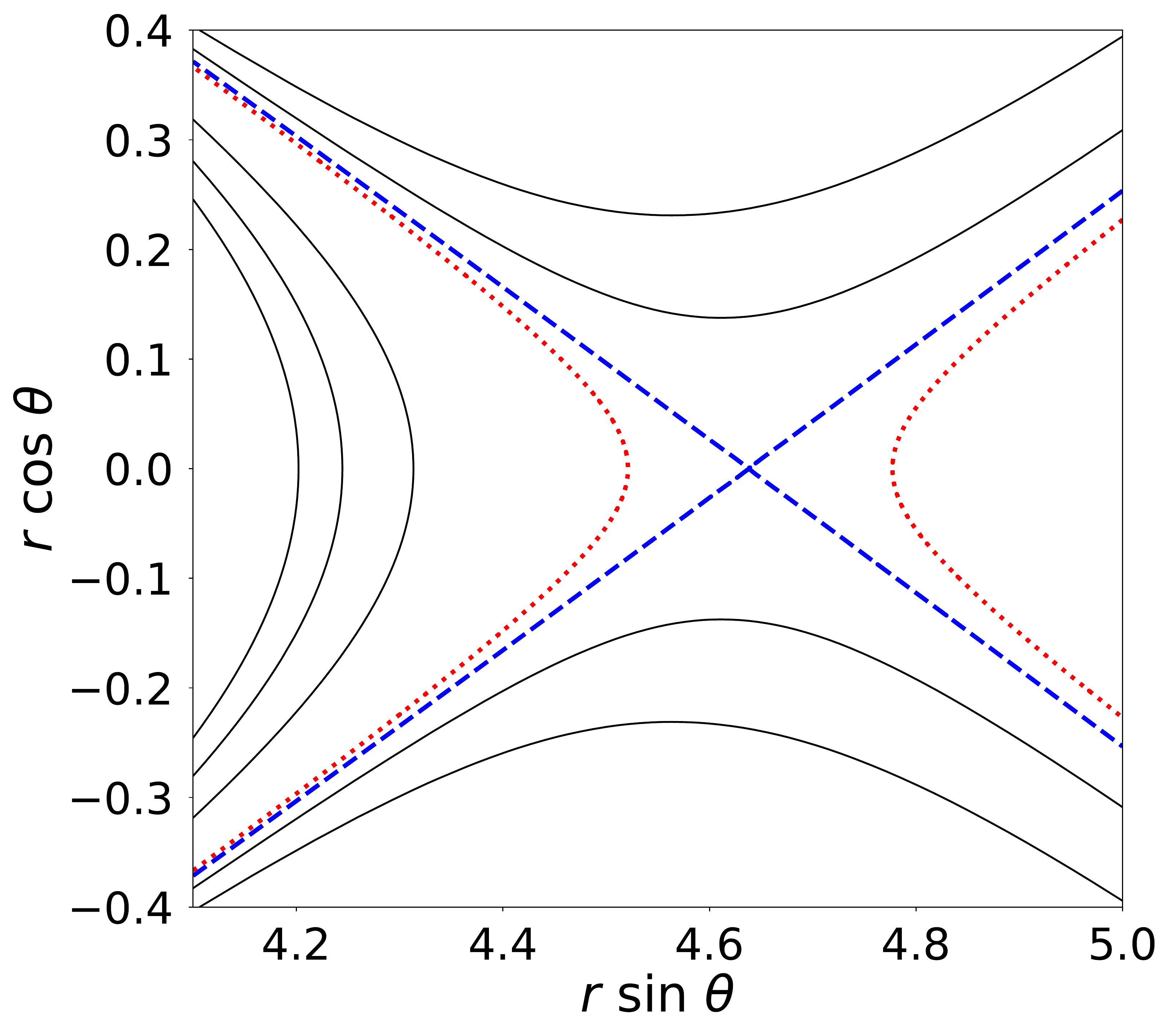}
\hspace{-0.2cm}
\includegraphics[scale=0.14]{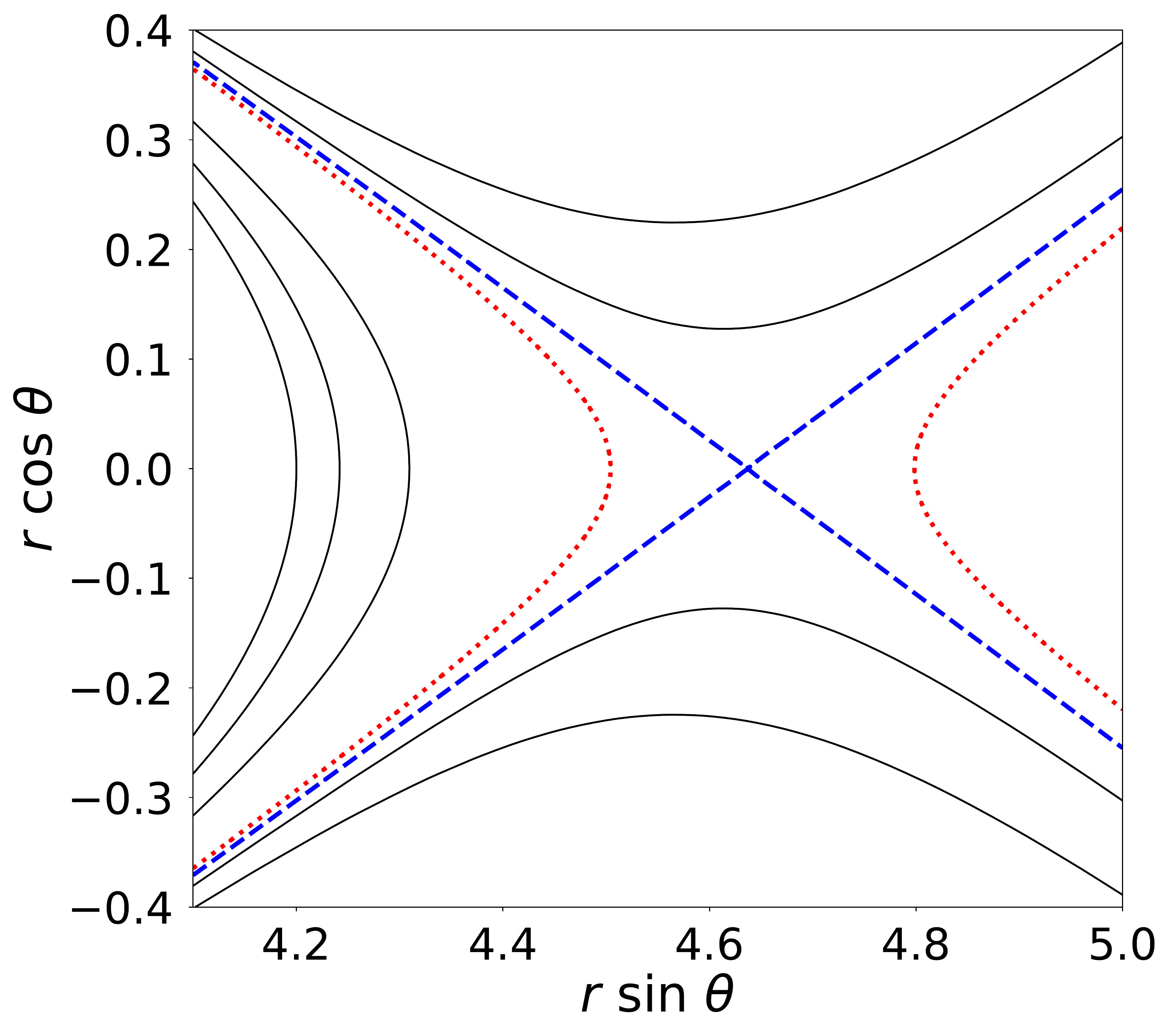}
\caption{Same as Fig.~\ref{isocontours_0039_3} but for $\beta_{\mathrm{m,c}} = 10^{-3}$. The three black isocontours in the left part of all plots represent values of the total pressure equal to $p_{\mathrm{t}_i} = ( p_{\mathrm{(0), max}} i - p_{\mathrm{(0), cusp}})/4$ for $i = 1, 2, 3$, where $p_{\mathrm{(0), max}}$ is the value of $p_{\mathrm{(0)}}$ at the maximum of the pressure.}
\label{isocontours_0039_4}
\end{figure*}

The change in pressure $\triangle p_{\rm cusp}$ at the newly formed cusp $\triangle r_{\rm cusp}$ of the magnetized disk in the presence of shear viscosity, as compared to the inviscid case, is determined in the following way,
\begin{eqnarray}
\triangle r_{\rm cusp} &=& \frac{r_{\rm cusp,new}-r_{\rm cusp}}{r_{\rm cusp}}\,,
\\
\triangle p_{\rm cusp} &=& \frac{p_{\rm t, cusp}-p_{(0),\rm cusp}}{p_{(0),\rm cusp}}\,,
\end{eqnarray} 
where $p_{\rm t}=p_{(0)}+p_{(1)}$ and $r_{\rm cusp,new}$ is the new location of the cusp due to  shear viscosity effects. Both $p_{\rm t}$ and $r_{\rm cusp,new}$ therefore contain all contributions from shear viscosity and spacetime curvature for various choices of input parameters $m_1$, $m_2$. The new position of the cusp at the equatorial plane corresponds to the location of the minimum of the total pressure $p_{\rm t}(r)$. We compute it by fitting the values of $p_{\rm t}$ using a third-order order spline interpolation. The values of $r_{\rm cusp,new}$ and $p_t(r_{\rm cusp,new})$ are obtained at the same time using this technique. For completeness, the locations of $r_{\rm cusp}$ and $p_{(0),\rm cusp}$ for an inviscid magnetized disk are reported in Table \ref{table-I}.

The allowed values of parameters $s_1$ and $m_2$ are reported in Table \ref{table-2} for all of our magnetized disk models. The range of variation of these parameters is $s_1=(0.001,0.005,0.01,0.05)$ and $m_2=(0,0.001,0.005,0.01,0.05)$. 
Forbidden values of $s_1$ and $m_2$ appear when $|\triangle p_{\rm cusp}|\gtrsim {\cal O} (1)$ (marked in boldface in Table~\ref{table-2}) implying ${p_{(1)}}/ {p_{(0)}} \sim {\cal O}(1) $. As $s_1$ (or $m_1$) and $m_2$ increase and the potential gap $\triangle W_{\mathrm{s}}$ decreases from $\triangle W_{\mathrm{s}}>0$ to $\triangle W_{\mathrm{s}} \approx 0$, the condition ${p_{(1)}} / {p_{(0)}} \sim {\cal O}(1) $ is more frequently satisfied. 
	 
A more concrete estimation of the allowed values of the parameters $s_1$ and $m_2$ with $\beta_{m,c}$ can be obtained from the 2D plot of $|\triangle p_{\rm cusp}|$  shown in figure \ref{deltapcusp_2D}. The black contour depicted in some of the plots in this figure indicates a cut-off value of $s_1$ and $m_2$ corresponding to $\log_{10}|\Delta p_{\mathrm{cusp}}|=0$. For low magnetized viscous disks ($\beta_{m,c}=10^{3}$, top panels), we find that the allowed values of $s_1$ and $m_2$ are large for $\Delta W_{\rm s}>0$ and that the permitted parameter space  of ($s_1, m_2$) appreciably decreases as the potential gap $\Delta W_{\rm s} \rightarrow 0$. This indicates that stationary magnetized disks with  $\Delta W_{\rm s} \approx 0$ do not allow for large shear viscosity and curvature effects in comparison to $\Delta W_{\rm s} > 0$. On the other hand, for highly magnetized disks ($\beta_{m,c} =10^{-3}$, bottom panels), stationary viscous models can be constructed over the entire choice of the parameter space and in  the considered regions of the potential gap i.e.~$\Delta W_{\rm s} > 0$ and $\Delta W_{\rm s} \approx 0$. Therefore, in order not to be in conflict with the adopted perturbative approach, our stationary models are restricted up to maximum values of $m_1 = s_1 = 0.05$ and $m_2 = 0.05$. Table~\ref{table-2} also shows that the changes in the location of the cusp positions are small for small values of $s_1$ and $m_2$. This behaviour remains the same for both low and high values of magnetization as well as for $\triangle W_{\rm s}>0$ and $\triangle W_{\rm s} \approx 0$.
 
Isocontours of the total pressure $p_{\rm t}$ of our stationary viscous tori are shown in figures \ref{isocontours_0039_3} and \ref{isocontours_0039_4} for low and high values of the central magnetization parameter, respectively. These figures concentrate on the regions close to the cusp of the disks since it is in those regions where the effects of the shear viscosity are most manifest. The self-intersecting contours of $p_{\rm t}$ possessing a cusp are depicted by the blue dashed curves in the figures for the values of $s_1$ and $m_2$ indicated in the captions. The red isocontours correspond to surfaces of constant pressure of magnetized ideal fluid disks which would self-intersect, had there been no dissipative effects in the disk. For a given value of $s_1$ and $W_{\rm s}$ we observe that when $m_2$ increases (from the top row to the bottom panels) the location of the newly formed cusp moves towards the black hole. At the same time the thickness of the cusp region in the disk also diminishes. This can be observed by looking at the change of location of the black isocontours located above and below the cusp region in figures \ref{isocontours_0039_3} and \ref{isocontours_0039_4}. These two iscontours correspond to the values of the total pressure $p_{\mathrm{t}} = 2p_{\mathrm{0}, \mathrm{cusp}}/3$ and $p_{\mathrm{t}} = p_{\mathrm{0}, \mathrm{cusp}}/3$. In particular, in Fig.~\ref{isocontours_0039_3} it can be seen that, in the bottom row and in the right column, the isocontour corresponding to $p_{\mathrm{t}} = 2p_{\mathrm{0}, \mathrm{cusp}}/3$, changes its position (from above and below the cusp, to the left and right of the cusp). This means that, for these cases, $p_{\mathrm{t, cusp}} < 2p_{\mathrm{0}, \mathrm{cusp}}/3$. In addition, the isocontour corresponding to $p_{\mathrm{t}} = p_{\mathrm{0}, \mathrm{cusp}}/3$ also moves significantly closer to the self-crossing surface.   
Therefore, within our framework based on causal relativistic hydrodynamics, the role of shear viscosity triggered by the curvature of the Schwarzschild black hole spacetime 
is apparent through a noticeable rearrangement of the constant pressure surfaces of magnetized viscous disks when compared to the purely inviscid case~\cite{lahiri2019toy}. In addition, the comparison of figures \ref{isocontours_0039_3} and \ref{isocontours_0039_4} shows that as the strength of the magnetic field increases the shift of the location of the cusp towards the black hole also  increases. This might have implications on the dynamical stability of constant angular momentum thick disks, mitigating the development of the so-called runaway instability that affects inviscid constant angular momentum tori~\cite{runaway,Font:2002}.

\section{Summary}
\label{summary}

We have discussed stationary solutions of magnetized, viscous thick accretion disks around a Schwarzschild black hole, neglecting the self-gravity of the tori and assuming that they are  endowed with a toroidal magnetic field and obey a constant angular momentum law. Our study has focused on the role of the spacetime curvature in the shear viscosity tensor and in the effects viscosity may have on the stationary solutions. This work is a generalization of a previous study for purely hydrodynamical disks presented in~\cite{lahiri2019toy}.

Following~\cite{lahiri2019toy} we have considered a simple framework to encapsulate the quantitative effects of the shear viscosity (neglecting any contributions of the heat flow) and the curvature of the background geometry. 
In this setup, both the shear viscosity and the curvature  have perturbative influences on the fluid, thereby allowing the fluid particles in the disk to undergo circular orbits.  In particular, the magnetic field distribution, the fluid pressure and the energy density (related to the pressure by a barotropic equation of state) are perturbatively modified due to dissipative effects. Our framework is based on causal relativistic hydrodynamics and uses the gradient expansion scheme up to second order such that the governing equations of motion of the fluid in the Eckart frame are hyperbolic. Within this approach the curvature of the background geometry, in which the accretion disk is situated, naturally appears in the equations of motion. In analogy with what was found in~\cite{lahiri2019toy} for unmagnetized tori, the present  work also shows that the viscosity and the curvature of the Schwarzschild black hole play some role on the morphology of magnetized tori.

The stationary models have been constructed by numerically solving the general relativistic momentum conservation equation using the method of characteristics. By varying the parameters $m_1$ and $m_2$ with two different choices of magnetization, we have studied the radial profiles of $p_{(0)}$ and $p_{(1)}$ to identify regions of the disk where shear viscosity and curvature are mostly casting their effects. Our results have revealed that the effects are most prominent near the cusp of the disk, which helped us focus our analysis on two regions of the potential gap, namely $\triangle W_{\mathrm{s}}>0$ and $\triangle W_{\mathrm{s}} \approx0$. Moreover, our study has allowed us to constrain the range of validity of the  second-order transport coefficients $m_1$ and $m_2$ (after setting $\tau_2 =1$). The allowed parameter space can be derived from figure  \ref{deltapcusp_2D} and from Table \ref{table-2}, where the bold-lettered values of $\triangle p_{ \rm cusp}$ for a given value of $W_{\mathrm{s}}$ mark the breakdown of the perturbative approach. Furthermore, the computations of $\triangle r_{\rm cusp}$ pinpoint the exact modification in the position of the cusp due to the shear viscosity and curvature effects.
 
The obtained isopressure contours of $p_{\rm t}$ corresponding to $\triangle W_{\mathrm{s}} >0 $ further divulge the cumulative effects of the viscosity and curvature on the magnetized disk. The self-intersection of these isopressure contours indicate new locations of the cusp as well as the formation of a new $p_{\rm cusp}$. We have found that for each magnetization and $\triangle W_{\rm s}$ considered, the location of cusps moves towards the black hole as parameter $m_2$ increases. Moreover, for higher magnetized disks the shift is even larger. Therefore, the combined effects of shear viscosity  and spacetime curvature  might help mitigate, or even suppress, the development of the runaway instability in constant angular momentum tori~\cite{runaway,Font:2002}, a conclusion that is at par with the assumptions of our setup.
 
The present work is a small step towards constructing stationary models of viscous magnetized tori based on a causal approach for relativistic hydrodynamics. Despite our simplistic approach we have shown here that the morphology of  geometrically thick accretion disks is non-trivially affected by viscosity and curvature. These effects, though small, should not be neglected.
In particular, they could
potentially alter the radiation profiles of magnetized accretion tori. As an example~\cite{2015A&A...574A..48V}
discussed magnetised Polish doughnuts
using Kommissarov's approach~\cite{Komissarov:2006} including radiation. However, they did not treat
dissipation or shear stresses from first principles as in the current work but used, instead, an ad-hoc parameterisation to allow the gas to be non-ideal.
It would be interesting to employ the second-order gradient approximation scheme discussed here to determine the temperature dependence in magnetized viscous tori from first principles and then examine the associated radiation spectra as the spectral properties are directly influenced by hydrodynamic and thermodynamic structures of the disks. Likewise, the intensity and emission lines of viscous magnetized tori are expected to show imprints of shear viscosity and curvature~\cite{2015A&A...574A..48V,Straub:2012uv}. Similarly, another system worth analysing would be a  thick disk with advection dominated flows, as discussed by \cite{2009MNRAS.400..422G}, since  the viscous heating rate might be modified when using the present form of the shear viscosity tensor. Investigating these various possibilities will be the target of future studies.

 Finally, to actually observe the consequences of dissipative flux quantities in detail, a more realistic construction is required. That would 
involve taking into account the contributions of the heat flux and of the radial velocity of the fluid. Ultimately,  considering dissipative flux quantities to behave as perturbations is an assumption that should also be relaxed.

\section*{Acknowledgements}
 The authors gratefully thank the anonymous referee for illuminating suggestions.
The work of SL is supported by the ERC Synergy Grant “BlackHoleCam: Imaging the Event Horizon of Black Holes” (Grant No. 610058). This work is further supported by the Spanish Agencia Estatal de Investigaci\'on (grants PGC2018-095984-B-I00 and PID2019-108995GB-C22), by the Generalitat Valenciana (grants PROMETEO/2019/071 and CIDEGENT/2018/021), and by the European Union's Horizon 2020 Research and Innovation (RISE) programme H2020-MSCA-RISE-2017 Grant No.~FunFiCO-777740.

\bibliography{references}
\bibliographystyle{apsrev4-1}

	\end{document}